\documentclass[aps,prd,twocolumn,twoside,superscriptaddress,groupedaddress,amsmath,amssymb,nofootinbib,preprintnumbers,letterpaper,floatfix]{revtex4-2}

\usepackage{ragged2e}
\usepackage{wrapfig}
\usepackage{dblfloatfix}
\usepackage{placeins}

\usepackage{amsmath,amssymb,amsfonts}
\usepackage{rviewport}
\usepackage{graphicx}
\usepackage[dvipsnames]{xcolor}
\usepackage{xspace}
\usepackage{physics}
\usepackage{pgffor}
\usepackage{afterpage}
\usepackage{comment}
\usepackage{dcolumn}
\usepackage{enumitem}
\usepackage{balance}
\usepackage{booktabs}
\usepackage{newtxtext,newtxmath}
\usepackage[normalem]{ulem}

\usepackage[hidelinks,bookmarks]{hyperref}
\hypersetup{
  bookmarks=true,
  bookmarksopen=true,
  bookmarksopenlevel=1,
  colorlinks=true,
  linkcolor=RoyalBlue,
  urlcolor=RoyalBlue,
  citecolor=violet,
}
\usepackage[nameinlink,capitalize]{cleveref}
\crefname{equation}{Eq.}{Eqs.}
\Crefname{equation}{Equation}{Equations}

\crefname{figure}{Fig.}{Figs.}
\Crefname{figure}{Figure}{Figures}

\crefname{section}{Sec.}{Secs.}
\Crefname{section}{Section}{Sections}

\crefname{table}{Table}{Tables}
\Crefname{table}{Table}{Tables}

\newcommand{\gcm}{\ensuremath{\text{g\,cm}^{-2}}\xspace}
\newcommand{\Xmax}{\ensuremath{X_\text{max}}\xspace}

\newcommand{\boldXmax}{\ensuremath{\boldsymbol X_\textbf{max}}\xspace}
\newcommand{\meanXmax}{\ensuremath{\langle X_\text{max}\rangle}\xspace}
\newcommand{\sigmaXmax}{\ensuremath{\sigma(X_\text{max})}\xspace}

\newcommand{\Xlow}{\ensuremath{X_\text{low}}\xspace}
\newcommand{\Xup}{\ensuremath{X_\text{up}}\xspace}

\newcommand{\lnA}{\ensuremath{\ln A}\xspace}
\newcommand{\meanlnA}{\ensuremath{\langle\lnA\rangle}\xspace}
\newcommand{\varlnA}{\ensuremath{\text{Var}(\lnA)}\xspace}
\newcommand{\lgE}{\ensuremath{\lg(E/\text{eV})}\xspace}
\newcommand{\lgErange}[2]{\ensuremath{#1 \leq \lg(E/\text{eV}) < #2}\xspace}
\newcommand{\PAO}{Pierre Auger Observatory\xspace}

\newcommand{\Sibyll}{\textsc{Sibyll\,2.3}d\xspace}
\newcommand{\EPOSR}{\textsc{EPOS\,LHC-R}\xspace}
\newcommand{\QGSJetIII}{\textsc{QGSJet\,III-01}\xspace}
\newcommand{\Sibylle}{\textsc{Sibyll\,2.3}e\xspace}
\newcommand{\Conex}{\textsc{Conex}\xspace}

\newcommand{\allCandidates}{5\,311\,333\xspace}
\newcommand{\allPresel}{281\,562\xspace}
\newcommand{\allFDEvents}{58\,887\xspace}
\newcommand{\allEASEvents}{58\,453\xspace}

\DeclareMathOperator{\med}{median}

\definecolor{darkgreen}{rgb}{0.0, 0.5, 0.0}

\def\Offline{\mbox{$\overline{\textrm{Off}}$\hspace{.05em}\protect\raisebox{.4ex}{$\protect\underline{\textrm{line}}$}}\xspace}

\begin{document}

\title{Depth of Maximum of Air-Shower Profiles above
  \texorpdfstring{10$^{\boldsymbol{17.7}}$\,eV}{10\textasciicircum{}{17.7} eV}\texorpdfstring{\\}{} Measured with the Fluorescence Detector of the Pierre Auger Observatory}


\collaboration{\hyperref[sec:PAcollaboration]{Pierre Auger Collaboration}}

\date{\today}

\begin{abstract}
We present measurements of the depth of shower maximum, $\Xmax$, for
cosmic-ray-induced extensive air showers recorded by the fluorescence
detector of the Pierre Auger Observatory over 17 years. The data set
covers primary energies from $10^{\smash{17.7}}$\,eV to beyond
$10^{\smash{19.6}}$\,eV.  With improved event reconstruction and an
exposure 2.4 times larger than in our previous analysis, this work
confirms and refines our conclusions on the mass composition at
ultra-high energies.  The energy evolution of the mean $\Xmax$
exhibits a pronounced break at around $10^{\smash{18.4}}$\,eV,
providing direct, model-independent evidence for a change in the
evolution of the mass composition. Independently, the observed
decrease of the $\Xmax$ fluctuations with energy indicates a
transition toward a heavier and less diverse primary mass
composition. No statistically significant declination dependence of the
$\Xmax$ distributions is observed within the exposure of the
Observatory, indicating an isotropic mass composition.

The mean and standard deviation of the $\Xmax$ distributions,
interpreted with air-shower simulations, yield the energy dependence
of the average and variance of the logarithmic mass of cosmic rays
arriving at Earth. Furthermore, energy-dependent fractional abundances
of four representative primary-mass groups (p, He, CNO, Fe) are
obtained by fitting the observed $\Xmax$ distributions in each energy
bin with a weighted sum of elemental templates.

These results provide strong evidence against a long-standing
assumption that ultra-high-energy cosmic rays are predominantly
protons: above ${\sim}10^{\smash{18.4}}$\,eV, the average cosmic-ray
mass increases, accompanied by a steadily decreasing diversity in the
elemental composition.
\end{abstract}

\maketitle

\section{Introduction}

The elemental composition of ultra-high-energy cosmic rays (UHECRs) is a fundamental observable for understanding their origin, providing key insights into the astrophysical sources and acceleration mechanisms that produce these rare, energetic particles (see, e.g.,\ Ref.~\cite{Globus:2025ftu} for a recent review).
However, determining the mass composition of UHECRs is challenging.
When a primary cosmic-ray particle enters the atmosphere, it interacts with air nuclei and initiates an extensive air shower.
The longitudinal particle profile of an air shower peaks once the cascade reaches a certain slant depth, at which the average electromagnetic energy is so low that most energy is lost through ionization rather than through the production of new particles.
This depth of shower maximum, $\Xmax$, is correlated with the mass of the primary cosmic ray \cite{Greisen:1960wc,Linsley:1981gh,Engel:1992vf,Matthews:2005sd,Kampert:2012mx,PierreAuger:2013xim}.
However, reconstructing the mass of individual events from $\Xmax$ alone is impossible because of large event-to-event fluctuations in the shower development.
Moreover, despite substantial progress in modeling hadronic interactions at ultra-high energies \cite{Ostapchenko:2010gt,Pierog:2013ria,Riehn:2019jet,Ostapchenko:2024myl,Pierog:2023ahq}, air-shower physics at energies beyond terrestrial accelerators still carries significant and difficult-to-quantify systematic uncertainties, see, e.g., Ref.~\cite{Albrecht:2025kbb}.

Experimentally, these particle cascades can be investigated using two
complementary detection methods: ground-based surface detectors, which
sample the secondary particles that reach the Earth's surface, and
fluorescence telescopes, which record the ultraviolet light emitted
when atmospheric nitrogen molecules return to their ground state after
being excited by secondary particles.  Thus, fluorescence telescopes
can directly observe the atmospheric development of air showers,
providing a unique advantage for mass-composition studies, despite the
limited duty cycle of ${\sim} 15\%$ for ground-based instruments
during clear, moonless nights.

In this paper, we present a detailed analysis of the distributions of the depth of shower maximum ($\Xmax$), recorded by the fluorescence telescopes of the Pierre Auger Observatory~\cite{PierreAuger:2015eyc}.
This study expands upon our previous mass-composition analyses with the fluorescence detector of the Observatory~\cite{PierreAuger:2010ymv,PierreAuger:2014sui,PierreAuger:2014gko}, incorporating improved event reconstruction methods and utilizing the full data set collected from 2004 to 2021, prior to the AugerPrime surface-detector upgrade~\cite{PierreAuger:2016qzd}.

The paper is structured as follows: \cref{sec:PAO} briefly describes
the Pierre Auger Observatory. The data analysis and event selection
are described in \cref{sec:data}, while detector effects including
acceptance, resolution, reconstruction bias, and the systematic
uncertainty on the \Xmax scale are characterized in
\cref{sec:detector}. The robustness of the analysis is assessed in
\cref{sec:crosschecks} through a series of validation
studies. Measurements of the $\Xmax$ moments and their interpretation
in terms of the moments of logarithmic cosmic-ray mass are presented
in \cref{sec:moments}. Finally, the $\Xmax$ distributions are fitted
in \cref{sec:fractions} to extract the energy-dependent fractional
composition of cosmic-ray primaries, and \cref{sec:conclusions}
summarizes the results.

\section{\label{sec:PAO} The Pierre Auger Observatory}

The \PAO, the largest cosmic-ray observatory ever built, spans an area of 3000\,km$^2$ near Malarg\"ue, Argentina, at an average altitude of 1400 meters above sea level~\cite{PierreAuger:2015eyc}.
This paper focuses on the data collected using the Surface Detector (SD)~\cite{PierreAuger:2007kus} and Fluorescence Detector (FD)~\cite{PierreAuger:2009esk} of the Observatory in their baseline configuration~\cite{PierreAuger:2004naf}.
In this configuration, the SD comprises 1600 water-Cherenkov stations arranged in a triangular grid with 1500\,m spacing. The FD consists of 24 fluorescence telescopes installed in groups of six at four sites around the SD, with an elevation coverage from $1.5^\circ$ to $30^\circ$. The data collected using low-energy enhancements, namely the SD region with a 750\,m grid size~\cite{Sanchez:2011zza} and the high-elevation telescopes with an elevation range from $30^\circ$ to $58^\circ$~\cite{Mathes:2011zz}, will be utilized in a future paper to extend $\Xmax$ measurements down to an energy of about $10^{17.2}$~eV.

To ensure a high-quality reconstruction of the shower energy and $\Xmax$, a reliable calibration of the fluorescence telescopes and a solid knowledge of atmospheric conditions at the time of observation are necessary.
An absolute calibration and nightly relative calibrations are required to estimate the number of photons at the aperture of the telescope from the ADC count in the PMTs. The absolute calibration is performed using a Lambertian light source of known intensity, while LEDs mounted on the mirrors are for the nightly relative calibration~\cite{Brack:2004af,Rovero:2008epr,Brack:2013bta}.
The Global Data Assimilation System (GDAS) provides atmospheric molecular density profiles on a 3-hour basis \cite{PierreAuger:2012jsu}.
The aerosol content of the atmosphere is monitored constantly during observations via vertical laser shots from which the density profiles of aerosols
above the array are derived for each fluorescence telescope site on an hourly basis~\cite{Fick:2006faa,PierreAuger:2010uyt}. Here we use the recently improved aerosol determination described in Ref.~\cite{PierreAuger:2023nbk}.
Finally, clouds are monitored both using an infrared camera at each FD site \cite{Chirinos:2013uty} and from the Geostationary Operational Environmental Satellites (GOES) \cite{PierreAuger:2013lgb}.

The ground-level impact point and arrival direction of each event are
reconstructed using a hybrid method~\cite{Sommers:1995dm,Dawson:1996ci}, combining the classical FD geometry fit~\cite{1968CaJPS..46..266B} with the position and time of particles recorded in the SD station nearest to the shower axis, yielding substantially improved geometric reconstruction~\cite{Dawson:1996ci}. The calorimetric
energy~\cite{Risse:2003fw} and the depth of shower maximum are
determined in a likelihood fit, based on Ref.~\cite{Unger:2008uq}, to
the photoelectrons detected by the FD telescopes.  A Gaisser-Hillas
function~\cite{gaisser_reliability_1977} is used to describe the
longitudinal energy-deposit profile. From this profile, we compute the
expected fluorescence and Cherenkov light, propagate it through the
atmosphere, and fold it with the optical response of the telescopes to
obtain the predicted photoelectrons which are compared with the measurements.
We employ the modified representation of the Gaisser-Hillas
function~\cite{Andringa:2011zz}, which allows for improved constraints
on the shape parameters of the longitudinal profile, as discussed in
Ref.~\cite{PierreAuger:2023att}.

The total energy of the shower is obtained by correcting the
calorimetric energy for the fraction carried away by neutrinos and
high-energy muons, following Ref.~\cite{PierreAuger:2019dhr}. The
combined uncertainties from calibration, atmospheric monitoring, and
event reconstruction contribute to a systematic uncertainty of 14\% on
the energy determination~\cite{PierreAuger:2020qqz}.

\section{\label{sec:data} Data selection}

The data set includes events recorded using the FD and SD between
December 1, 2004, and December 31, 2021.  The event selection and
analysis are based on our previous publication~\cite{PierreAuger:2014sui}, which
provides more comprehensive details on all key steps.  The
data set contains \allCandidates air-shower candidates, from which we
select \allEASEvents high-quality air-shower events using the cuts listed in
\cref{tab:selection} and described below.

\begin{table}[!t]
\centering
  \caption{\label{tab:selection} Selection criteria, number of events passing the selection, and efficiency $\varepsilon$ with respect to the previous step. The last row reports the number of unique air showers obtained after merging high-quality FD stereo events.}
  \renewcommand{\tabcolsep}{10pt}
  \begin{tabular}{lrc}
  \toprule
   & \multicolumn{1}{l}{events} & $\varepsilon$ / \%\\ \midrule
  air-shower candidates & \allCandidates & -- \\
  \multicolumn{3}{l}{\emph{pre-selection:}}\\
  hardware status & 4\,283\,450 & 80.6\\
  aerosols &  3\,479\,008 &  81.2 \\
  hybrid geometry & 1\,249\,849 & 35.9 \\
  profile reconstruction & 1\,038\,782 & 83.1\\
  clouds & 769\,699 & 74.1\\
  $E \geq 10^{17.7}$\,eV & \allPresel & 36.6\\
  \multicolumn{3}{l}{\emph{quality and fiducial selection:}}\\
  $P(\text{hybrid})$ & 277\,378 & 98.5\\
  \Xmax observed & 218\,904 & 78.9 \\
  quality & 156\,015 & 71.3\\
  fiducial field of view & 60\,189 & 38.6 \\
  profile &  \allFDEvents & 97.8 \\
  \midrule
  unique air-shower events & \allEASEvents &\\ \bottomrule
\end{tabular}
\end{table}

\subsection{Pre-selection}
The first stage of the pre-selection focuses on the hardware status of
the FD telescopes.  Candidate events are retained only for time
periods in which the telescope optics are properly aligned, the PMT
gains of the cameras are well calibrated, and the FD-SD timing is
sufficiently synchronized to allow an accurate hybrid
reconstruction. These conditions are verified using the electronic
logs and slow-control data associated with each event.

Next, the atmospheric conditions are assessed, since poor atmospheric
quality can introduce significant uncertainties in the correction for
fluorescence-light attenuation.  If no aerosol measurement from the
laser facilities is available within one hour of the event time, the
candidate is rejected. Otherwise, the vertical aerosol optical depth
(VAOD), integrated from ground level up to 3\,km altitude, must be
below 0.1, ensuring a vertical transmission of about 90\%.  Higher VAOD values lead to substantial and
difficult-to-correct attenuation of the fluorescence
light~\cite{PierreAuger:2010uyt}.

An event is retained only if at least one SD station is triggered within the timing and geometric window consistent with the FD observation, ensuring that a hybrid geometry reconstruction is possible.
Events rejected by this criterion are
typically low-energy showers with insufficient ground-level particle
density to trigger the SD-1500 array~\cite{PierreAuger:2011dqc}, or
meteorological events that satisfy the FD trigger but cannot be
reconstructed. A successful reconstruction of the longitudinal
energy-deposit profile is also required.

To exclude air showers for which the light may be obscured or
reflected by clouds, cloud coverage is monitored using the devices
described in \cref{sec:PAO}\@. Events are retained if no clouds are
detected in the shower direction in either the telescope projection
(cloud infrared camera) or the ground-level projection (GOES), or if the
average cloud fraction measured by the LIDARs is below
25\%~\cite{Harvey:2021vut}.

Finally, the analysis is restricted
to events with energies above $10^{17.7}$\,eV. Below this
threshold, the enlarged field of view of the high-elevation telescopes is needed, which will be addressed in a separate paper.

After this pre-selection, \allPresel{} air-shower events remain for
further analysis.

\subsection{Quality selection}
The next selection cut aims to reduce a possible bias arising from the
different probabilities for a single-station SD trigger in proton- and
iron-induced air showers.  For each event, this single-station probability is
estimated from its energy, zenith angle, and core location using the
data-driven lateral trigger probability described in
Ref.~\cite{PierreAuger:2011dqc}. This probability is required to exceed 95\%.

Furthermore, we accept only events whose longitudinal profiles include
the shower maximum within the field of view of the telescopes, since
\Xmax cannot be reliably determined when only the rising or falling
edge of the profile is observed.

The next step removes events with large expected uncertainties in the
\Xmax reconstruction. The expected uncertainty, $\hat{\sigma}$, is
computed from the event energy and geometry, and only events with
$\hat{\sigma} \leq 40\,\gcm$ are retained. In addition, showers for
which the minimum observation angle, $\alpha_\text{min}$, with respect
to the shower axis is smaller than $20^\circ$ are discarded. The
significant contribution of direct Cherenkov light and the short time
duration of the detected light at such small viewing angles can lead
to substantial \Xmax biases.

After the fiducial cuts described below in \cref{sec:fidselect} are applied, an additional final set of quality cuts is applied to the reconstructed energy-deposit profile to ensure an accurate determination of \Xmax and energy.
The minimum
observed profile length must exceed 200\,\gcm, and any gaps within the
profile, such as those arising in telescope-crossing showers, must
constitute less than 20\% of the total profile length. Profiles
distorted by residual cloud or aerosol contamination cannot be fitted
reliably.
To remove such cases, the goodness-of-fit of the Gaisser-Hillas
profile is evaluated using the standard-normal transformation of the
fit $\chi^2$, $z = (\chi^2 - \mathrm{ndf}) / \sqrt{2\,\mathrm{ndf}}$.
Events with $z \gtrsim 2.7$, where a non-Gaussian tail of poor-quality
fits is observed, are excluded. This cut differs from that used in our
previous analysis~\cite{PierreAuger:2014sui} because of the change in the
Gaisser--Hillas parametrization described in
Ref.~\cite{PierreAuger:2023att}.

\begin{figure}[t] \centering
    \includegraphics[width=\linewidth]{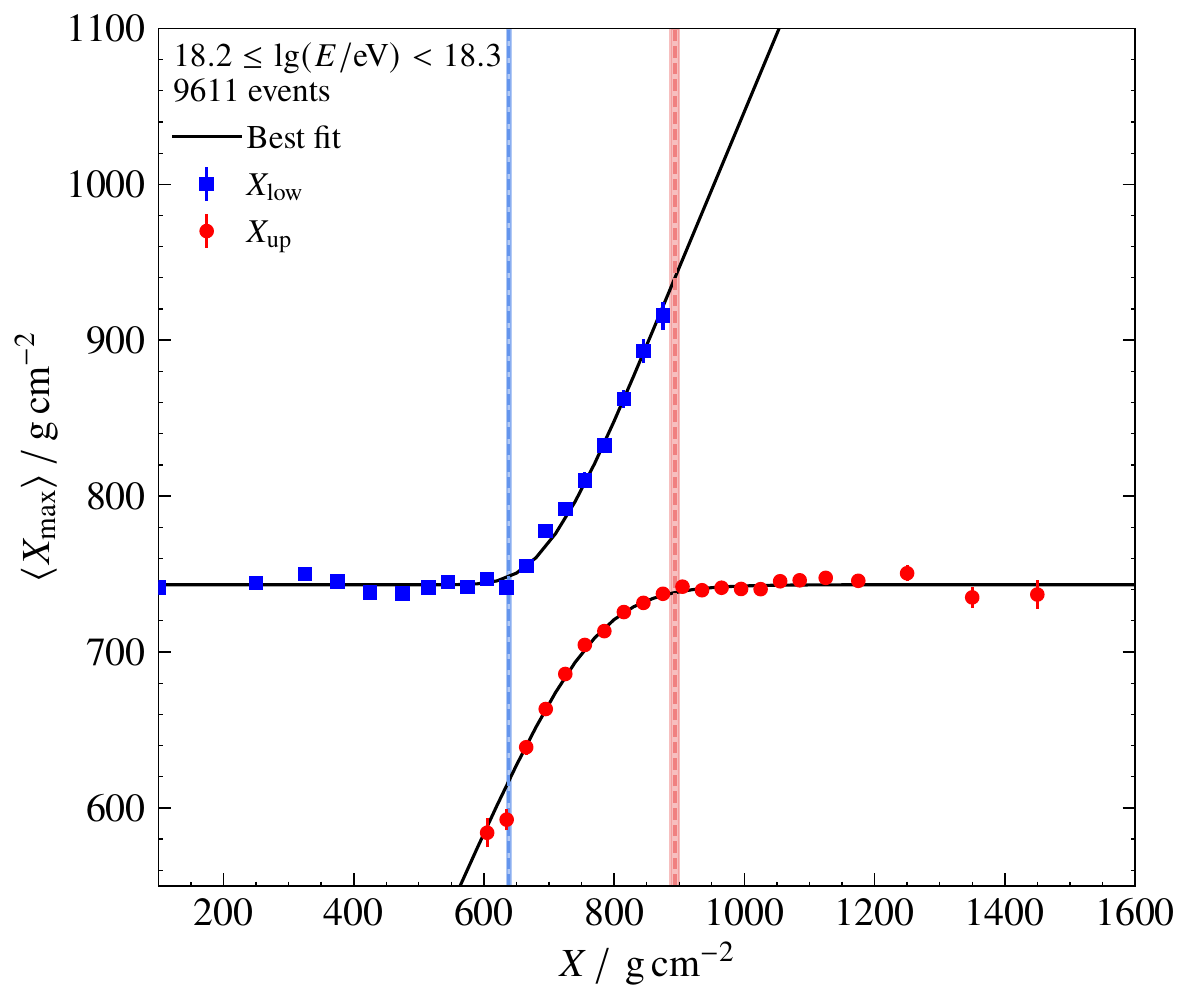}
    \caption{Determination of the fiducial field-of-view limits
      ($\Xlow^\mathrm{cut},\,\Xup^\mathrm{cut}$), marked by the
      vertical lines, for the energy bin
      $18.2\leq\lg(E/\mathrm{eV})<18.3$. Data points show the truncated
      mean of the \Xmax distribution above $\Xlow$ and below $\Xup$.}
    \label{fig:fidFoV}
\end{figure}

\subsection{Fiducial selection}\label{sec:fidselect}

With the selection described so far, the data set consists only of
events with a high-quality reconstruction. However, as discussed in
Ref.~\cite{PierreAuger:2014sui}, the field of view of the FD telescopes introduces a
selection bias that depends on the shower \Xmax, core position,
arrival direction, and energy. To minimize this bias, a slant-depth
interval $(\Xlow, \Xup)$ is calculated for each event such that any
\Xmax lying within that interval would satisfy the dominant \Xmax
quality requirements, $\hat{\sigma}<40\,\gcm$ and
$\alpha_\text{min}>20^\circ$. An event is accepted only if its
$(\Xlow, \Xup)$ interval is sufficiently large to accommodate the bulk
of the \Xmax distribution in the data~\cite{PierreAuger:2014sui}.

\begin{figure}[t] \centering
    \includegraphics[width=\linewidth]{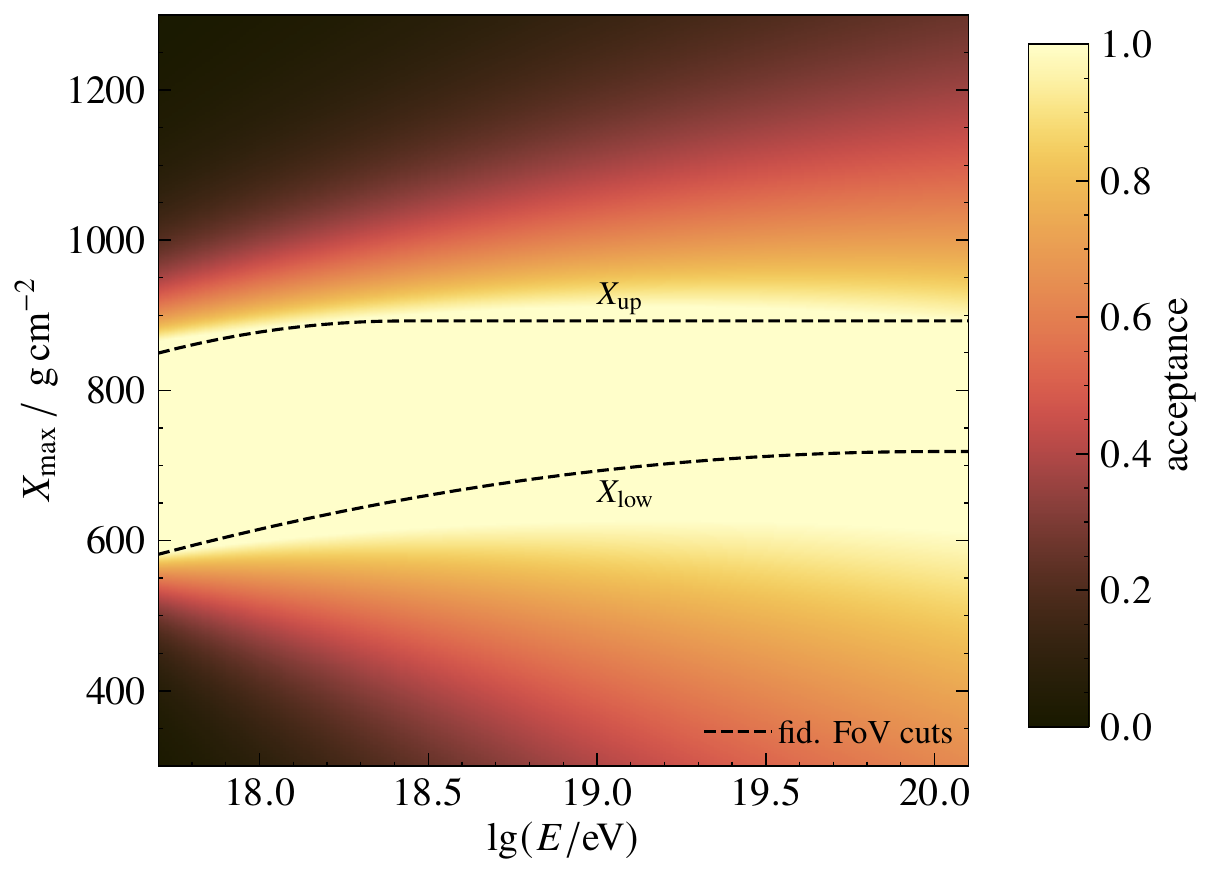}
    \caption{Fiducial field-of-view boundaries from \cref{eq:fidfov} (black dashed lines) and $\Xmax$ acceptance (color scale) as a function of energy.}
    \label{fig:acceptance_E}
\end{figure}

Since the true \Xmax distribution is not known at this stage, the
fiducial field-of-view boundaries are determined by studying the bias
on \meanXmax as a function of $(\Xlow, \Xup)$ for each energy bin, an
example of which is shown in \cref{fig:fidFoV}. Once the field of view
begins truncating the \Xmax distribution, the observed \meanXmax
deviates from its unbiased asymptotic value. The fiducial field-of-view
boundaries, indicated by the vertical lines in
\cref{fig:fidFoV}, correspond to a \meanXmax bias of 5\,\gcm. The
energy dependence of these boundaries is parameterized as:
\begin{equation}
    \label{eq:fidfov}
    X_\mathrm{up/low}^\mathrm{fid}(E) =
    \begin{cases}
        X_0 + K\,\lg^2(E/E_0) & \text{if } E \leq E_0, \\
        X_0 & \text{otherwise}.
    \end{cases}
\end{equation}
The fit of $(X_0$, $K$, $\lg E_0)$ yields values of $(892.51,$ $-80.39,$
$18.43)$ for $\Xup^\mathrm{cut}$ and $(718.56,$ $-25.88,$ $20.00)$ for
$\Xlow^\mathrm{cut}$, where $X_0$ and $K$ are given in units of \gcm,
and $E_0$ in eV.  An event of energy $E$ is retained if $\Xlow <
\Xlow^\mathrm{cut}(E)$ and $\Xup > \Xup^\mathrm{cut}(E)$. The behavior
of the field-of-view boundaries as a function of energy is shown by
the black dashed lines in \cref{fig:acceptance_E}.

\subsection{Final data set}

\begin{figure*}[t]
    \centering
    \includegraphics[width=0.8\linewidth]{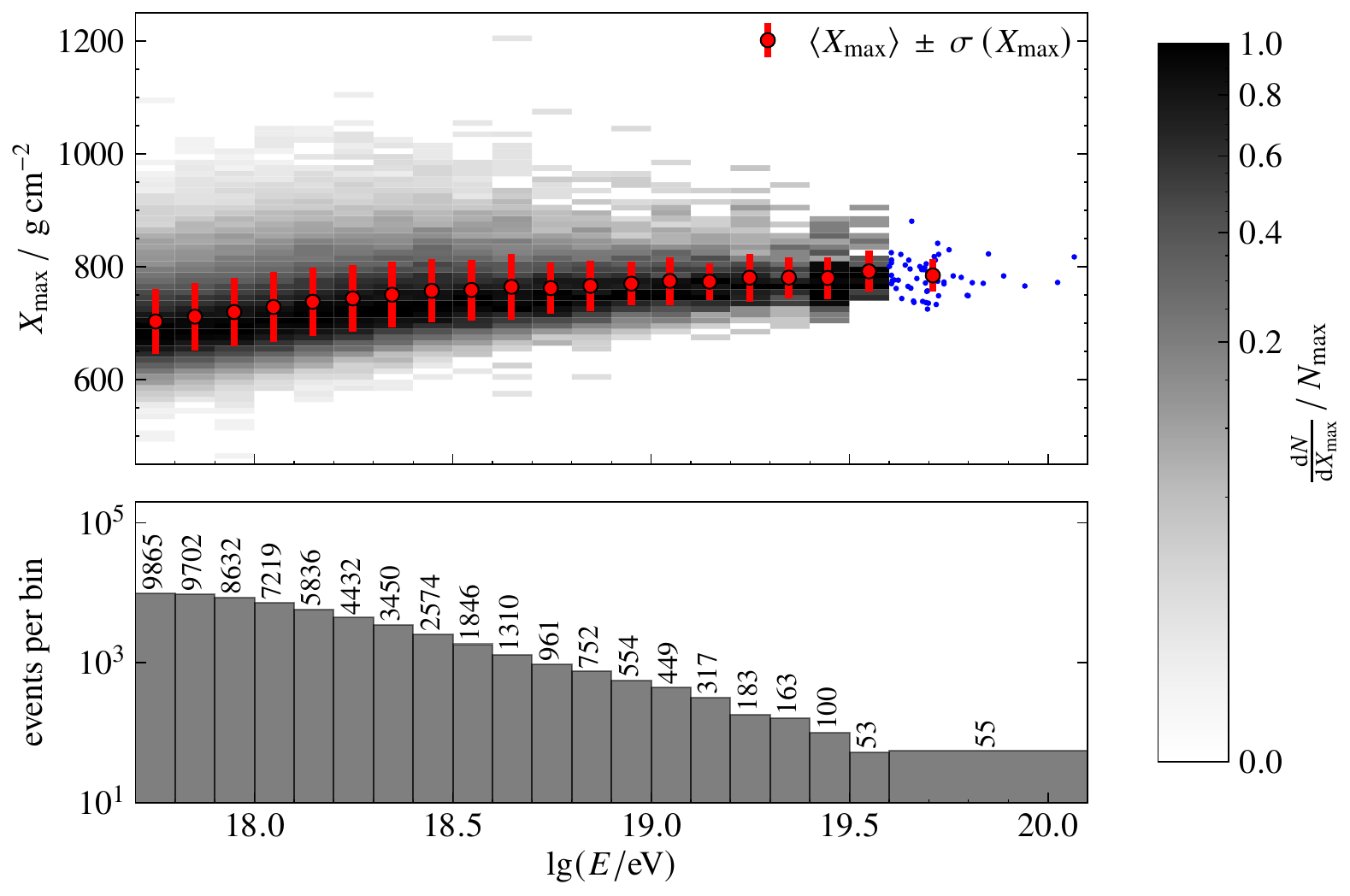}
    \caption{Top panel: Measured $\Xmax^{}$ and energy of the events
      selected for this analysis.  The event density is shown in
      gray scale, normalized to the entries of the maximum bin at each
      energy. Individual events above $10^{19.6}$\,eV are displayed as
      blue points. The large red points represent the mean of
      the $\Xmax^{}$ distribution in each energy bin, and the error bars
      indicate its standard deviation. Bottom panel: Number of
      selected events as a function of energy.}
    \label{fig:final_selection}
\end{figure*}

After this rigorous selection process, \allFDEvents{} events from individual FD
sites remain in the final data set, corresponding to about 1\% of the
initial air-shower candidates. Events observed by more than one FD
site are merged by calculating uncertainty-weighted averages of
their $\Xmax$ and energy.  This procedure yields a final high-quality data set of
\allEASEvents{} unique air-shower events.  The $\Xmax$ and energy
distributions of these events are shown in
\cref{fig:final_selection}. For illustration, the $\meanXmax$ and
$\sigmaXmax$ values computed later in this paper are also
displayed. The distribution of \Xmax in energy bins of 0.1 in \lgE
are displayed in \cref{fig:Xmax_distributions}. The last energy range
contains all events above $10^{19.6}$\,eV.  The energy of the
highest-energy event is $(1.17 \pm 0.07\,(\text{stat})) \times
10^{20}$\,eV and its $\Xmax$ is $817 \pm 25\,(\text{stat})$\,\gcm.

A list of all selected events is available from Ref.~\cite{supplementaryMaterial}.

\section{\label{sec:detector} Detector effects and systematics}

\begin{figure*}[b!]
    \centering
    \includegraphics[width=.8\linewidth]{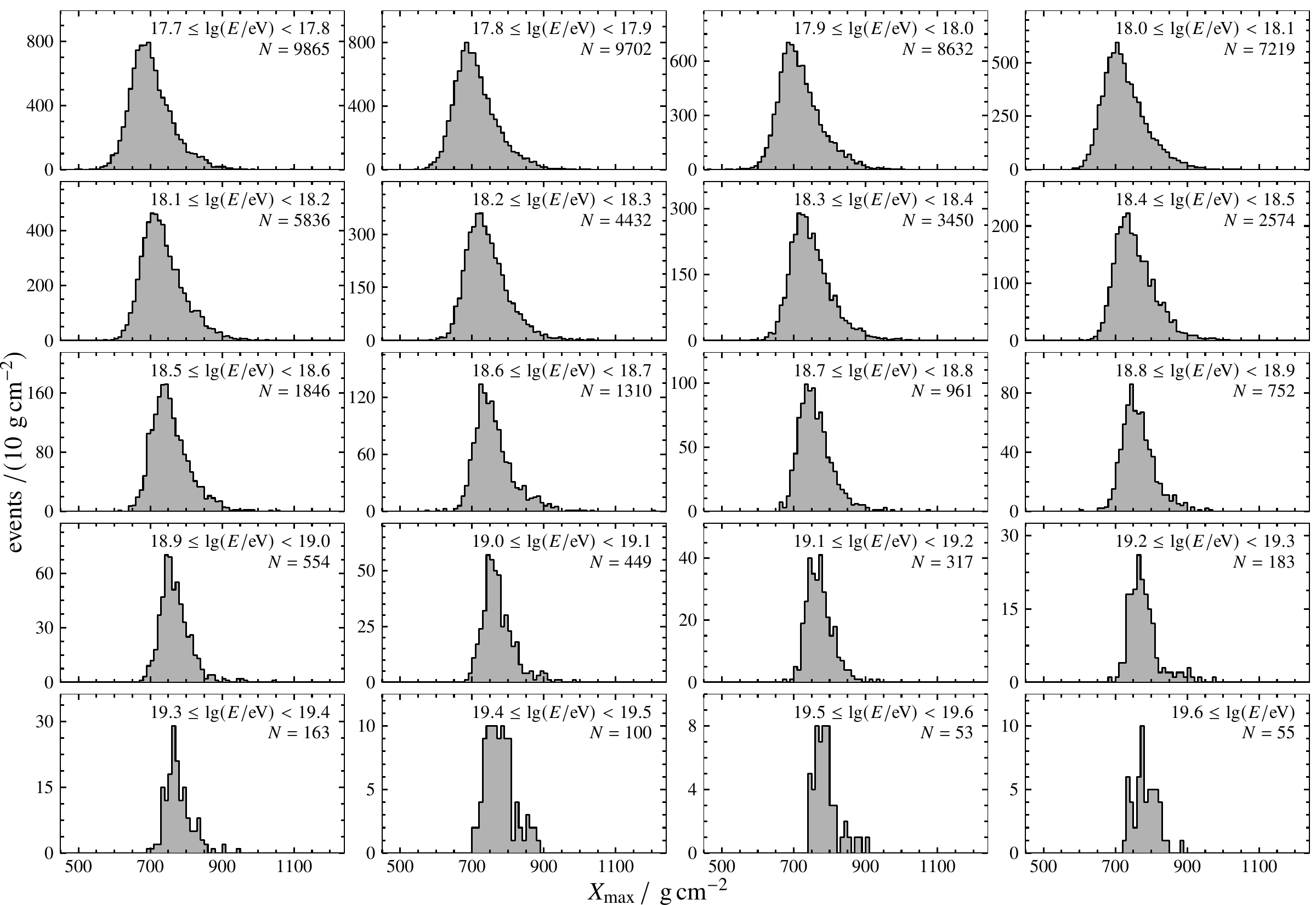}
    \caption{$\Xmax^{}$ distributions in different energy bins.}
    \label{fig:Xmax_distributions}
\end{figure*}

The $\Xmax^{}$ distributions shown in \cref{fig:Xmax_distributions}
differ from the true distributions of air showers in the atmosphere
due to residual sampling and reconstruction biases, and the finite $\Xmax^{}$
resolution. The quality and fiducial selection have been designed to
keep these distortions small, but they are not negligible for a
precise determination of the moments of the distribution or for a fit
with composition templates. In the following, we characterize these
measurement effects.

The relation between the true and observed $\Xmax^{}$ distributions,
$f(\Xmax^{})$ and $f^{}_\mathrm{obs}(\Xmax^\mathrm{rec})$ respectively,
is given by the convolution
\begin{multline} \label{eq:Xmax_distrib}
    f^{}_\mathrm{obs}(\Xmax^\mathrm{rec}) = \\ \int_{0}^{\infty} f(\Xmax^{}) \,
    R(\Xmax^\mathrm{rec} - \Xmax^{} + \Delta)\, \varepsilon(\Xmax^{}) \, \mathrm{d}\Xmax^{},
\end{multline}
where $\varepsilon(\Xmax^{})$ is the probability of observing and
selecting an event with a given shower maximum, denoted as ``acceptance'', and $R(\Xmax^\mathrm{rec} - \Xmax^{})$ is the detector
response, including resolution and bias. $\Delta$ denotes
a systematic shift in the $\Xmax^{}$ scale that is usually unknown, but
whose plausible range, commonly referred to as the \emph{systematic uncertainty}, can be estimated.

The acceptance and resolution are evaluated using detailed simulations
that replicate the time-dependent state of the SD, FD, and
atmosphere~\cite{PierreAuger:2010swb,PierreAuger:2023cqe}. Longitudinal air-shower profiles are
generated with the \Conex program~\cite{Bergmann:2006yz} using the hadronic
interaction model \Sibyll~\cite{Riehn:2019jet}. The detector simulation and event reconstruction are carried out using the \Offline framework of the
\PAO~\cite{Argiro:2007qg}. The simulated showers are reconstructed and
selected in the same manner as the measurements.

The acceptance and resolution derived from the simulation are to good
approximation independent of the choice of hadronic interaction model
or mass composition, because longitudinal air shower profiles are
universal for a given \Xmax and
energy~\cite{Andringa:2011zz,PierreAuger:2018gfc}. To achieve model
independence, it is therefore sufficient to weight the simulated
events such that the generated energy spectrum follows the measured
one~\cite{PierreAuger:2020qqz} and the \Xmax distribution of the
simulated sample reproduces the observed one.

\subsection{\label{sec:acceptance} Acceptance}

The acceptance is defined as the ratio of the number of selected events
to the number of generated events as a function of $\Xmax$.
To obtain the \emph{relative} acceptance at a particular energy, we normalize the acceptance to its maximum value.
The resulting relative acceptance can be well parameterized by a constant with exponentially rising and falling edges,
\begin{equation}
  \label{eq:accept}
  \varepsilon(\Xmax^{}) =
  \begin{cases}
    \mathrm{e}^{+(\Xmax^{} - X_1)/\lambda_1}, & \Xmax^{} < X_1; \\
    1, & X_1 \leq \Xmax^{} \leq X_2; \\
    \mathrm{e}^{-(\Xmax^{} - X_2)/\lambda_2}, &  X_2 < \Xmax^{} .
  \end{cases}
\end{equation}

\begin{figure}[t!] \centering
  \includegraphics[width=\linewidth]{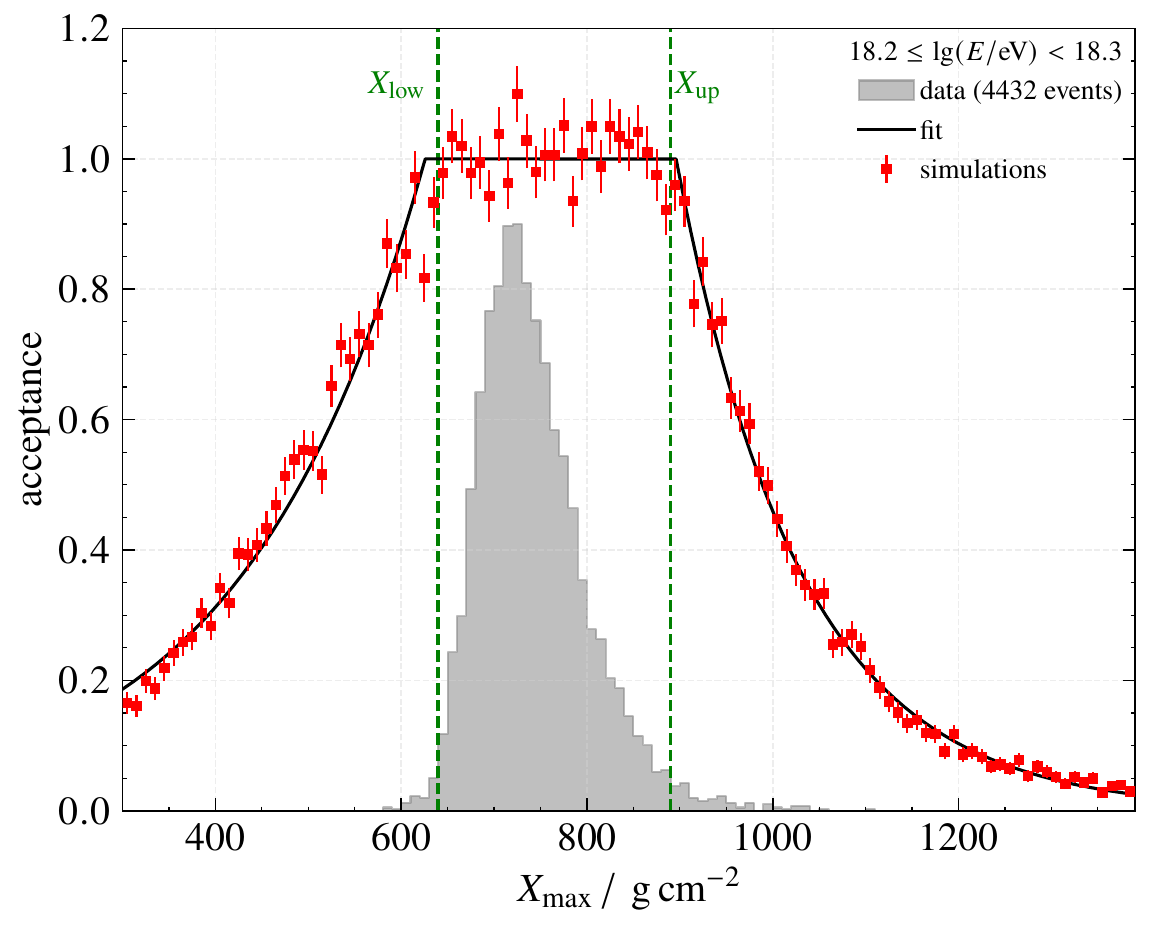}
    \caption{Acceptance as a function of $\Xmax^{}$ in simulations (points)
      in the energy interval \lgErange{18.2}{18.3}. The fit
      with \cref{eq:accept} is shown as a dashed line. For comparison,
      the distribution of $\Xmax^{}$ of the data in this energy range is
      shown as a gray histogram (arbitrary normalization).}
    \label{fig:acceptance}
\end{figure}

The parameters $(X_1, X_2, \lambda_1, \lambda_2)$ are energy-dependent
as specified in \cref{app:acceptance_E}. An example of the acceptance
using simulations in the energy interval \lgErange{18.2}{18.3}, fitted
with \cref{eq:accept}, is presented in \cref{fig:acceptance} along
with the $\Xmax^{}$ distribution in the data. By construction, most of the
observed events fall within the fiducial $\Xmax^{}$ range,
$(\Xlow^\mathrm{fid}, \Xup^\mathrm{fid})$, displayed by vertical dashed lines.  Within this
slant depth range, the acceptance is found to be constant,
demonstrating the desired effect of the fiducial-field-of-view
selection to ensure an equal-probability sampling of air showers with
a shower maximum within this range. The energy dependence of the
acceptance is displayed in \cref{fig:acceptance_E}. As can be seen,
the fiducial $\Xmax^{}$ range is well contained in the flat part
of the acceptance, i.e.,\ $X_1 \leq \Xlow^\mathrm{fid}$ and   $\Xup^\mathrm{fid} \leq X_2 $.

\subsection{\label{sec:biasreso}\texorpdfstring{$\boldXmax$}{Xmax} bias and resolution}

\subsubsection{Detector response from simulations}
\label{sec:resoFromSim}
The detector response $R(\Xmax^\mathrm{rec} - \Xmax^{})$ of the
reconstructed $\Xmax^\mathrm{rec}$ of simulated showers after detector
simulation can be well described empirically by the sum of two Gaussian distributions, $G$,
\begin{equation} \label{eq:Xmaxreso}
    R(\Xmax^\mathrm{rec} - \Xmax^{}) = f \, G(\mu, \sigma_1) + (1-f) \, G(\mu, \sigma_2).
\end{equation}
This double-Gaussian parameterization provides an empirical approximation of the detector response obtained after averaging over showers with different geometries, for which the \Xmax resolution varies, and thereby describes the resulting non-Gaussian shape. Here $f$ denotes the relative contribution of the first Gaussian, and $\sigma_1 < \sigma_2$, the widths of the first and second Gaussians, respectively. The mean of $R$ is the
$\Xmax^{}$ reconstruction bias $\mu_R$ and its standard deviation $(f \,
  \sigma_1^2 + (1-f) \, \sigma_2^2)^{\frac{1}{2}}$ is the $\Xmax^{}$ resolution. The
resolution quantifies the broadening of the $\Xmax^{}$ distribution and the
bias of its offset.

The fitted parameters of $R$ depend slightly on
$\Xmax^{}$~\cite{PierreAuger:2025avg}, but for the purpose of this analysis
we use an effective response averaged over the $\Xmax^{}$
distribution. This is obtained by weighting the simulated showers to
match the observed $\Xmax^{}$ distribution.

\begin{figure}[t!] \centering
    \includegraphics[width=\linewidth]{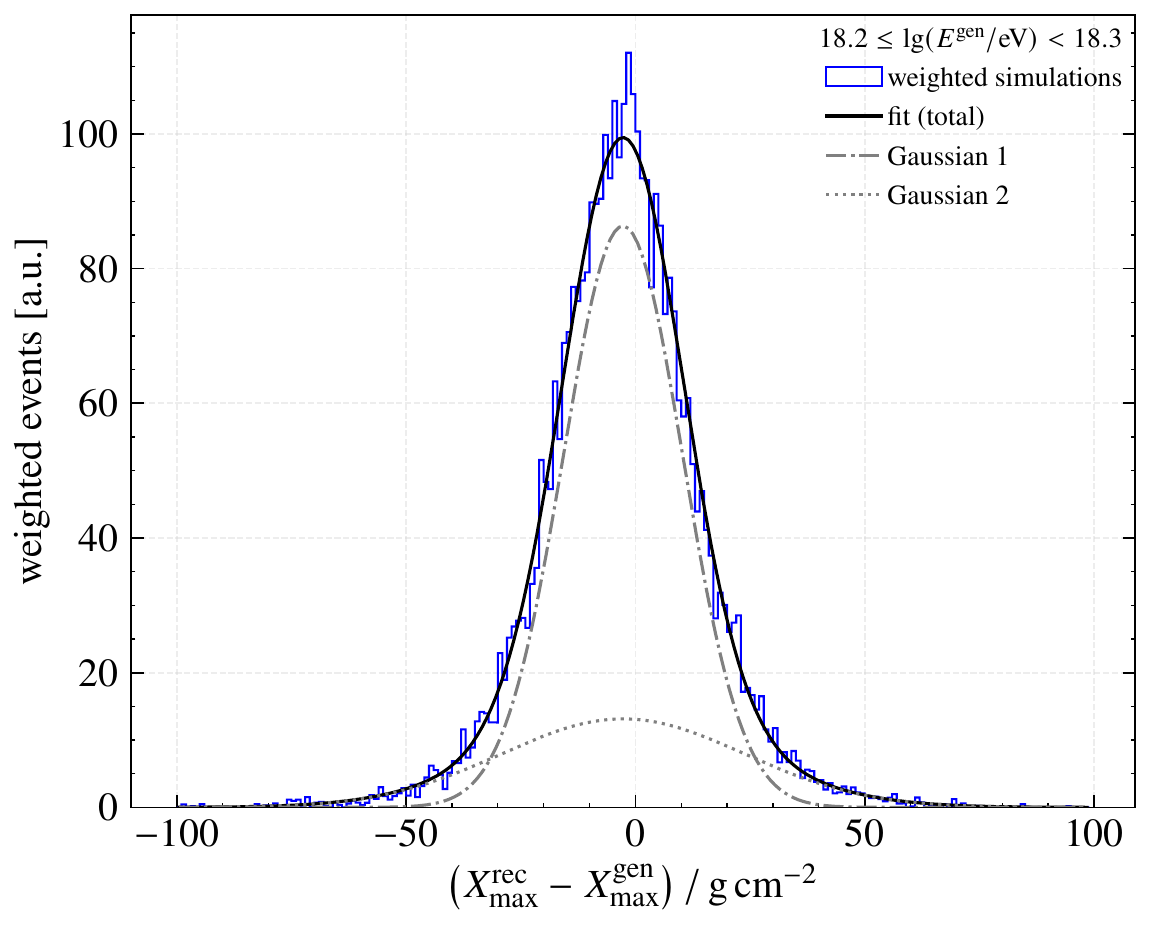}
    \caption{Distribution of the difference between reconstructed and
      generated $\Xmax$ for simulated showers in the energy interval
      \lgErange{18.2}{18.3}. The events are weighted to match
      the $\Xmax$ distribution in the data. The fit
      with \cref{eq:Xmaxreso} is shown with a blue dashed line, while the
      broad and narrow Gaussians that compose it are represented with
      dotted and dash-dotted lines, respectively.}
    \label{fig:Xbias}
\end{figure}

In \cref{fig:Xbias}, an example of $R$ in the energy
interval \lgErange{18.2}{18.3} is shown, along with a fit
to \cref{eq:Xmaxreso}. Most of the detector response
is described by the narrow Gaussian with width $\sigma_1$, whereas the
broad Gaussian with width $\sigma_2$ describes the tails of the
response.

\begin{figure}[t!] \centering
    \includegraphics[width=\linewidth]{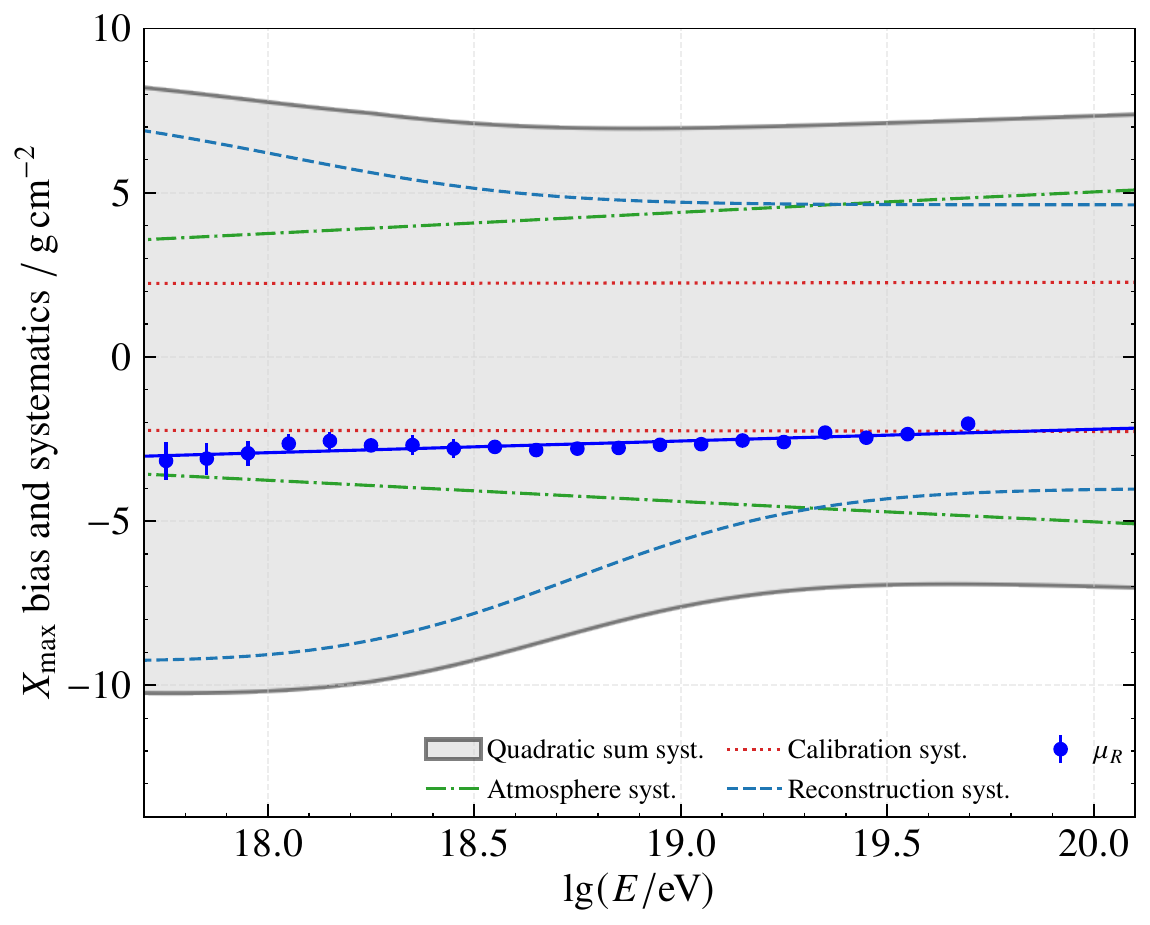}
    \caption{Reconstruction bias (points) and systematic uncertainties
      in the \Xmax scale from the detector calibration, reconstruction,
      and atmosphere (lines) as a function of energy.}
    \label{fig:Xmax_bias_E}
\end{figure}

The energy evolution of the \Xmax bias is shown in
\cref{fig:Xmax_bias_E}. The fitted $\mu_R$ value is shown with blue
points, and its parametrization, $\mu_R / (\text{\gcm}) = -2.9 + 0.4
\, \lg(E/\mathrm{EeV})$, is shown as a blue line. It varies from
approximately $-3$\,\gcm at the lowest energies to  $-2$\,\gcm at
$10^{20}$\,eV. We correct each event for this average bias by
subtracting $\mu_R(E)$ from the reconstructed \Xmax.

The resolution derived from simulation captures the statistical
fluctuations of the reconstructed \Xmax due to the finite number of
photo-electrons detected by the photomultipliers of the fluorescence
telescopes. It therefore includes contributions from the uncertainty
in the shower geometry and from the fit to the longitudinal profile.
It decreases from ${\sim}25$\,\gcm at $10^{17.6}$\,eV to
${\sim}12$\,\gcm at $10^{19.0}$\,eV.

\subsubsection{Additional contributions to the \Xmax resolution}
\label{sec:moreReso}
The detector response is further broadened by the following contributions
that are not included in the simulations:

\paragraph{Telescope alignment}
This analysis utilizes data recorded by 24 individual
fluorescence telescopes. The angular alignment of each telescope was
determined using two methods: observations of UV star tracks and
using the independent reconstruction of the shower geometry available
for high-quality SD events. The reconstruction of the data with either set of alignment
constants yields no difference in \Xmax on average, but the standard
deviation of the distribution of \Xmax differences between the two
reconstructions can be used as an estimate of the possible influence
of the telescope-alignment uncertainty on the \Xmax
resolution. We find a standard deviation of $\delta = \left(5.07 +
0.87 \, \lg(E/10^{17.55}\,\text{eV})\right)$\,\gcm. We take
$\delta/2$ as an estimate of the alignment contribution
to the \Xmax resolution with a systematic uncertainty of $\pm
\delta/2$.

\paragraph{Aerosols}
The statistical uncertainty of the measurements of aerosols
constitutes an additional contribution to the overall \Xmax
resolution. This uncertainty originates from fluctuations in the
night-sky background, the number of photons detected from laser shots,
and the variability of aerosol content relative to the hourly average
values. It is estimated from the standard deviation of quarter-hourly
VAOD
measurements~\cite{PierreAuger:2013dlv}.
A second contribution arises from the non-uniformity of the aerosol
layers over the array. It is assessed using observations from
different FD sites~\cite{PierreAuger:2010uyt}.

\paragraph{Molecular atmosphere}
Another contribution arises from the uncertainty in the atmospheric
density profile at the time of an event.  It is evaluated by comparing
\Xmax values reconstructed using GDAS with those obtained from balloon
soundings available for a fraction of the data~\cite{PierreAuger:2012jsu}.

\subsubsection{Full \texorpdfstring{$\Xmax$}{Xmax} resolution}
The additional contributions are assumed to be Gaussian, and their
widths are added in quadrature to the double-Gaussian parametrization
obtained from simulations. The resulting parameters are listed
in \cref{app:resotable}.

\begin{figure}[t] \centering
    \includegraphics[width=\linewidth]{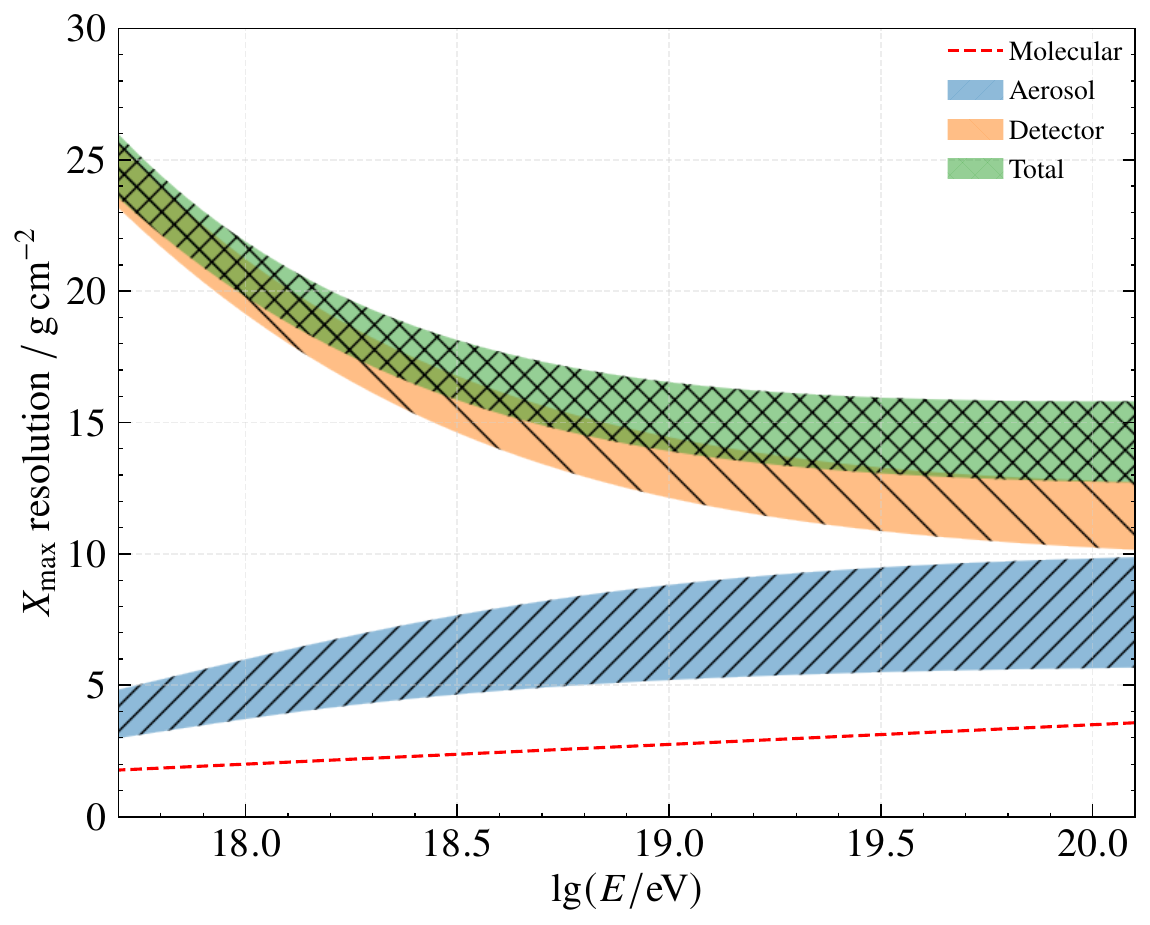}
    \caption{The total \Xmax resolution and the individual contributions
      from detector, aerosols, and density profiles as a function of
      energy. Shaded bands indicate the systematic uncertainties.}
    \label{fig:Xmax_resolution_E}
\end{figure}

The standard deviation of the full $\Xmax$ response function is shown
in \cref{fig:Xmax_resolution_E}. It improves from 25\,\gcm at
$10^{17.7}$\,eV to 16\,\gcm at $10^{19.0}$\,eV, remaining
approximately constant at higher energies. The quadratic sum of
the statistical resolution (derived from simulations) and the telescope
alignment is shown in \cref{fig:Xmax_resolution_E} with an orange
band. The decrease of this contribution with energy is mainly due to
the increased brightness of air showers at higher energies and due to
the larger fraction of events observed at more than one FD
site. The contribution from aerosols, shown with a blue band,
increases with energy, as high-energy showers can be viewed at larger
distances and, correspondingly, the correction for aerosol transmission
increases. The uncertainty of the molecular atmosphere is shown as a
dashed red line and remains below 4~\gcm at all energies.

\subsection{\texorpdfstring{$\boldXmax$}{Xmax} scale uncertainty}
\label{sec:xmaxscalesys}
The estimated systematic uncertainty on the \Xmax scale
$\Delta$ in \cref{eq:Xmax_distrib} is displayed
in \cref{fig:Xmax_bias_E}. It changes from $^{+8.1}_{-10.2}$\,\gcm at
$10^{17.7}$\,eV to about $\pm7$\,\gcm at $10^{20}$\,eV and is
attributed to three main contributions discussed in the following.

\paragraph{Detector calibration} The uncertainty associated with the relative timing between the FD and SD is
estimated to be $\pm 2~\mathrm{g/cm^2}$ from dedicated reconstructions with a
systematic shift of $\pm 100$~ns applied to the relative timing constants.
The calibration of the relative PMT responses adds an additional uncertainty
of less than $\pm 1~\mathrm{g/cm^2}$, as estimated from comparisons of data
reconstructed with different versions of the calibration database.

\paragraph{Reconstruction} Part of the uncertainty arising from
shower-profile fitting is determined by varying the constraints of the
shape parameters of the Gaisser-Hillas function within one standard
deviation of their central values~\cite{PierreAuger:2023att}. This
contribution is $^{+4.6}_{-4.0}$\,\gcm and remains constant with energy.

Additionally, an incomplete understanding of the interplay between the
telescope point-spread function and the lateral
distribution of the Cherenkov and fluorescence light
leads to an energy-dependent \Xmax uncertainty. In the data, we find a
larger fraction of light outside the collection angle than
expected from simulations of the optical system and lateral light
distribution from air showers. This mismatch depends on shower
age, and its origin is not fully understood. The corresponding
contribution to the \Xmax systematic uncertainty is found to be
$^{+5.1}_{-8.3}$\,\gcm at $10^{17.7}$\,eV, decreasing to
$^{+0.1}_{-0.6}$\,\gcm at $10^{20}$\,eV. The total reconstruction uncertainty
amounts to $^{+6.9}_{-9.2}$\,\gcm at
$10^{17.7}$\,eV and $^{+4.6}_{-4.0}$\,\gcm at $10^{20}$\,eV.

\paragraph{Atmosphere} The 4\% uncertainty in the absolute yield of
the fluorescence light production~\cite{AIRFLY:2012msg} introduces an
energy-independent \Xmax uncertainty of $\pm 0.4$\,\gcm. An
additional contribution of $\pm 0.2$\,\gcm arises from the
uncertainty in the wavelength dependence of fluorescence
yield. Different parametrizations of multiple scattering of fluorescence light introduce another uncertainty of $\pm
2$\,\gcm~\cite{Louedec:2013caa}.  At the highest energies,
where showers can be observed at greater distances from the FD
telescopes, the dominant contribution to atmospheric uncertainties
arises from measurements of the aerosol
content~\cite{PierreAuger:2010uyt,PierreAuger:2013dlv}.
Recently, stereo observations of air showers
\cite{PierreAuger:2023nbk} were used to cross-check and correct the
VAOD impact on \Xmax~and energy measurements. Overall, the VAOD
contribution to the \Xmax systematic uncertainty amounts to $\pm
2.9$\,\gcm at $10^{17.7}$\,eV and $\pm4.6$\,\gcm at $10^{20}$\,eV. The total uncertainty due to atmospheric effects increases with energy from $\pm
3.6$\,\gcm to $\pm 5.0$\,\gcm.

\begin{figure}[!t]
\centering    \includegraphics[width=\linewidth]{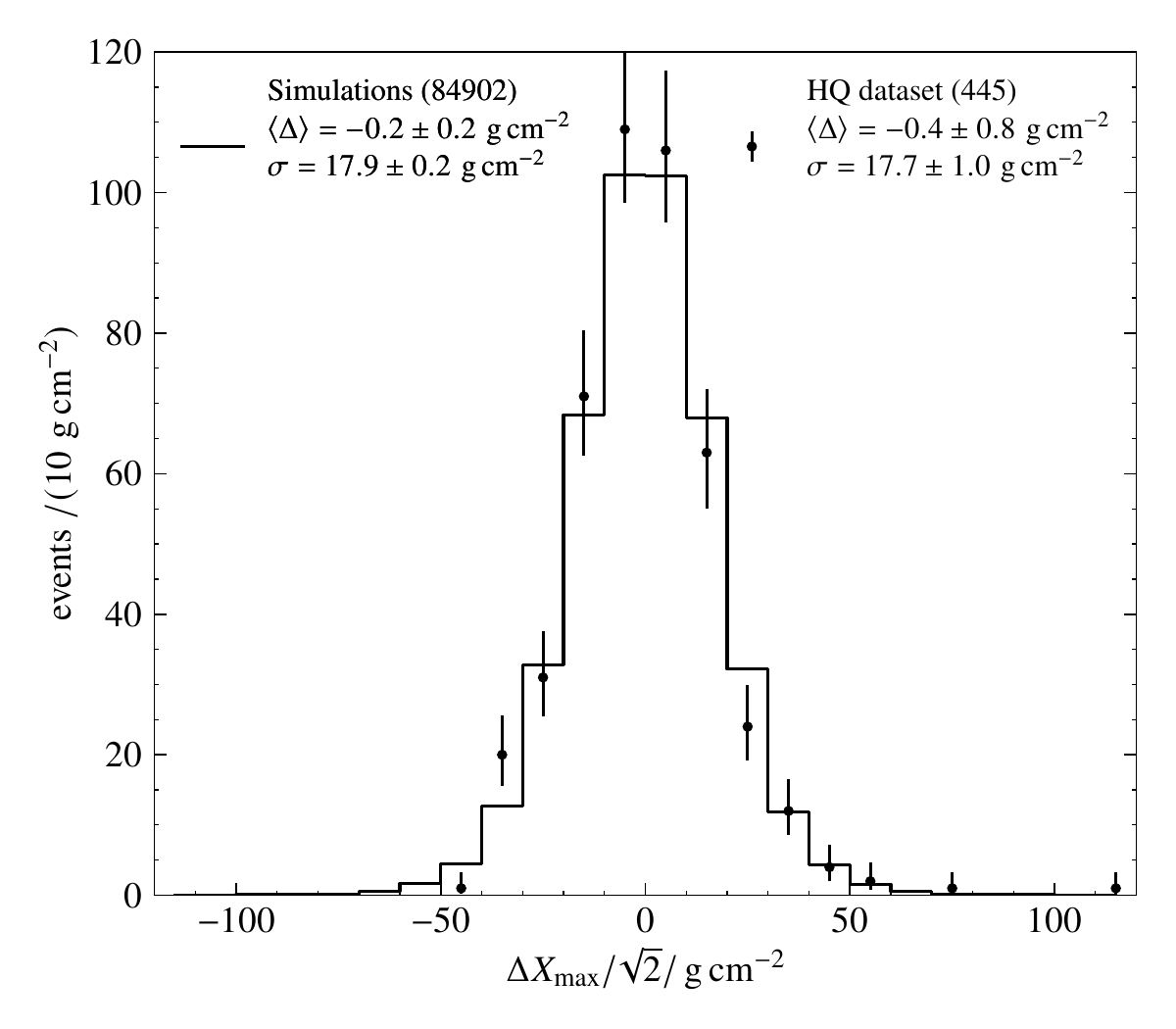}
\caption{Distribution of $\Xmax$ difference for events observed at more than one FD site for data and simulation.}
    \label{fig:stereo_check}
\end{figure}

\begin{figure}
    \centering
    \includegraphics[width=0.49\textwidth]{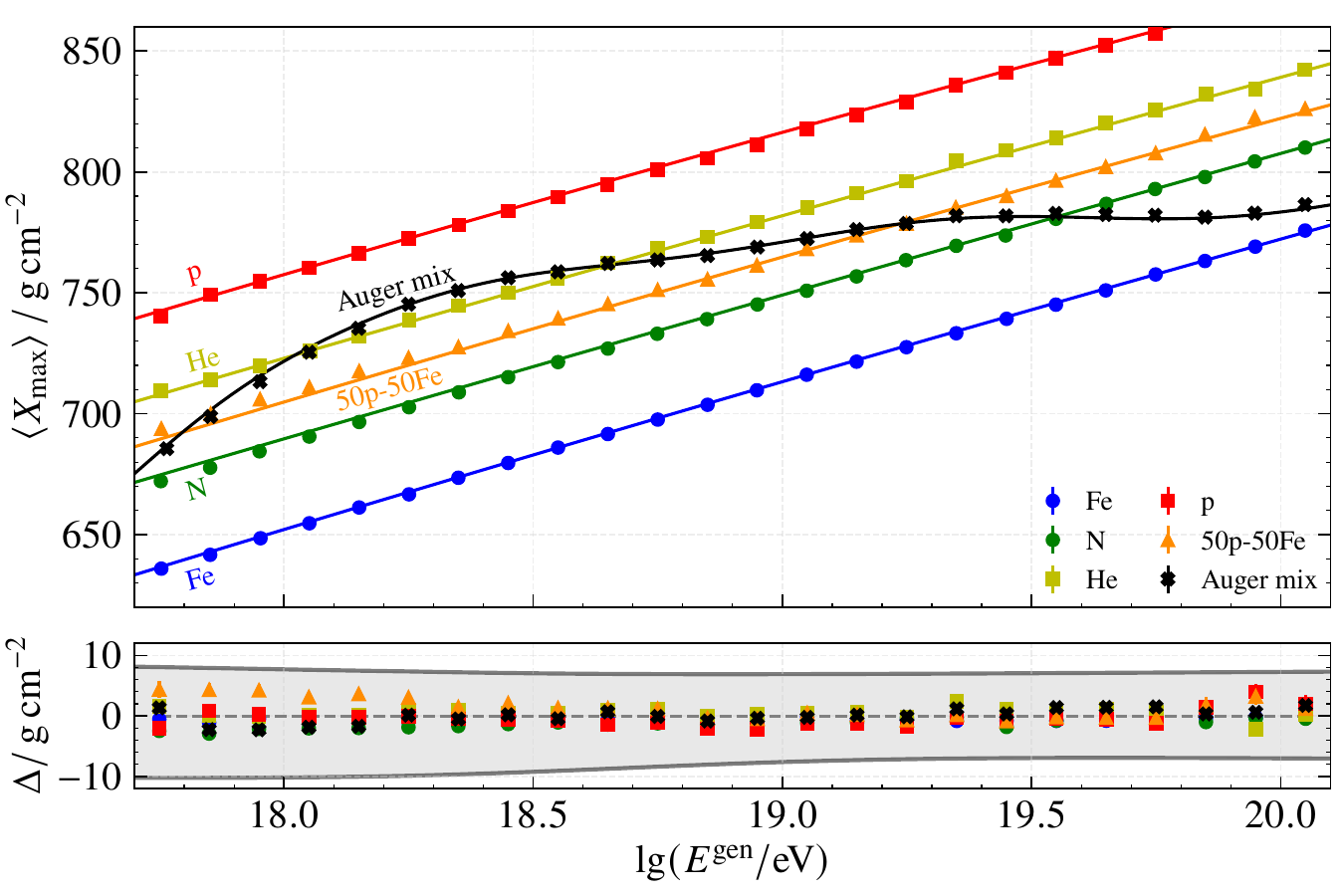}
    \includegraphics[width=0.49\textwidth]{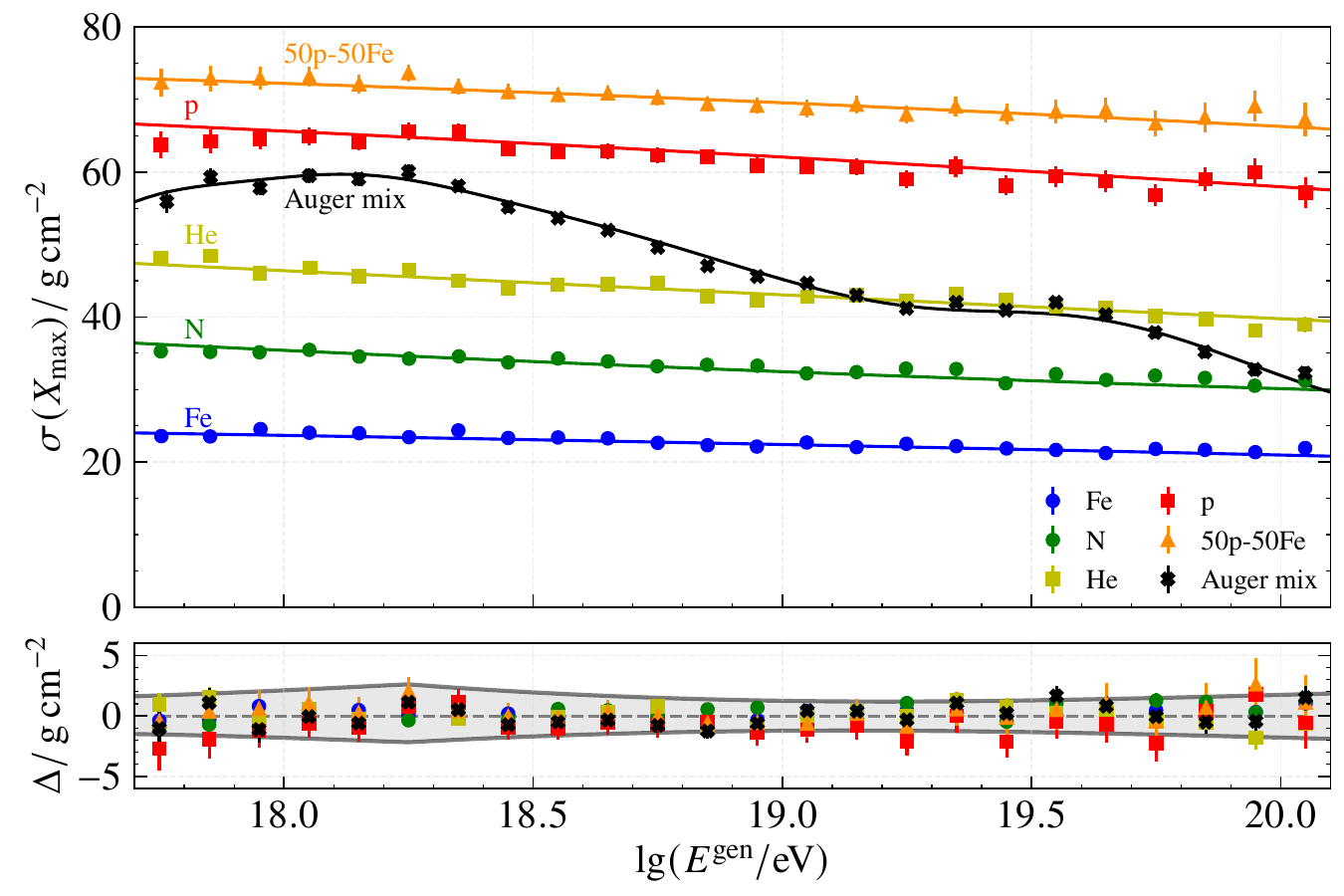}
    \caption{$\meanXmax$ and $\sigmaXmax$ for simulated pure proton, helium, nitrogen, and iron primaries, a 50/50 proton-iron mixture, and the 4-component fractional composition derived in \cref{sec:fractions} (Auger mix).
    The line indicates the true moments of the generated data set, while points show the reconstructed values.
    The lower panel of each figure shows the difference $\Delta$ between reconstructed and generated values.
        Simulations use the hadronic model \Sibyll.}
    \label{fig:analysis_check}
\end{figure}

\def\figh{0.2}
\begin{figure*} \centering
    \includegraphics[height=\figh\textheight]{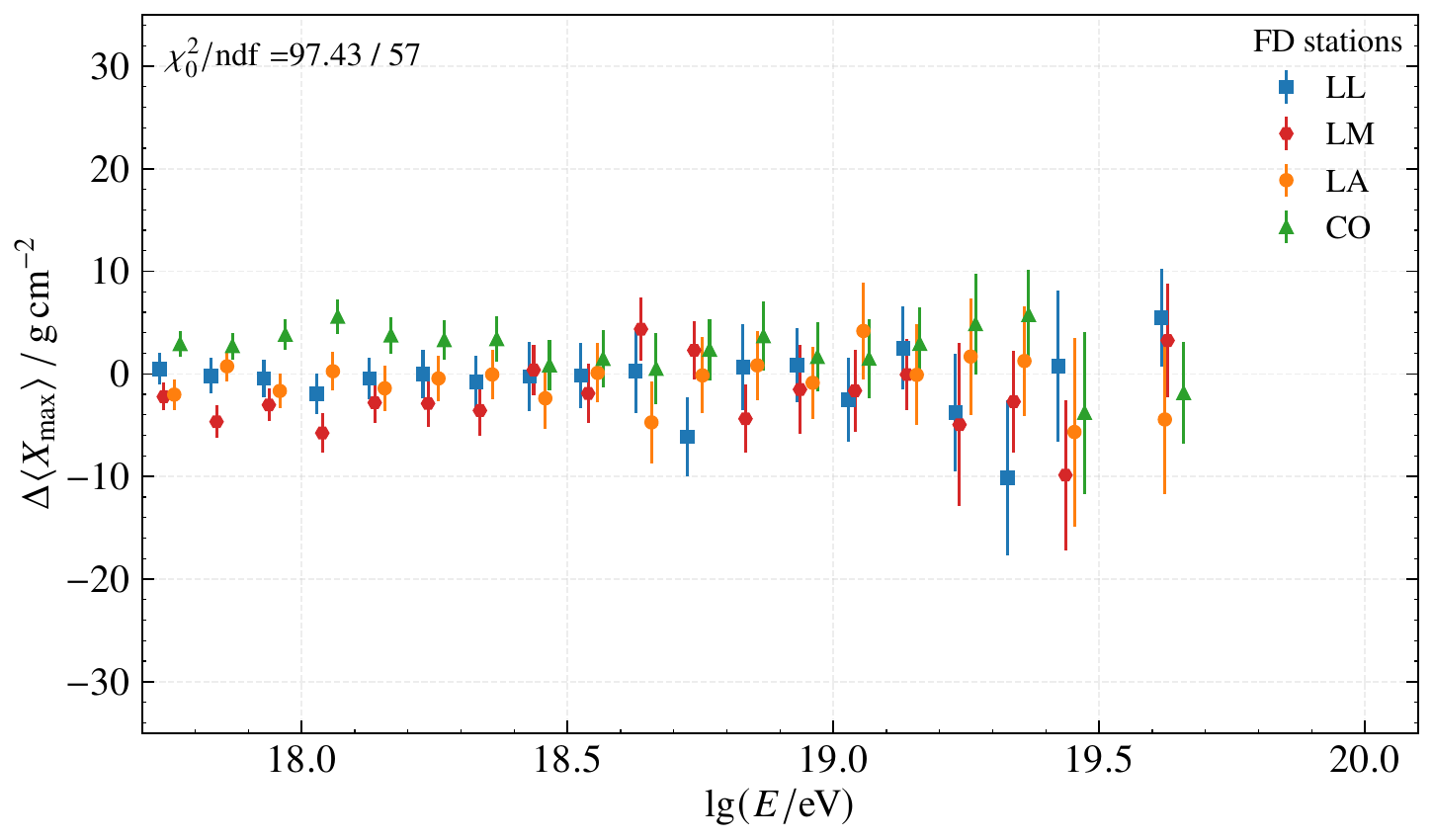}
    \includegraphics[height=\figh\textheight]{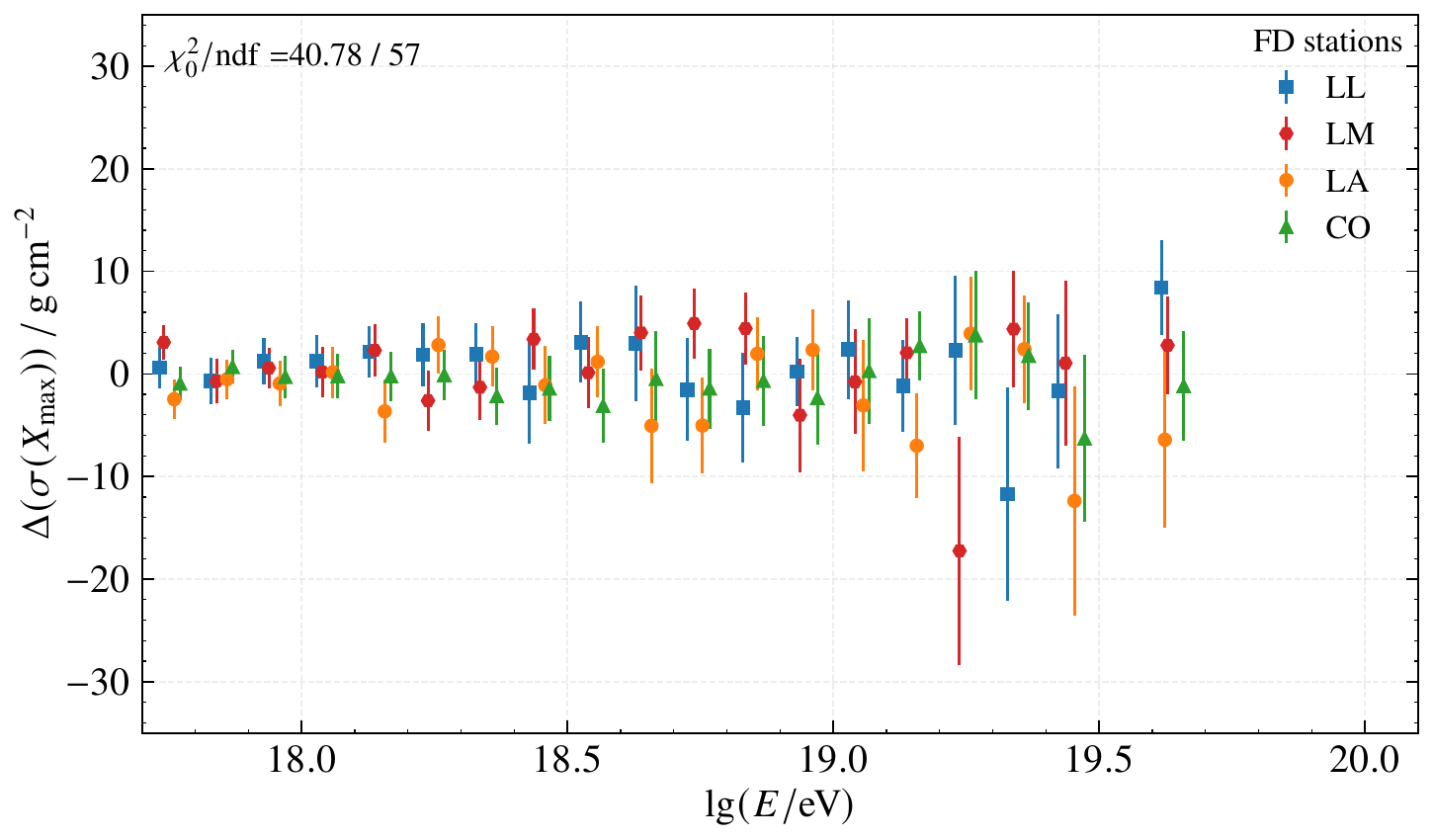}\\
    \includegraphics[height=\figh\textheight]{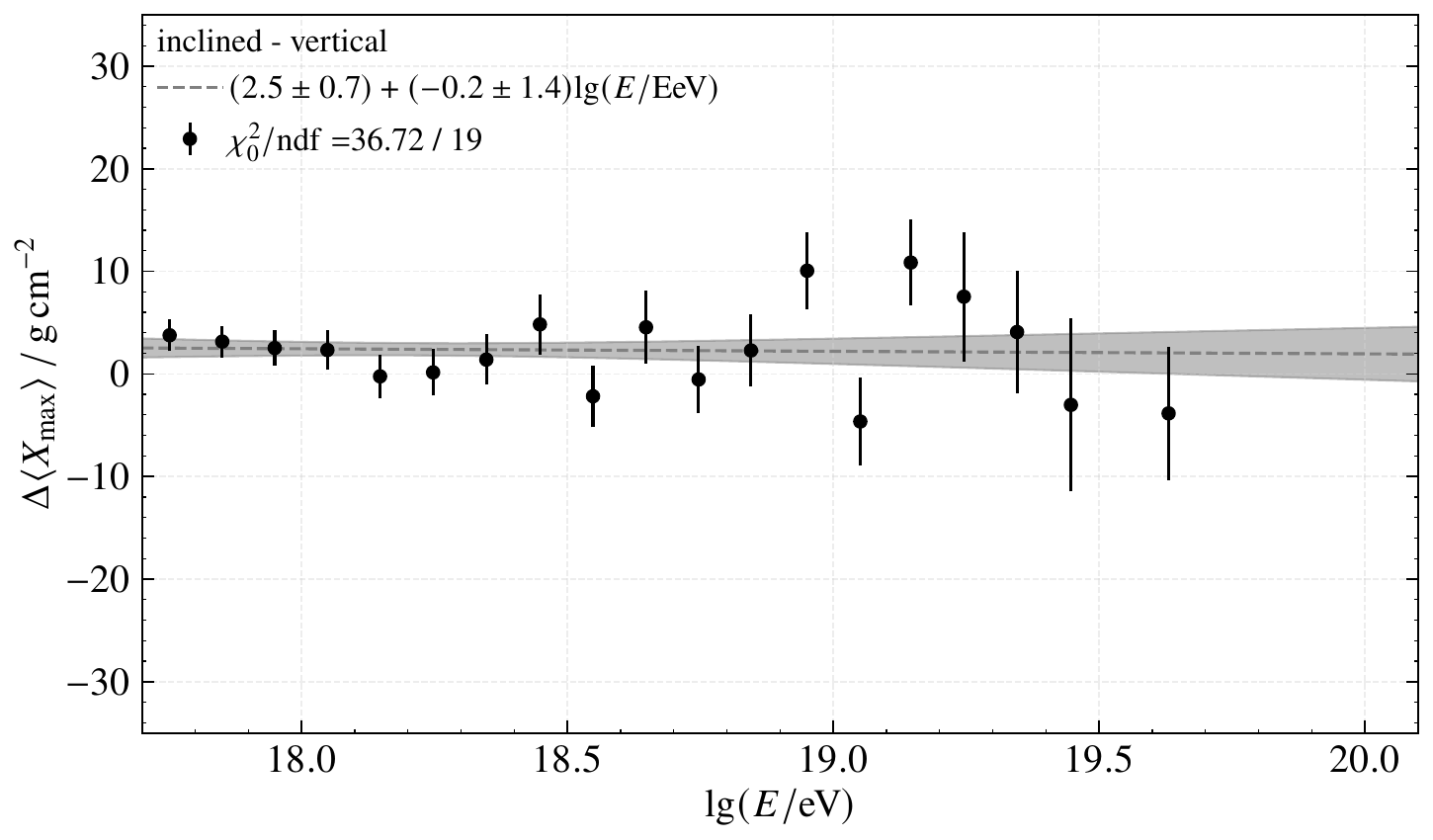}
    \includegraphics[height=\figh\textheight]{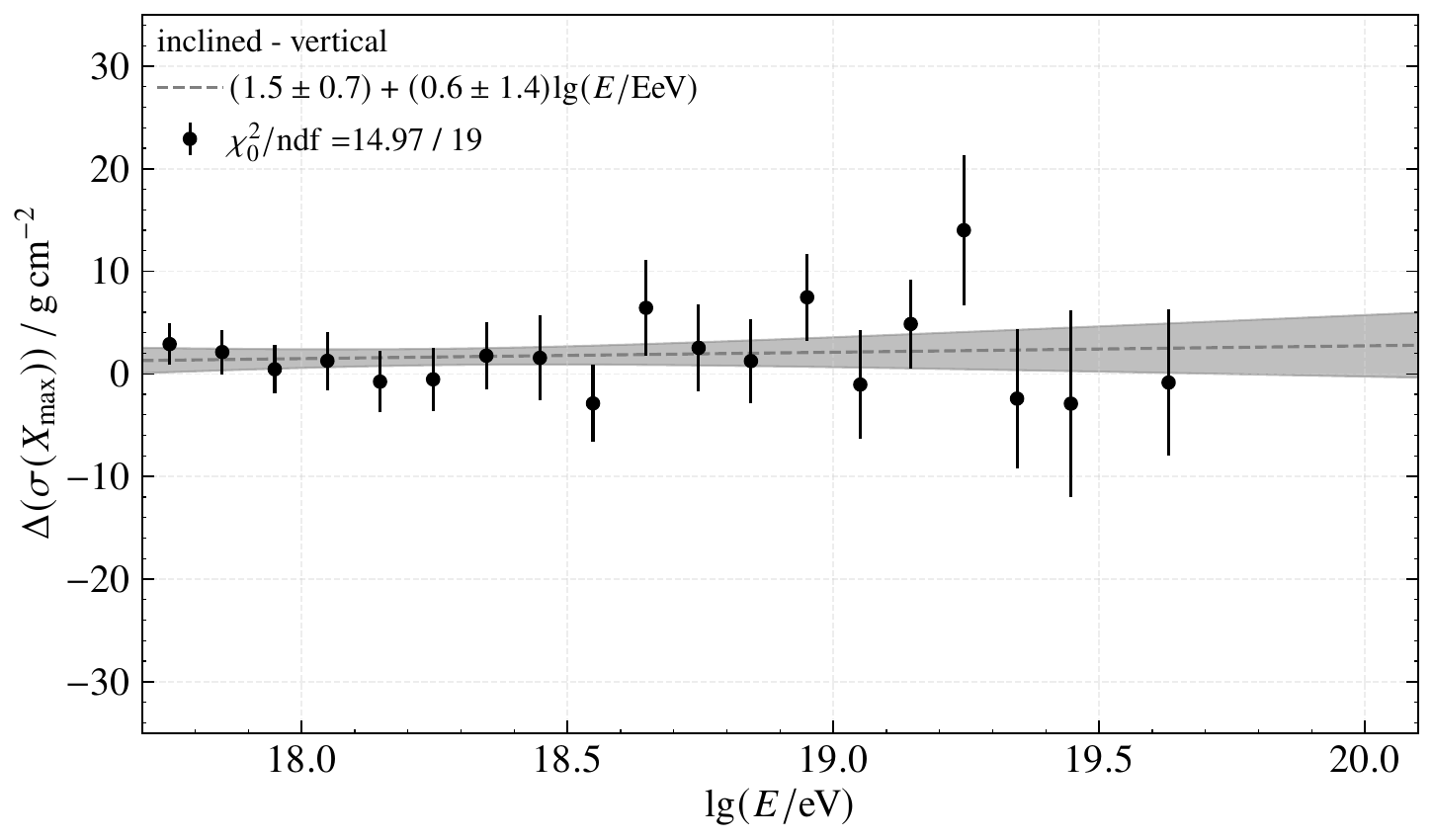} \\
    \includegraphics[height=\figh\textheight]{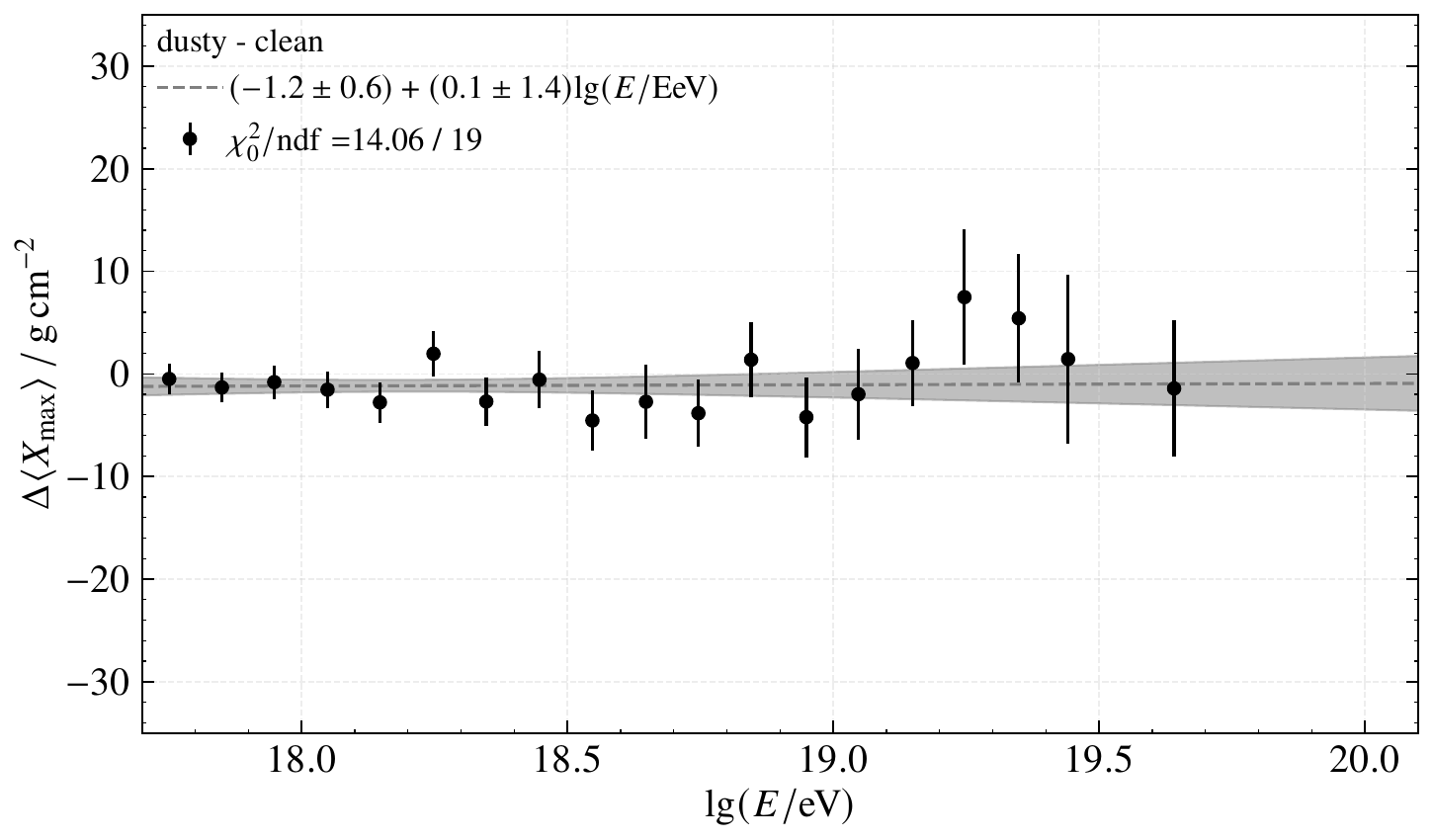}
    \includegraphics[height=\figh\textheight]{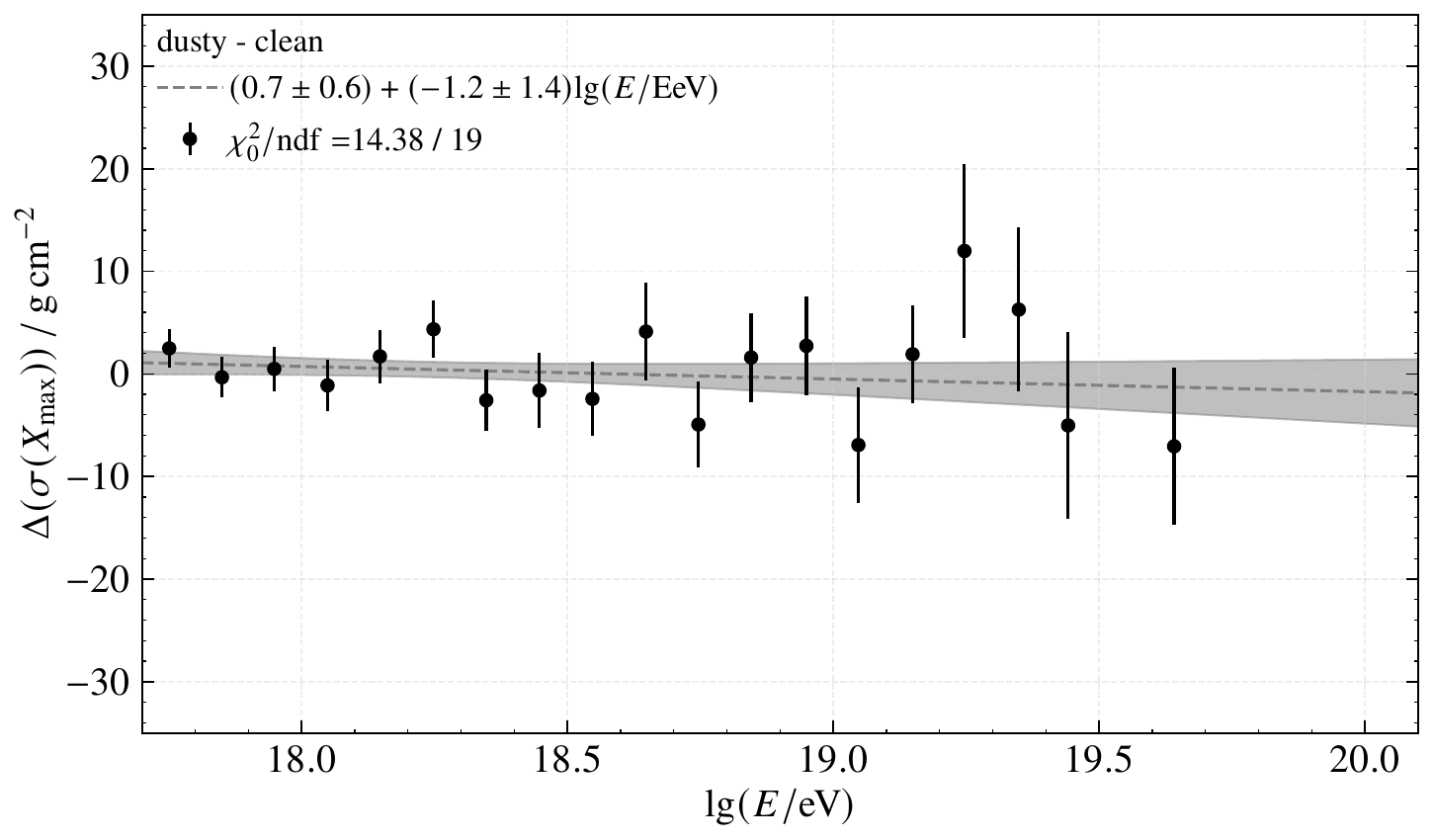} \\
    \includegraphics[height=\figh\textheight]{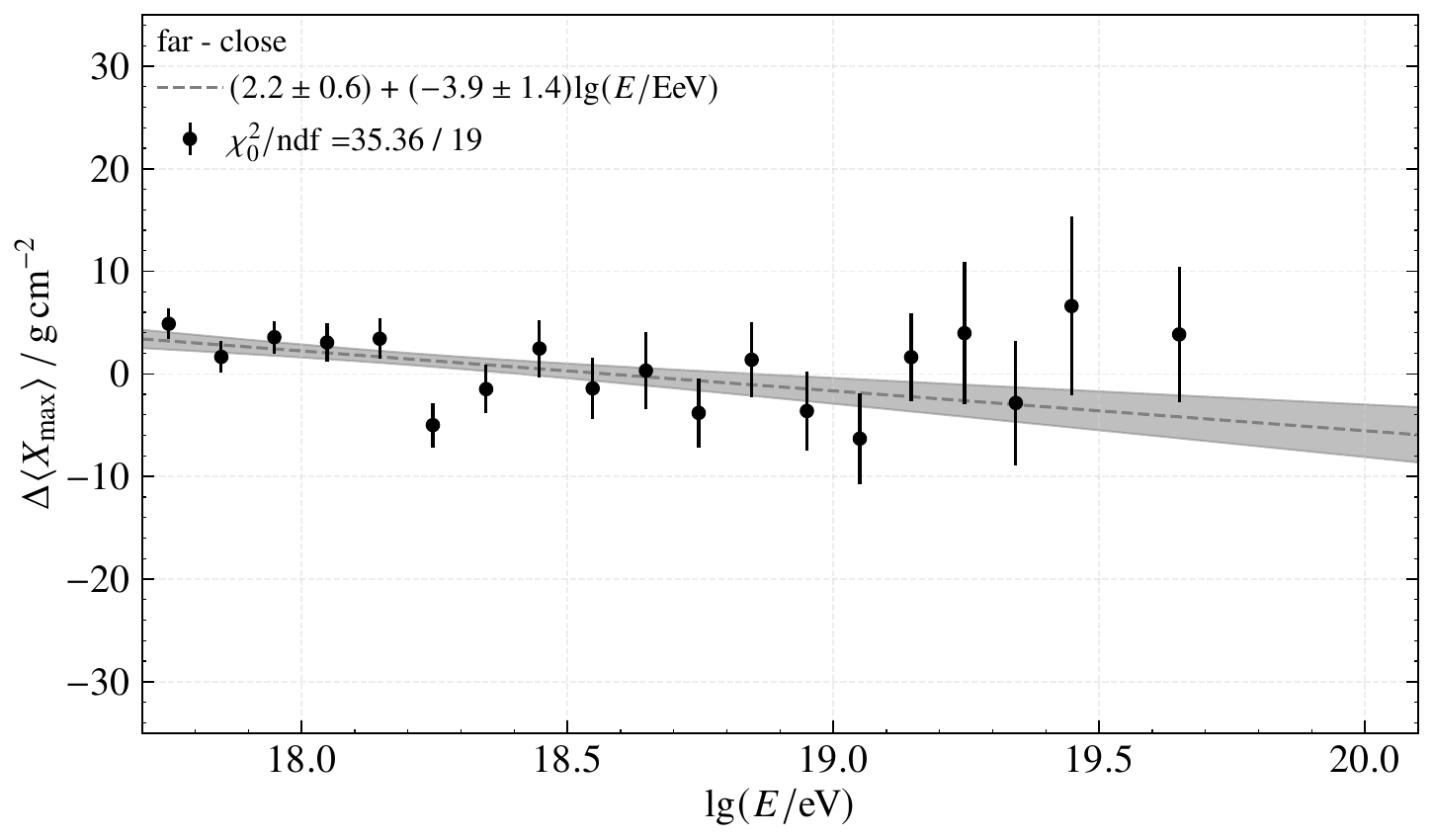}
    \includegraphics[height=\figh\textheight]{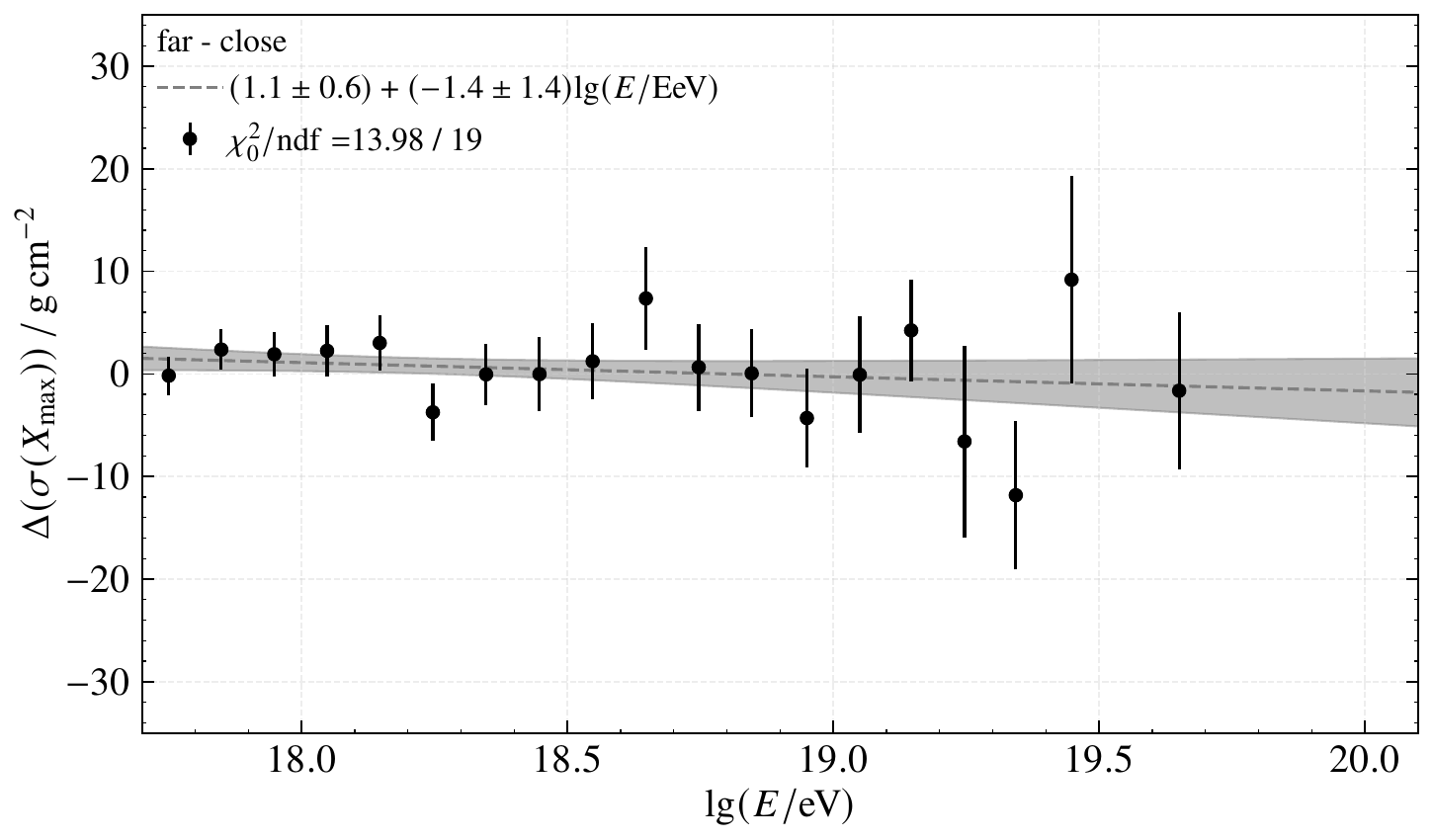}
    \caption{Differences in the mean and standard deviation of the
      \Xmax distribution for various cross-checks. From top to bottom:
      events split by FD site (LL, LM, LA, CO); near-vertical versus inclined
      showers; dusty versus clean nights; and events close to the FD
      site versus those at larger distances.
      A linear fit to the data is shown as a dashed line with a gray uncertainty contour.
      The $\chi^2_0/\mathrm{ndf}$ values for compatibility with zero are given, as are the linear fit values.}
    \label{fig:cross_checks}
\end{figure*}

\section{\label{sec:crosschecks}Cross-checks and comparisons of data-subsets}

The systematic uncertainties discussed in the previous section have
been thoroughly validated through extensive cross-checks of the
stability of the results under detector, geometric, and atmospheric
variations, and through tests of data-simulation consistency.  The
most relevant studies are summarized below.

\subsection{Detector resolution}

Events reconstructed by multiple FD sites can be used to validate the
detector resolution obtained from simulation in \cref{sec:detector}.
The distributions of the differences between the reconstructed $\Xmax$
values at each participating site, $\Delta \Xmax$, for data and
simulation are shown in \cref{fig:stereo_check}.  The two
distributions agree statistically, with a Baker-Cousins
$\chi^2/\mathrm{ndf} = 26.1/23$~\cite{Baker:1983tu}, indicating that
the \Xmax resolution estimated from simulations is reliable.  Two
events with $\Delta \Xmax \geq 100$\,\gcm appear as outliers.  Such
events occur at a rate of 0.45\% in the stereo data set, compared to 0.20\% in simulations.

\subsection{Analysis of simulated data}

The analysis method can be validated end-to-end using simulations by comparing the reconstructed moments to the true moments at the generator level.
The profiles of air showers induced by protons as well as helium-, nitrogen-, and iron-nuclei are simulated with \Conex using the hadronic model \Sibyll.
As in \cref{sec:detector}, the simulations replicate the time-dependent state of the SD, FD, and atmosphere and are subjected to the same selection described in \cref{sec:data}.

Three checks are performed: one against each of the four simulated primary species, one against a 50/50 proton-iron mixture, and one against the fractional mix presented below in \cref{sec:fractions} (the ``Auger mix''), which closely follows the behavior of the measurements.
Primary-specific acceptance and resolution are computed and used for the tests against single-element injection. For the mixed-composition tests, we use the acceptance and resolution from \cref{sec:detector}, since the corresponding simulations, by construction, closely reproduce the measured distributions.

Results for $\meanXmax$ and $\sigmaXmax$ are shown in
\cref{fig:analysis_check}.  The points show the reconstructed values,
and the lines show the generated ones. The difference $\Delta$ between reconstructed and generated values is shown in the lower panel of each figure, with the systematic uncertainty in the reconstruction shown in gray.
Deviations are generally of order 2 to 5\,\gcm for all compositions. These discrepancies stem
from uncertainties in the acceptance, bias, and resolution corrections
and are well within the systematic uncertainties discussed in
\cref{sec:xmaxscalesys}.  We conclude that the analysis method
performs well.

\begin{figure*}[t!] \centering
    \includegraphics[width=0.5\linewidth]{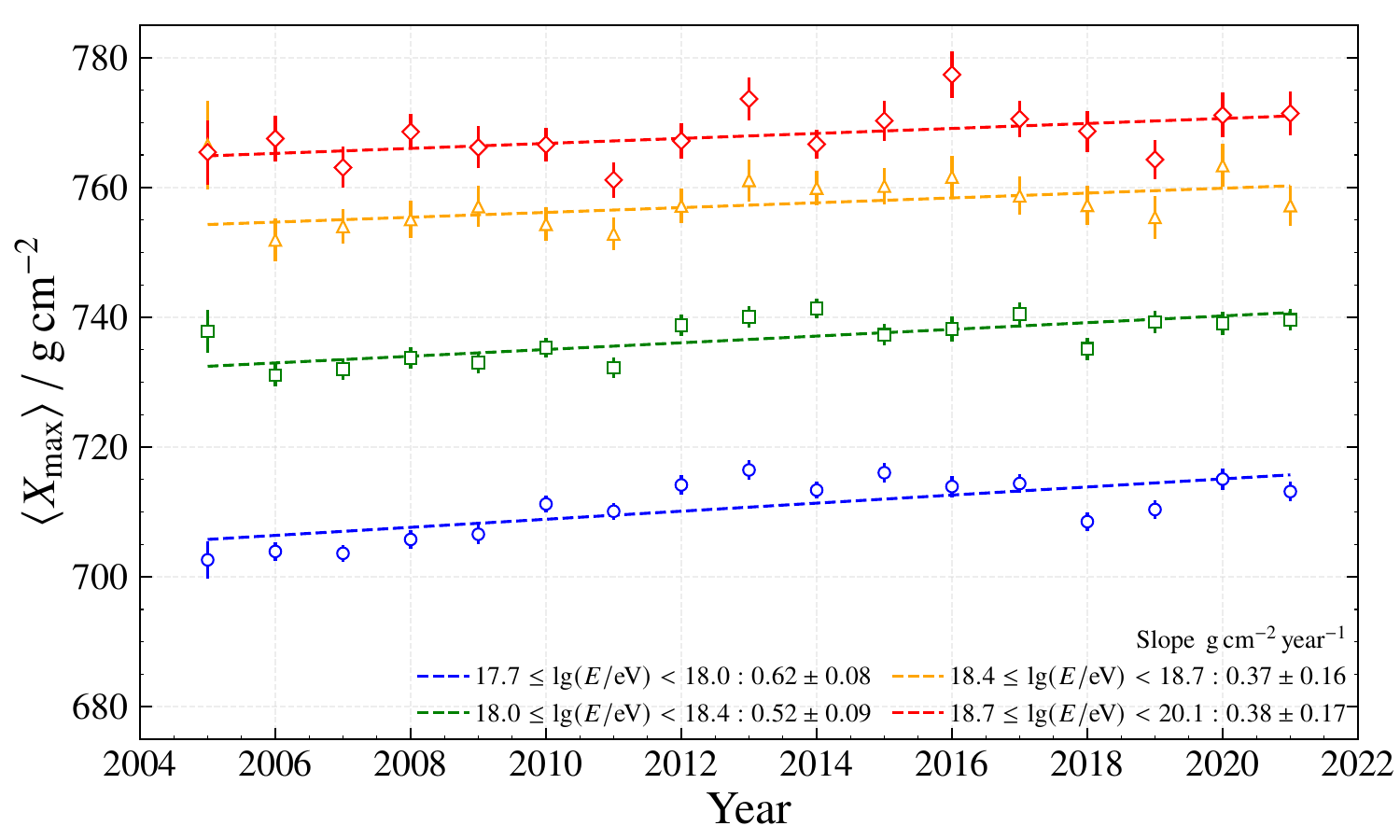}\includegraphics[width=0.5\linewidth]{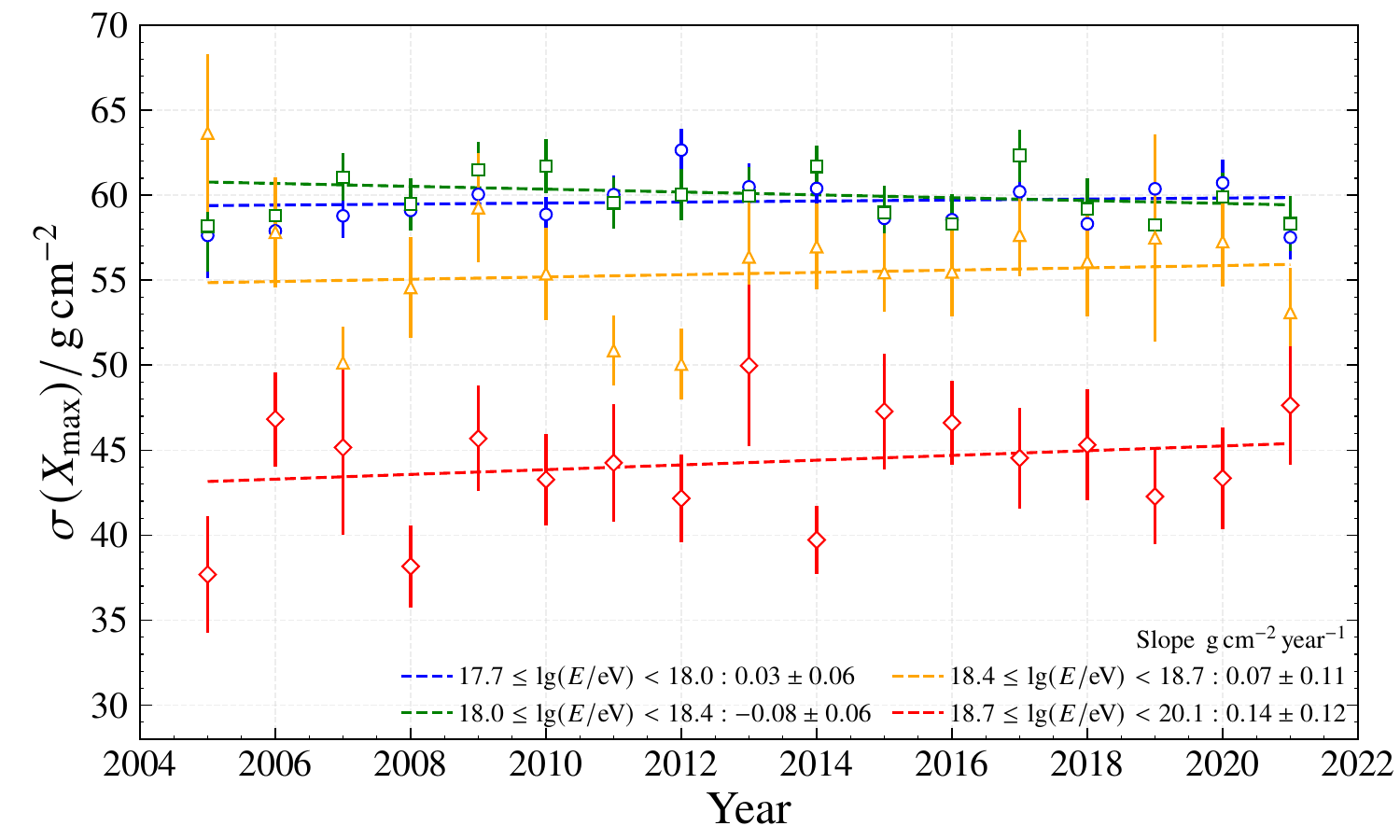}
    \caption{Time evolution of the mean and standard deviation of
      $\Xmax$ for four different energy bins. The dashed lines
      indicate linear fits, and the corresponding slopes are given in
      the legend.}
    \label{fig:time_check}
\end{figure*}
\begin{figure*}[t!] \centering
    \includegraphics[width=0.5\textwidth]{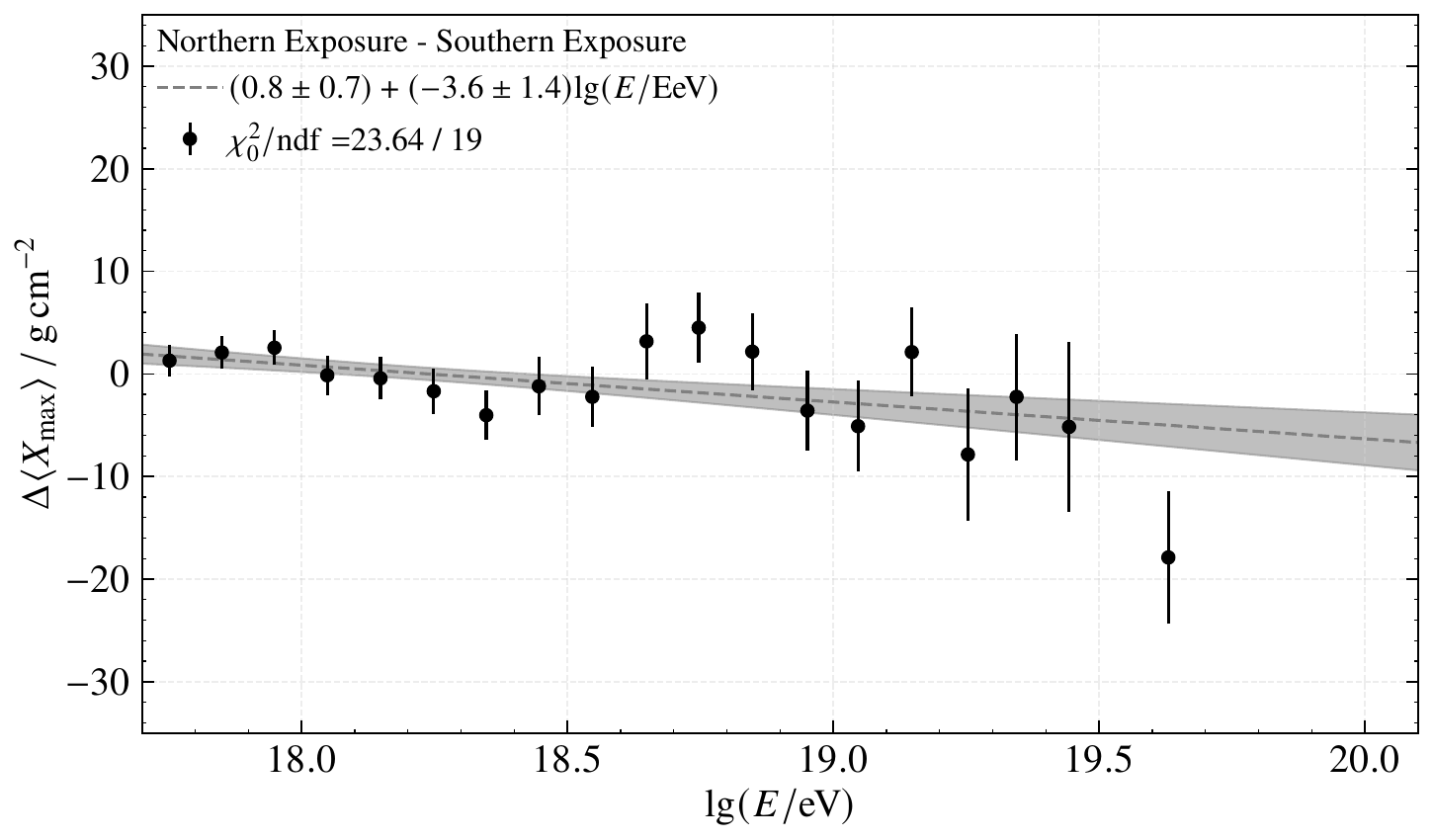}\includegraphics[width=0.5\textwidth]{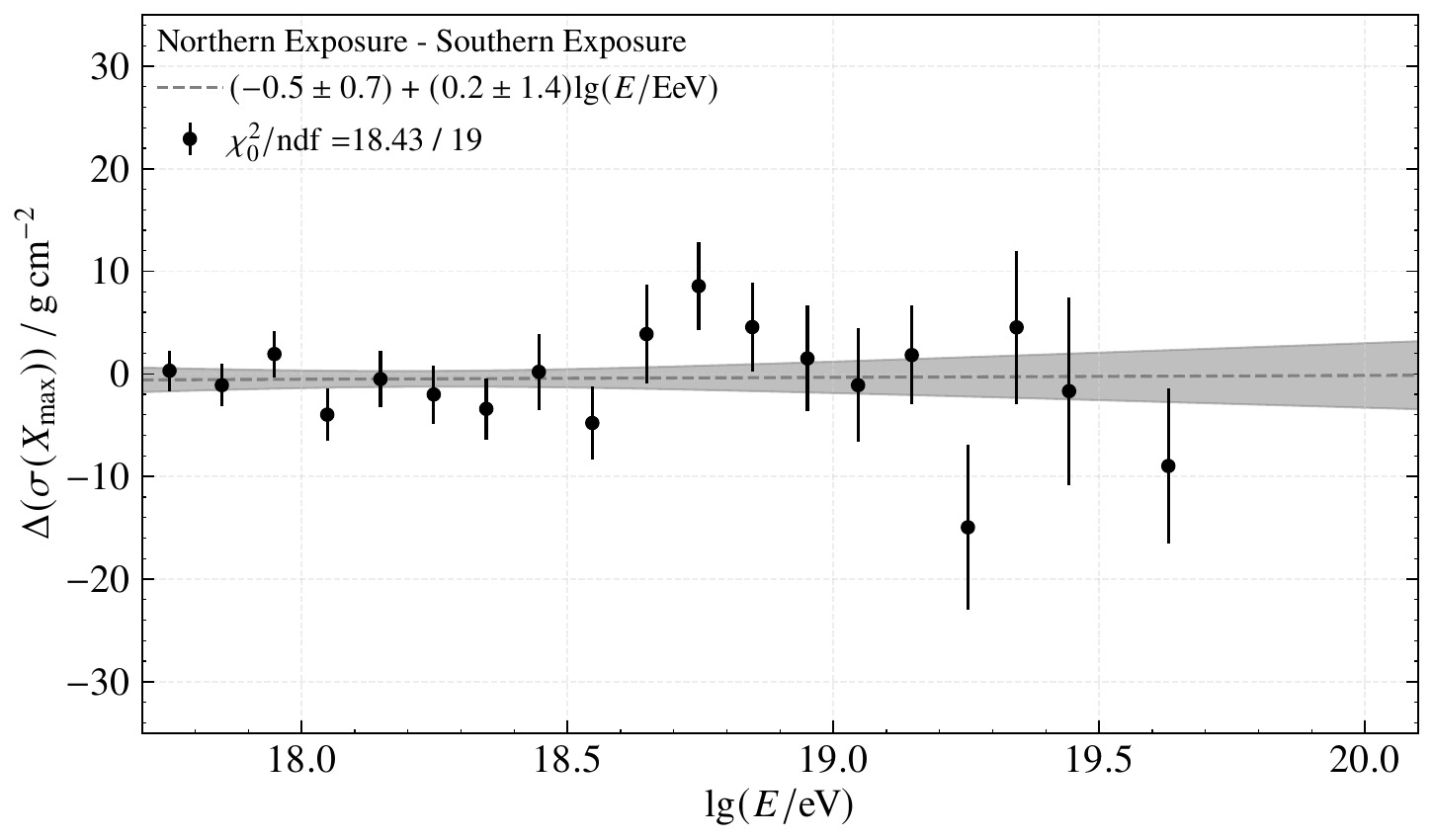}
    \caption{Comparison of events arriving from northern ($\delta \geq -15.7^\circ$) and southern ($\delta < -15.7^\circ$) portions of the FD exposure.
    Left: The difference (north - south) in \meanXmax.
    Right: The difference (north - south) in \sigmaXmax.
    A linear fit to the data is shown as a dashed line with a gray uncertainty contour.
    The $\chi^2_0/\mathrm{ndf}$ values for compatibility with zero are given, as are the linear fit values.}
    \label{fig:PAOTACompareMoments}
\end{figure*}

\subsection{FD sites}

The four FD sites are Los Leones (LL), Los Morados (LM), Loma Amarilla
(LA), and Coihueco (CO). Each is constructed similarly, though there
are site-to-site differences in placement, alignment, electronics, and
other factors that can affect the reconstruction of $\Xmax$. The top
panel of \cref{fig:cross_checks} shows, for each site, the differences
in $\meanXmax$ and $\sigmaXmax$ relative to the values obtained from
the combined data set. Testing the compatibility of these differences
with zero for all sites simultaneously yields $\chi^2 = 97$ for
$\meanXmax$ and $\chi^2 = 40$ for $\sigmaXmax$, both with 57 degrees
of freedom. The incompatibility of $\Delta\meanXmax$ with zero is largely driven by a systematic offset of a few \gcm for both CO and LM as compared to LL and LA in the energy range from $10^{17.7}$\,eV to $10^{18.5}$\,eV, likely related to site-dependent differences in the energy scale, as discussed in Ref.~\cite{Dawson:2020bkp}.

\subsection{Geometry and atmospheric quality}

The data set has been split by zenith angle, atmospheric aerosol
content, and distance to the event to probe different geometric and
atmospheric effects. The differences between the $\meanXmax$ values of
the two subsets for each case are shown on the left of
\cref{fig:cross_checks}, while the differences in $\sigmaXmax$ are
shown on the right. The corresponding $\chi^2$ values for
compatibility with zero, indicating no observable dependence, are also
shown.  For all three data splits, the differences are well contained
within the systematic uncertainties for the atmosphere and detector
presented in \cref{sec:detector}.

\paragraph{Inclined vs.\ vertical showers}
Events are divided into \emph{inclined} and \emph{vertical} subsets at $\med(\cos\theta) = 0.793 - 0.085\,\lg(E/\mathrm{EeV})$.
The altitude at which a given $\Xmax$ occurs increases with zenith angle as inclined showers reach the same slant depth higher in the atmosphere.
As a result, the detector acceptance and resolution differ between inclined and vertical events, potentially introducing a bias.
This effect is evaluated in the second row of \cref{fig:cross_checks}, which shows that the inclined subset is consistent with the vertical to better than 3\,\gcm $\meanXmax$.
$\Delta\sigmaXmax$ is compatible with zero.

\paragraph{Dusty vs.\ clean atmospheres}
Events are divided into \emph{clean} and \emph{dusty} subsets at $\med(\text{VAOD}) = 0.027 + 0.004 \lg(E/\text{EeV})$.
On nights with more atmospheric aerosols, the attenuation of fluorescence light on its way to the telescope is increased, and thus, potential biases in the aerosol determination are more important. This effect is investigated in the third row of \cref{fig:cross_checks}, and shows that both $\Delta \meanXmax$ and $\Delta \sigmaXmax$ are compatible with zero.

\paragraph{Close versus distant showers}
Events are divided into \emph{close} and \emph{far} subsets at $\med{d} = 10 + 9.6 \lg(E/\text{EeV})$\,km. This cross-check probes the understanding of the transmission through
the atmosphere, which is more important for the more distant data set.
The result of this test is shown in the last row of \cref{fig:cross_checks}, which shows that the differences in $\meanXmax$ in the far subset are within 2.5\,\gcm at low energies, and 3.5\,\gcm at the highest energies. $\Delta\sigmaXmax$ is compatible with zero.

\subsection{Time evolution}

Detector aging could introduce a time-dependent bias in the $\Xmax$
reconstruction.
\cref{fig:time_check} shows the yearly $\meanXmax$
and $\sigmaXmax$ values obtained from events at all energies, with
linear fits applied to quantify the trends.
The values of $\sigmaXmax$ are found to be very stable in time within
uncertainties, but a small time dependence of $\meanXmax$ is observed,
of about $+0.50\,\gcm/\text{year}$. Part of this effect can be
attributed to changes in the optical response of the telescopes. Therefore, it
should be regarded as part of the 14\% energy-scale
systematic uncertainty, as changes in energy translate into shifts of $\meanXmax$
via the elongation rate, see \cref{sec:moments}. After applying the
time-dependent response correction from~\cite{PierreAuger:2024neu},
which is compatible with recent measurements of the optical
efficiency~\cite{PierreAuger:2023blc}, the remaining change in
$\meanXmax$ amounts to about $+0.25\,\gcm/\text{year}$. The origin of
this residual effect is not known, but its magnitude is well within
the overall systematic uncertainties. We therefore conclude that over
the 17 years covered by this analysis, the \Xmax scale remains stable
to within $\pm 2\,\gcm$.

\subsection{Declination dependence of \texorpdfstring{$\boldXmax$}{Xmax}}

The statistics of the FD data set are sufficient to support an independent analysis of the $\Xmax$ distributions obtained from the northern and southern portions of the exposure of the Observatory.
The split between the northern and southern regions is made at a declination of $\delta = -15.7^\circ$, which was chosen as it represents the southernmost extent of the FD hybrid composition sensitivity of the majority of Telescope Array analyses~\cite{TelescopeArray:2018xyi}.
This data split results in 37\,956 events in the southern sub-sample and 20\,497 events in the northern sub-sample.

To prepare the $\Xmax$ measurements from each subset of the sky, the corrections described in \cref{sec:detector} have been calculated separately for the northern and southern declination bands.
The differences in reconstruction bias, resolution, and other systematics were found to be negligible.
Minor differences in $\Xmax$-dependent event acceptance were observed.
The systematic uncertainties in the comparison between the two regions arising from these corrections were found to be less than 2\,\gcm{} in any bin.

After these corrections, the first two moments of the \Xmax
distributions in the two declination bands were calculated and
compared in \cref{fig:PAOTACompareMoments}.
Overall, good agreement is observed between the two regions.
In the highest-energy bin, however, where statistics are limited, a larger difference does appear, with the northern portion of the sky having, on average, shallower $\Xmax$ values.
To evaluate any possible significance, a detailed statistical comparison of the two samples was performed using the Kolmogorov--Smirnov~\cite{massey1951kolmogorov} and Anderson--Darling~\cite{scholz1987k} two-sample tests in each energy bin. No statistically significant difference was found in any bin.

These results indicate that, within the exposure of the Observatory, there is no evidence for a declination dependence in the mass composition of UHECRs. This conclusion is consistent with the findings of the joint composition working group with the Telescope Array~\cite{PierreAuger:2023yym} and aligns with the absence of declination-dependent features in the UHECR energy spectrum reported by us in Ref.~\cite{PierreAuger:2025eun}.

\subsection{Assessment of validation studies}

The extensive validation studies demonstrate excellent stability and
reliability of the $\Xmax$ reconstruction and analysis.  The detector
resolution derived from simulation agrees with the resolution measured
in stereo events, and the full end-to-end analysis of simulated events
reproduces generated $\meanXmax$ and $\sigmaXmax$ values within the
assigned systematic uncertainties.  Site-to-site differences,
geometric and atmospheric dependencies, and temporal effects are small
and consistent with expectations and known uncertainties.  The
comparison of northern and southern declination bands shows no
statistically significant differences. Overall, the cross-checks
indicate that no unaccounted systematic effects are present.

\section{\label{sec:moments} Moments of \texorpdfstring{$\boldXmax$ and $\boldsymbol{\ln A}$}{Xmax and lnA}}

After applying the event selection and correcting for biases in \Xmax, the observed \Xmax distributions exhibit minimal biases with respect to the true distributions.

To enable comparison to the unbiased moments obtained directly from
the simulated showers at generator level, the residual bias of the
acceptance on the mean and standard deviation is corrected for using
the $\Lambda_\eta$ method as detailed in Ref.~\cite{PierreAuger:2014sui}.  In this
method, the observed \Xmax distribution is used directly in the region
of constant acceptance, while its tails are described by exponential
functions fitted with the acceptance taken into account; here, $\eta$
denotes the fraction of events in a given tail used for the fit, with
$\eta=0.20$ for the deep tail and $\eta=0.15$ for the leading
edge. The moments are then obtained from this combined distribution. This correction to the moments
from the residual acceptance bias is found to be small,  reaching at
most $3$\,\gcm for \meanXmax and $5$\,\gcm for \sigmaXmax, depending
on energy. The \Xmax resolution is then subtracted from $\sigmaXmax$ in quadrature.  Statistical uncertainties are calculated using a parametric bootstrap method described in~\cite{PierreAuger:2014sui}. The numerical values of \meanXmax and \sigmaXmax are listed in \cref{tab:Xmax_moments} (\cref{app:xmaxmoments}).

\begin{figure}[t] \centering
    \includegraphics[width=\linewidth]{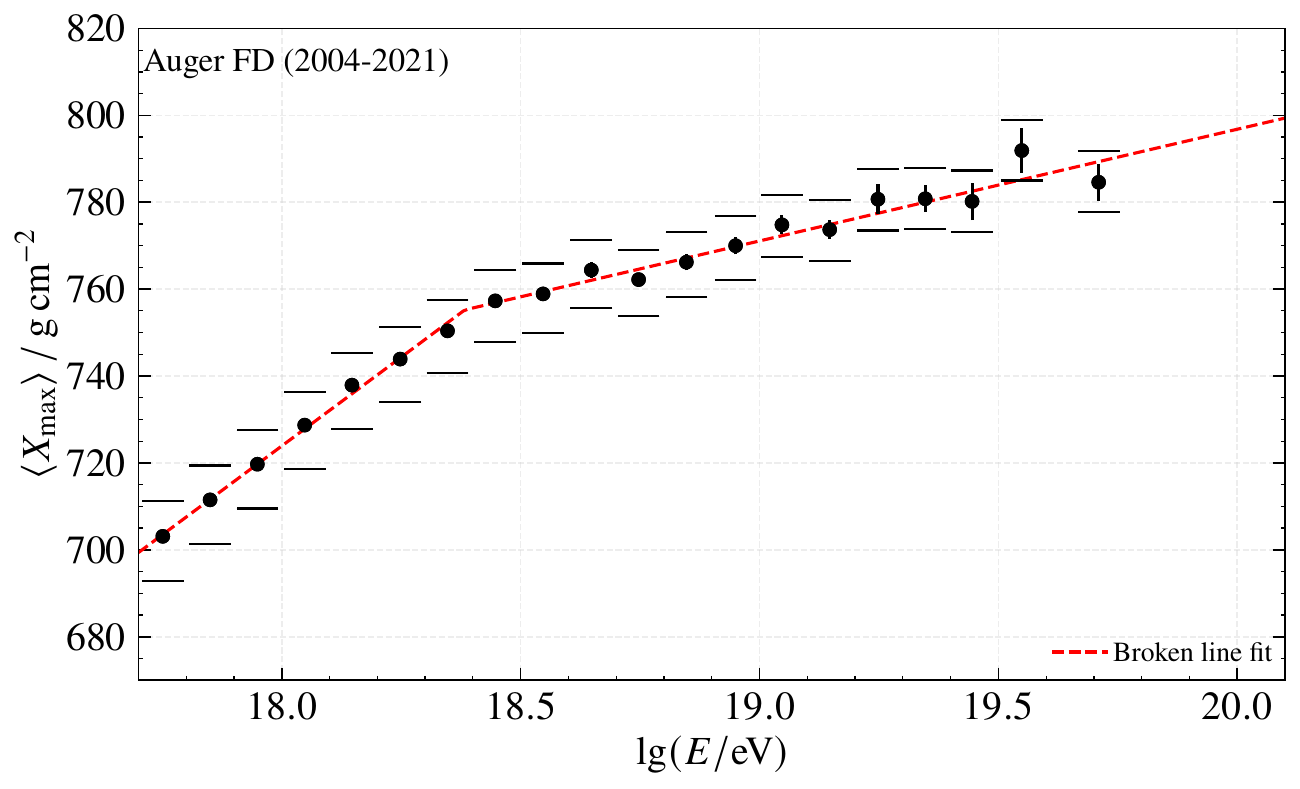}
    \caption{Energy evolution of \meanXmax. Statistical
      uncertainties are displayed using vertical error bars, while systematic
      uncertainties are indicated by horizontal caps above and below the markers. The red line shows a broken-line fit to the data. }
    \label{fig:meanXmaxFit}
\end{figure}

The energy evolution of \meanXmax is shown in \cref{fig:meanXmaxFit}.
The \emph{elongation rate},
\begin{equation}
  \label{eq:er}
  D_{10} = \dv{\meanXmax}{\lg E},
\end{equation}
exhibits a distinct transition from
\begin{equation}
    \label{eq:erLow}
    D_{10} = 81.8 \pm 1.7\, \text{(stat.)} \, ^{+1.9}_{-3.0} \,\text{(sys.)}\,\gcm/\text{decade}
\end{equation}
below $\lg(E_\text{break}/\text{eV}) = 18.38 \pm 0.02
\text{(stat.)}^{+0.01}_{-0.01} \text{(sys.)}$ to
\begin{equation}
    \label{eq:erHigh}
        D_{10} = 25.7 \pm 1.9 \,\text{(stat.)} \,^{+5.3}_{-1.5} \,\text{(sys.)}\,\gcm/\text{decade}
\end{equation}
above the break, in good agreement with the value of $24.1\pm 0.12\, \text{(stat.)}~\text{\gcm/decade}$
derived from SD data above $10^{18.5}$~eV~\cite{PierreAuger:2024flk}.
Note that the uncertainty quoted on $\lg(E_\text{break})$ does not include the
energy scale uncertainty of 14\%. The best-fit average shower maximum
at $\lg(E_\text{break})$ is $\Xmax^\text{break} = 755.2 \pm 1.5 \, \text{(stat.)}
\, ^{+5.3}_{-10.6} \, \text{(sys.)}$ \gcm.  The corresponding fit is shown as a red
dashed line in \cref{fig:meanXmaxFit} and results in a goodness of fit
of $\chi^2/\text{ndf} = 18.4/16$ ($p$-value${}=0.30$), indicating a
good agreement with the data. In contrast, a model using a single line
is decisively rejected by the data ($\chi^2/\text{ndf} =
512/18$).  Including two additional breaks at higher energies, as observed with the surface detector~\cite{PierreAuger:2024flk}, does not significantly improve the fit quality.

The observed elongation rates provide robust, model-independent
evidence with high statistical significance that the composition is
changing in the observed energy range, since for a constant
composition the elongation rate would remain
constant~\cite{Linsley:1981gh,Watson:2025vpu}.

Since \meanXmax depends linearly on \meanlnA, \cref{eq:er} implies that $D_{10}$ is linear in $\text{d}\meanlnA/\text{d}\lg E$.
Air shower simulations using the latest models of hadronic
interactions at ultra-high energies predict an elongation rate in the
range of 55 to 60\,\gcm/decade for a constant composition.
Therefore, the large elongation rate measured below the break, \cref{eq:erLow}, implies a transition from heavier to light composition.
Likewise, the smaller $D_{10}$, above the break,
\cref{eq:erHigh}, indicates a transition from light to a heavier composition.

\begin{figure*}[b] \centering
  \includegraphics[width=0.48\linewidth]{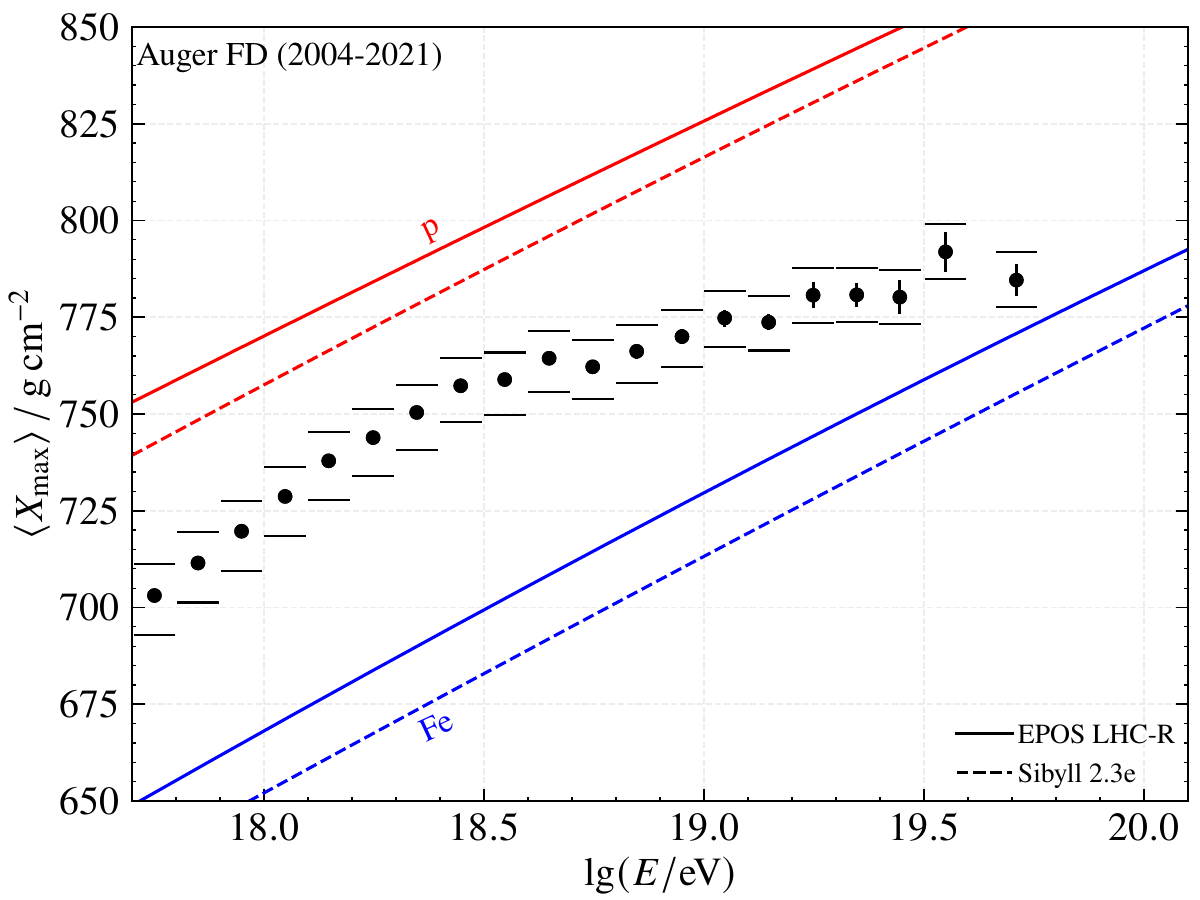}\hfill
  \includegraphics[width=0.48\linewidth]{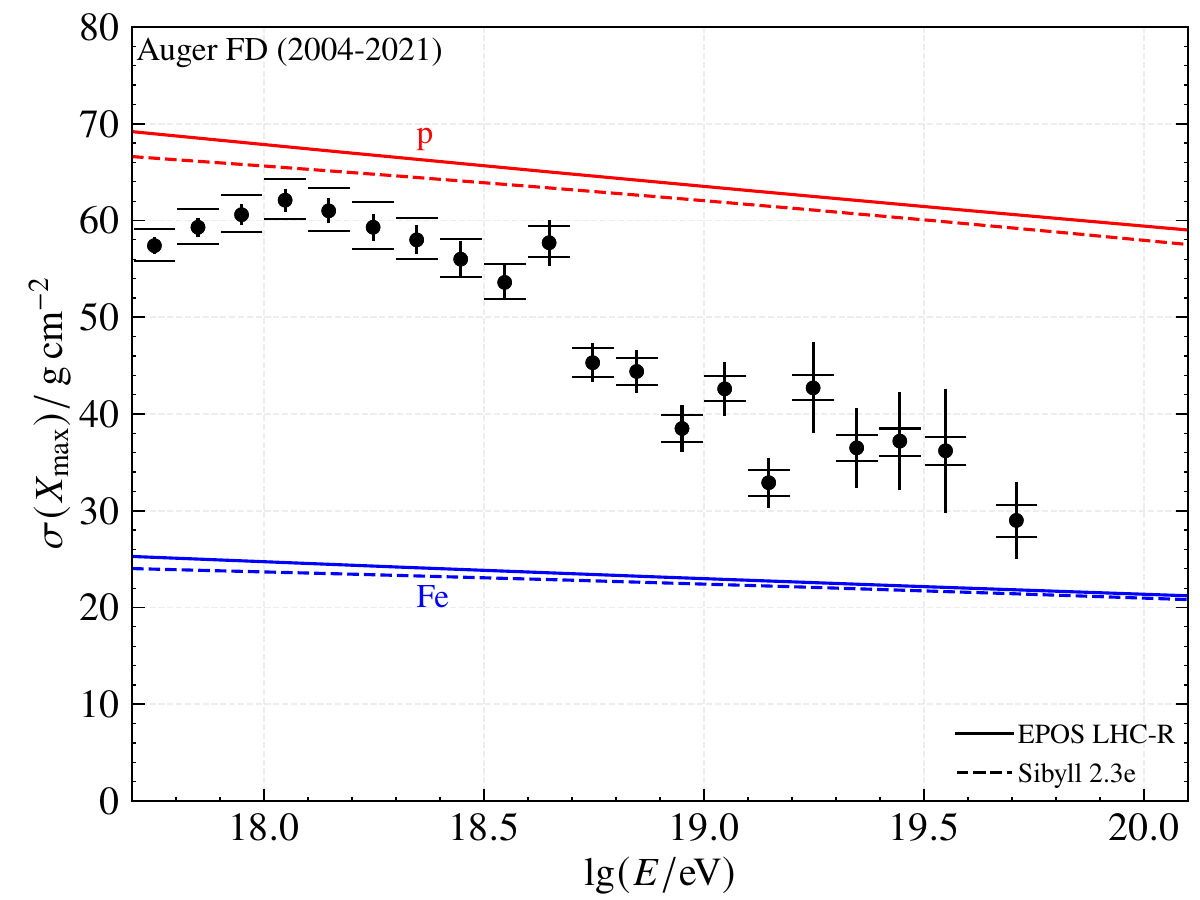}
    \caption{Energy evolution of \meanXmax and \sigmaXmax compared to
      the predictions from air-shower simulations using the hadronic interaction models \EPOSR and  \Sibylle for proton and iron primaries. Statistical
      uncertainties are displayed by vertical error bars, while systematic
      uncertainties are indicated by horizontal caps above and below the markers.}
    \label{fig:Xmax_moments}
\end{figure*}

Along with this change of the mean \Xmax, we also observe a decrease
in the standard deviation of the \Xmax distribution, \sigmaXmax.  It
is shown in the right panel of \cref{fig:Xmax_moments}. Whereas a
large value of \sigmaXmax can indicate either a light or mixed
composition, the pronounced decrease of \sigmaXmax from 60 to 30~\gcm is
another model-independent indicator for a transition to predominantly
heavy cosmic-ray nuclei at ultra-high energies.

In the energy bin \lgErange{18.6}{18.7}, the \sigmaXmax value is
considerably larger than expected from an interpolation between the
neighboring energy bins. The local significance of the deviation of
\sigmaXmax in this bin from the overall trend, evaluated by
excluding the bin from a fit with a smooth curve to the data, is
$3.4\sigma$. The global significance of observing such a fluctuation
in any energy bin is $2.4\sigma$.

It is worth noting that an event with an exceptionally deep \Xmax of
about 1200\,\gcm was recorded in this interval. Excluding this event
reduces \sigmaXmax by only 1\,\gcm, indicating that it does not drive
the observed increase. The event is consistent with an extreme
fluctuation of a proton-initiated shower; further details are given in
\cref{sec:extreme_shower}.

\begin{figure*}[b] \centering
    \includegraphics[width=0.48\linewidth]{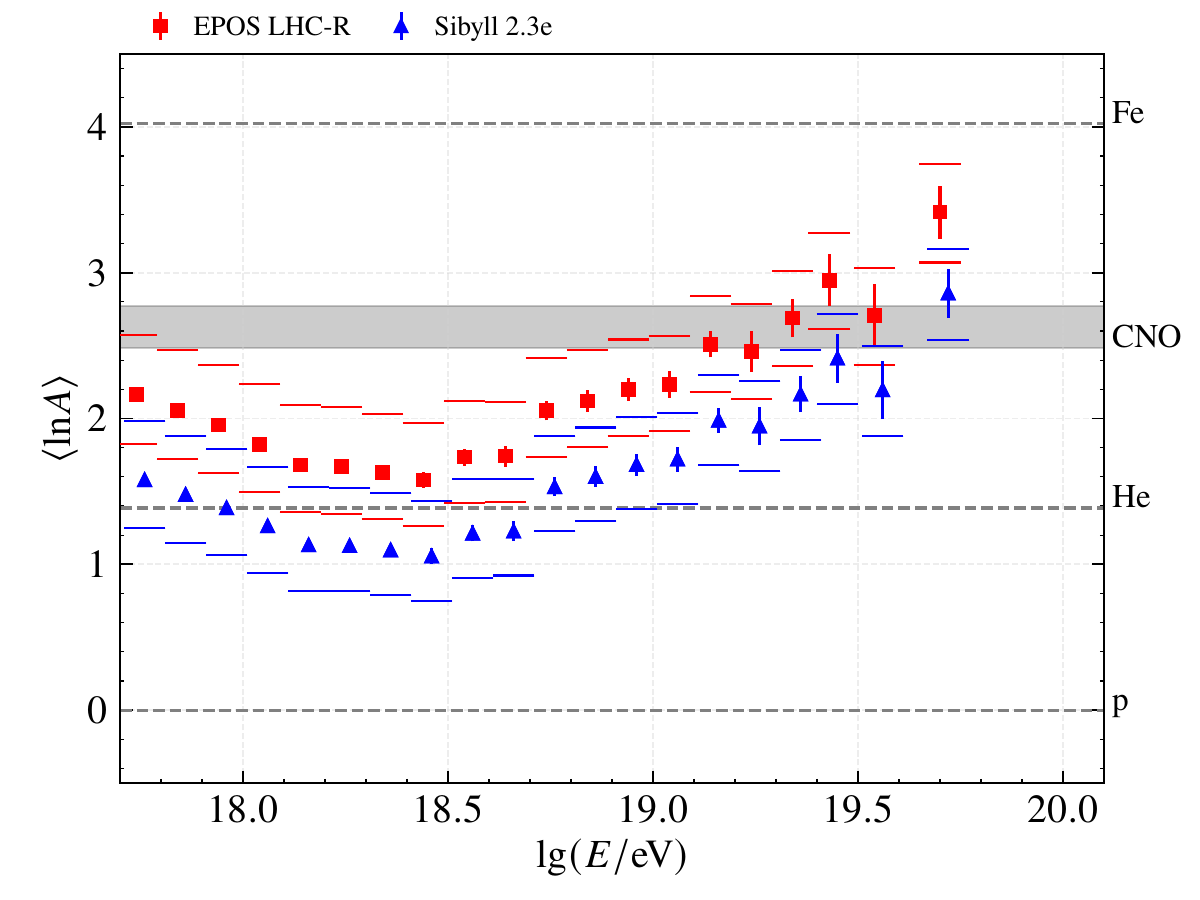}\hfill
    \includegraphics[width=0.48\linewidth]{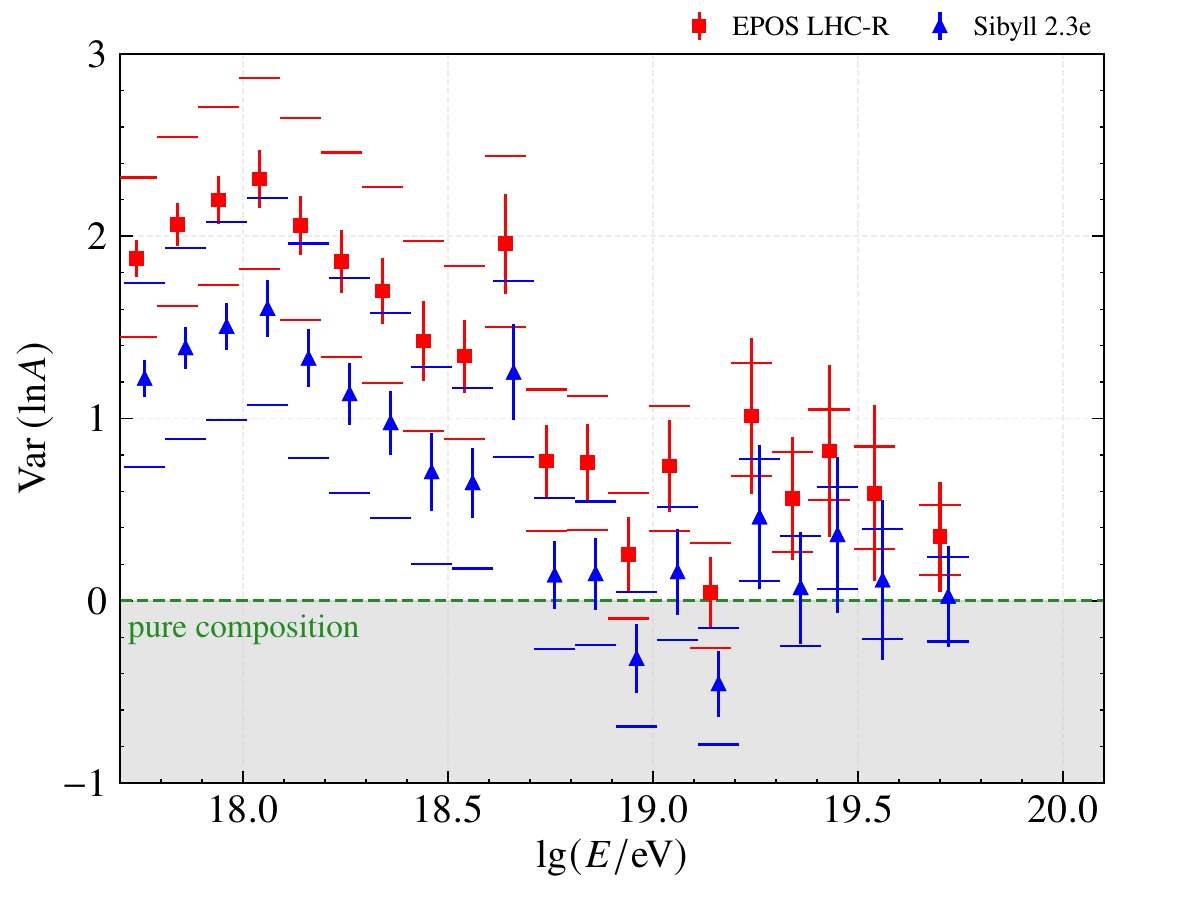}
    \caption{Energy evolution of \meanlnA and \varlnA estimated using
      air-shower simulations with the hadronic interaction models
      \EPOSR and \Sibylle. Statistical uncertainties are
      represented by vertical error bars, and systematic uncertainties are
      indicated by horizontal caps above and below the markers.}
    \label{fig:lnA_moments}
\end{figure*}

For further, but more model-dependent, insights into the cosmic-ray
composition, we compare in \cref{fig:Xmax_moments} the measured
\meanXmax and \sigmaXmax to the predictions of
\Sibylle~\cite{Riehn:2019jet} and \EPOSR~\cite{Pierog:2023ahq}
for protons and iron\footnote{We defer comparisons with the newest version of the QGSJet model (\QGSJetIII~\cite{Ostapchenko:2024myl}) to future work, pending resolution of a technical issue in the current implementation that affects simulated
nucleus-nucleus interactions (S. Ostapchenko, private communication).}. This allows an interpretation of how
the absolute values of \meanXmax and \sigmaXmax evolve between proton-
and iron-induced showers. The observed \meanXmax is close to predictions for helium at $E_\text{break}$. At the highest energies,
both \meanXmax and \sigmaXmax evolve toward the predictions for iron,
although within the current statistics also a somewhat lighter,
intermediate average mass of cosmic rays is compatible with the data.

The measured \meanXmax and \sigmaXmax in each energy bin can be
converted into the first two central moments of the logarithmic mass,
using the method described in Refs.~\cite{Linsley:1983ch,PierreAuger:2013xim}, and the parametrizations from~\cite{Evoli:2026bea}, see \cref{fig:lnA_moments}. Due to the small variations in the model
predictions of the \meanXmax elongation rate, a similar energy
dependence is observed in \meanlnA for both interaction
models. Below $E_\text{break}$ we obtain
\begin{multline*}
      \dv{\langle \ln A \rangle}{\lg E}
      = -0.94 \pm 0.07\,(\text{stat.})
      \; ^{+0.14}_{-0.07}\,(\text{sys.})
      \; \pm 0.05\,(\text{model}),
\end{multline*}
and above $E_\text{break}$
\begin{multline*}
      \dv{\langle \ln A \rangle}{\lg E}
      = +1.26 \pm 0.08\,(\text{stat.})
      \;^{+0.10}_{-0.21}\,(\text{sys.})
      \; \pm 0.01\,(\text{model}).
\end{multline*}

The spread among the model predictions of \meanXmax\ translates into a difference of about $0.5$ in \meanlnA, with \EPOSR yielding a heavier and \Sibylle a lighter average mass composition. At the
highest energies, the estimated mean logarithmic masses from both
models are close to the CNO group. The variances of \lnA are similar
and indicate an increase in mass mixing below
$10^{18.1}$\,eV, where they reach their maximum values. Above this
energy, \varlnA decreases until approximately $10^{18.8}$\,eV.

At higher energies, \varlnA shows no clear trend and remains
approximately constant at $0.3-0.5$, consistent with expectations for
equal mixtures of adjacent mass groups (p-He, He-CNO,
CNO-Fe). However, pure or more broadly mixed compositions cannot be
ruled out within the current uncertainties.

\section{\label{sec:fractions} Mass fractions}

The mean and variance of \lnA cannot be interpreted unambiguously in
terms of the relative fractions of primary nuclei, i.e.\ the mass
fractions. Instead, the observed \Xmax distributions can be fitted
with a superposition of simulated \Xmax distributions for different
nuclei using their relative fractions as fit parameters. In our
previous work~\cite{PierreAuger:2014gko}, we demonstrated that satisfactory fits
to the data can be achieved using at least four species representing
light, intermediate, and heavy mass groups. Specifically, we used
protons and three representative nuclei (helium, nitrogen, and iron),
nearly equally spaced in \lnA, and investigated the energy evolution
of the mass fractions of these four species.

\subsection{Templates for Monte Carlo simulations of \texorpdfstring{$\boldsymbol \Xmax$}{Xmax}}

The $\Xmax$ distributions used for comparison with the data are
generated using the \Conex air-shower program~\cite{Bergmann:2006yz} for two
hadronic interaction models: \EPOSR and \Sibylle. To
account for the effects of detector acceptance and resolution, these
distributions are modified according to \cref{eq:Xmaxreso,eq:accept}.  The
resulting composition \emph{templates} are binned in $\Xmax$ over a
range from 0 to 2000\,\gcm, with a bin
width of 10\,\gcm. The energy
distribution in each energy bin follows a power law $E^{-\alpha}$,
where the spectral index $\alpha$ is estimated using the energy
distribution of the selected events
(cf.\ \cref{fig:final_selection}). We use $\alpha=1.1$ for energies below $10^{18.0}$\,eV and $\alpha=2.2$ for higher
energies. Above $10^{19.6}$\,eV, the spectral
index is estimated as $\alpha=4.7$.

Because both the hybrid exposure~\cite{PierreAuger:2010swb} and the
fiducial selection efficiency increase with energy, these values of
$\alpha$ for the observed energy distribution differ significantly
from the spectral indices of the UHECR energy spectrum.

\subsection{Fitting procedure}

\begin{figure}[t] \centering
  \includegraphics[clip,rviewport=0 0 1 1,width=0.8\linewidth]{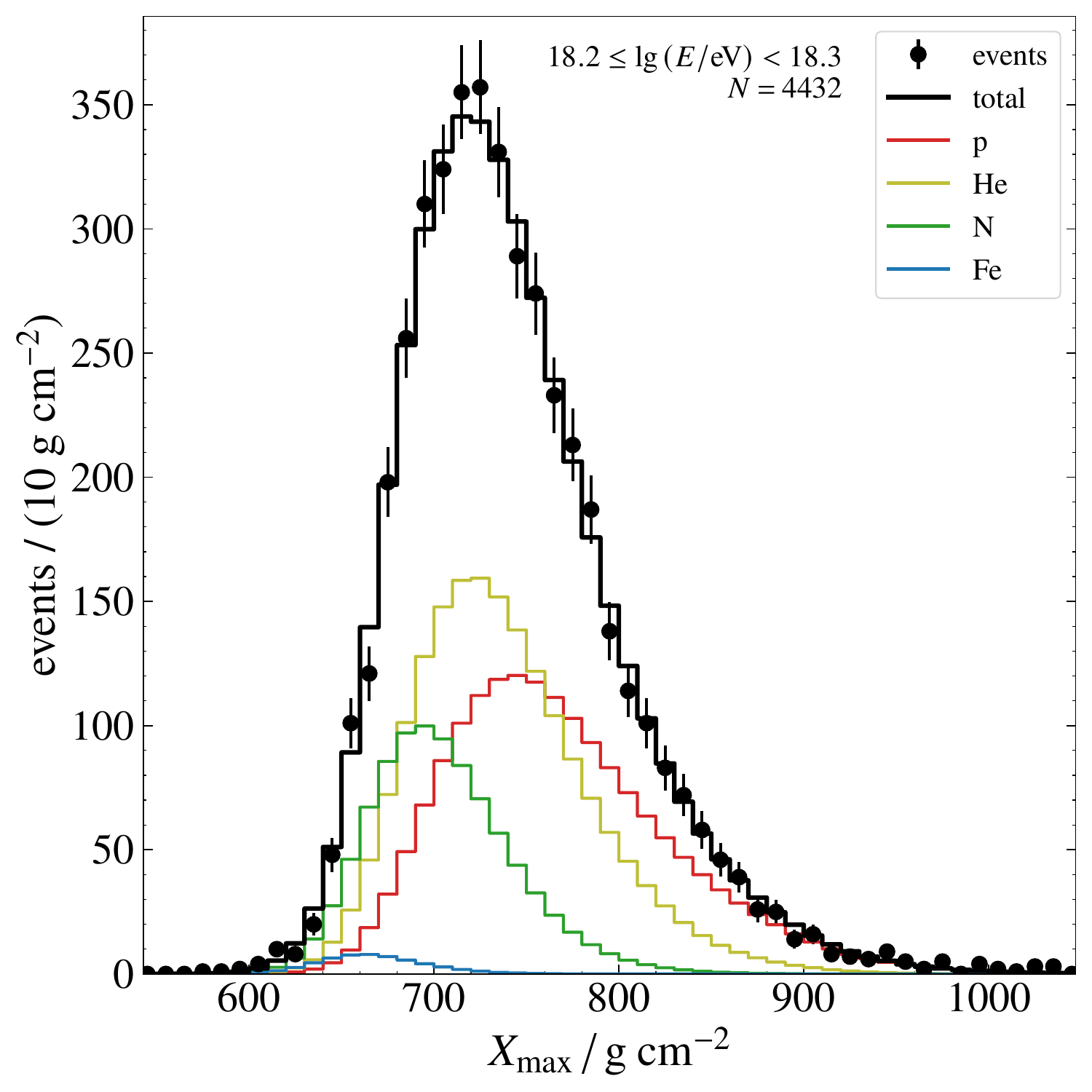}
    \caption{Example of the fit of composition fractions of the mass
      groups (H, He, N, Fe) with \Sibylle in the energy range
      \lgErange{18.2}{18.3}: \Xmax distribution of the data
      (points with error bars) and composition templates
      (histograms) scaled by their fitted fractions.}
    \label{fig:XmaxFit}
\end{figure}

\begin{figure}[t] \centering
 \includegraphics[width=\linewidth]{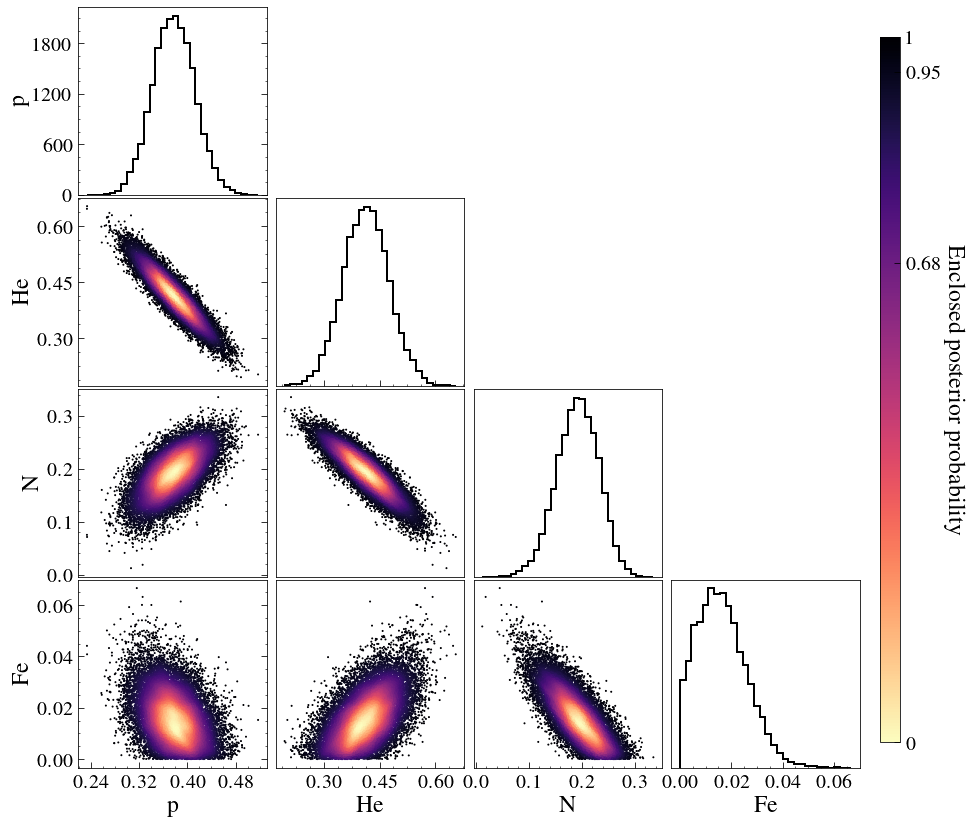}
\caption{Corner plot of the posterior mass-fraction parameters for
  H, He, N and Fe obtained with \Sibylle in the energy range
  \lgErange{18.2}{18.3}. The diagonal panels show the one-dimensional
  marginalized posteriors for each group, while the off-diagonal panels
  show the two-dimensional joint posteriors (pairwise correlations). Colors indicate the enclosed posterior probability, with the lighter colors corresponding to regions of higher posterior density.}
\label{fig:Corner_plot}
\end{figure}

A Poissonian binned maximum-likelihood approach is
used~\cite{Baker:1983tu} to determine the best-fitting combination of
primary species, with the likelihood function described in
Ref.~\cite{PierreAuger:2014gko}. To estimate the posterior distribution of the
mass fractions, we apply a Bayesian Markov Chain Monte Carlo method,
using the sampler described in Ref.~\cite{Foreman-Mackey:2012any}.
The mass fractions are restricted to the interval $[0,1]$, subject to
the constraint that their sum equals 1.  A symmetric
$\operatorname{Dirichlet}(1,1,1,1)$ prior is adopted for the
fractions, corresponding to a uniform prior on the allowed fraction
simplex.  We use the maximum a-posteriori estimator to provide
numerical values that characterize the posterior distribution from the
fit.  The statistical uncertainties on the estimated composition
fractions are expressed using the 68\% highest-posterior-density
interval, representing the 1-$\sigma$ confidence range. An example of
a fit and the one- and two-dimensional posterior probabilities are
shown in \cref{fig:XmaxFit,fig:Corner_plot}. The fitted \Xmax
distributions for all energies and both interaction models are
presented in \cref{app:xmaxdistrfits}. As can be seen, because of the
large overlap of the \Xmax distribution of different mass groups, the
fractions of neighboring groups are negatively correlated.

To evaluate the fit quality, the \Xmax distributions for the best-fit
mass fractions are used to generate mock samples of the same size as
the data. The likelihood of each mock sample with respect to the
best-fit model is compared to the likelihood found for the data to
count the fraction of worse fits. This fraction, the $p$-value,
is a statistical measure of how well the best-fit model represents the
underlying data.

\subsection{Systematic uncertainties}

Alongside the statistical uncertainties, we evaluate the systematic
uncertainties on the fitted mass fractions, primarily arising from the
systematic uncertainty in $\Xmax$. This contribution is evaluated by
varying the $\Xmax$ scale within its uncertainty range using steps of
$0.1(\sigma_\mathrm{low}$ + $\sigma_\mathrm{up})$, where
$\sigma_\mathrm{low}$ and $\sigma_\mathrm{up}$ represent the lower and upper bounds of the
systematic-uncertainty intervals. This approach is necessary due to
the non-monotonic relationship between the fractions and changes in
the $\Xmax$ scale.  We re-fit the shifted and re-binned data, taking
the largest and smallest obtained fractions as the limits of the lower
and upper systematic uncertainties. Similarly, the smallest and
largest $p$-values identified in this procedure serve as systematic
uncertainties on the $p$-values. Another source of systematic
uncertainty arises from the energy scale, which is evaluated by
re-fitting the data after shifting the energy scale by
$\pm14\%$. As a cross-check, the mass fractions were also evaluated within the frequentist approach previously used in Ref.~\cite{PierreAuger:2014gko}. The results were found to be consistent with the Bayesian method.
Finally, we estimated the contribution from uncertainties
in the parametrization of the detector acceptance and resolution by
re-fitting the data using extreme values of the parametrizations.


\begin{figure*}[p] \centering
    \includegraphics[clip,rviewport=-0.0 0 1 1,width=\textwidth]{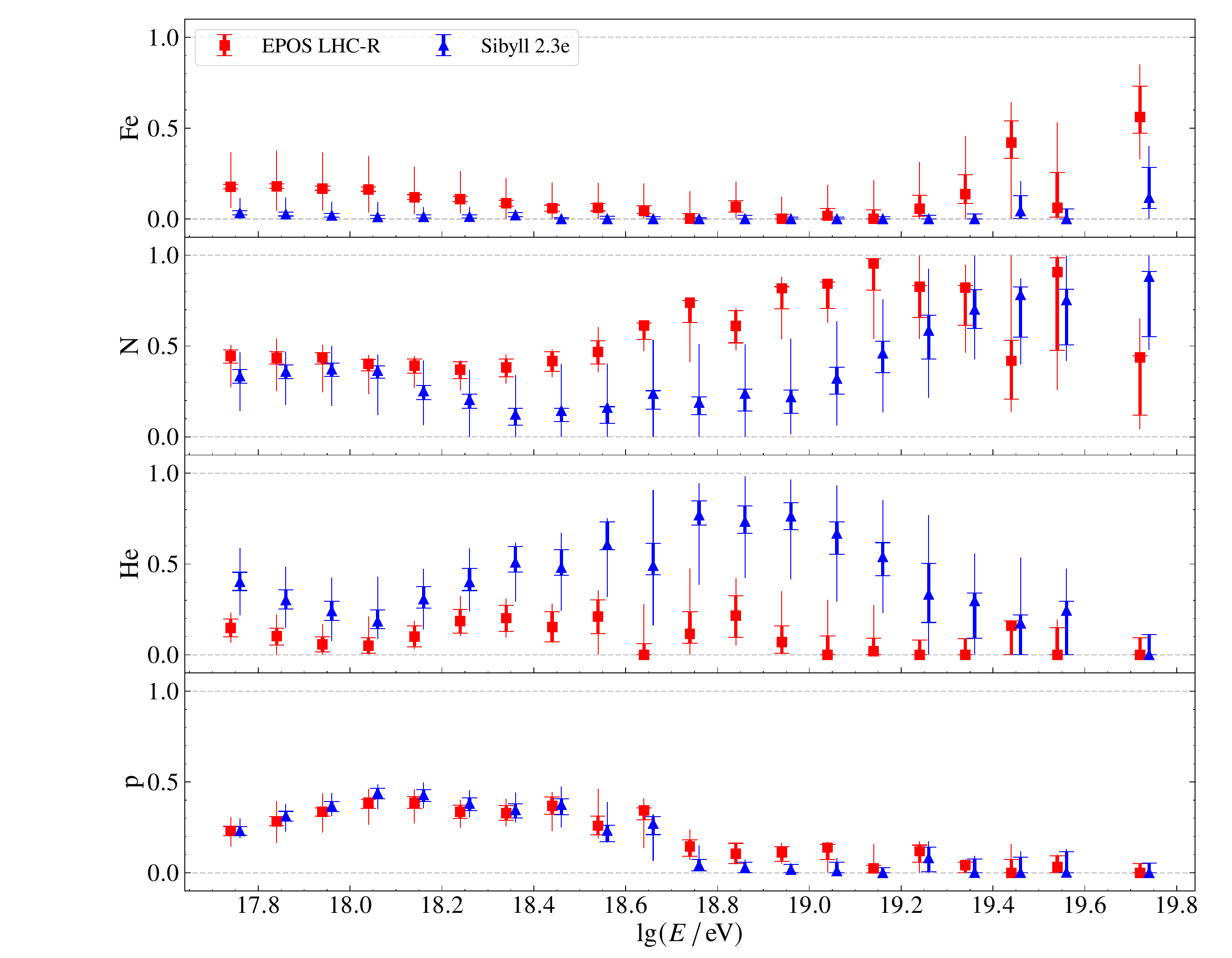}
    \includegraphics[width=\textwidth]{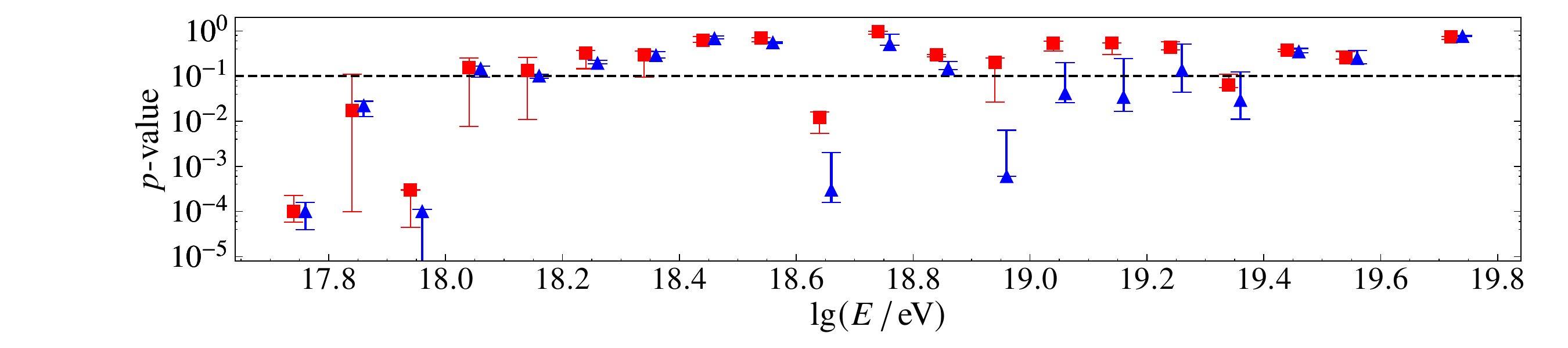}
    \caption{Top four panels: best-fit fractions of protons, helium,
      nitrogen, and iron nuclei using air-shower simulations with the \Sibylle and \EPOSR hadronic interactions models. The error bars denote statistical (thick lines)
      and total (thin lines) uncertainties. Bottom panel: $p$-values of the fit. The error bars denote systematic uncertainties.}
\label{mass_composition_5panels}
\end{figure*}

\subsection{Results}

The results for the determined mass fractions of protons, helium,
nitrogen and iron nuclei are shown in
\cref{mass_composition_5panels}, using \EPOSR and \Sibylle. The $p$-values are shown in the lower panel and indicate
an overall good agreement between the data and the model predictions.
Over most of the energy range, the $p$-values are compatible with $p
\gtrsim 10^{-1}$, except in three energy bins where they drop by
several orders of magnitude. We verified that, in these bins, the
$p$-values improve by at most one order of magnitude when excluding
the ``outliers'' discussed in \cref{sec:extreme_shower} or when adding
an intermediate heavy mass group, represented by silicon, to the fit.

The cosmic-ray mass composition obtained with
\Sibylle contains an approximately equal proton-helium-nitrogen mixture at the lowest energies, with helium becoming dominant above $10^{18.5}$ eV followed by a transition to nitrogen dominance above $10^{19.0}$ eV. The iron contribution remains consistent with zero within uncertainties at all
energies. These results align qualitatively and
quantitatively with our previous estimates~\cite{PierreAuger:2014gko}. In contrast, the composition inferred using \EPOSR is heavier owing to its deeper \Xmax scale (cf.\ \cref{fig:Xmax_moments,fig:lnA_moments}). The nitrogen fraction increases from approximately 50\% at the lowest energies to above 80\% above the ankle, before the iron contribution becomes significant at the highest energies, accounting for about half of the total mix. The helium fraction derived from \EPOSR remains at a level of
5-10\% across the entire energy range and shows no systematic trend
with energy.

The qualitative behavior of the proton fraction is consistent for both models. This can be understood in terms of the evolution of the \Xmax fluctuations, which, like the proton fraction, first increase and then decrease as the deep tails of the \Xmax distributions die out (\cref{fig:Xmax_moments,fig:Xmax_distributions}).

\section{\label{sec:conclusions}Summary and Conclusion}

In this paper, we have presented measurements of the \Xmax
distributions with minimal selection bias using the hybrid data of the
\PAO. The data set comprises \allEASEvents high-quality events
with energies above $10^{17.7}$~eV. The most energetic event has an energy of about $1.17 \times 10^{20}$\,eV.

The mean and standard deviation of the $\Xmax$ distributions have been
corrected for the residual biases, enabling direct comparison to
current and future predictions from air-shower simulations.  From
these measurements, we draw two key observations that are independent
of theoretical uncertainties of the modeling of hadronic interactions
at ultra-high energies\footnote{These conclusions assume that no
  qualitatively new high-energy physics processes alter air-shower
  development at ultra-high energies. For speculative scenarios in
  which this assumption could be violated, see, e.g.,
  Refs.~\cite{Farrar:2019cid,Pavlidou:2018yux}.}. First, the rate at
which \meanXmax evolves with the logarithm of energy exhibits a marked
change at $\approx10^{18.4}$\,eV, providing evidence that the trend
toward lighter mass composition reverses at this energy. Second, the
\Xmax distributions become narrower above approximately the same
energy, indicating a transition to a heavier and less-mixed mass
composition.  A comparison of the $\Xmax$ distributions in the
northern and southern declination bands of the Observatory's exposure
shows no statistically significant difference, indicating no
directional dependence of the mass composition.

For a more direct interpretation of these findings, we have converted
\Xmax moments into $\meanlnA$ and $\varlnA$ using the interaction
models \EPOSR and \Sibylle. The universality of model
predictions for the \meanXmax elongation rate leads to similar
\meanlnA energy evolution for both models, with
\(\mathrm{d}\langle \ln A \rangle/\mathrm{d}\lg E\) changing from $-0.94$ to $+1.26$ at the \meanXmax break.
At the highest energies, the
\meanlnA values inferred with both models suggest an average mass
approaching that of the CNO group. The spread of the primary masses
increases until $10^{18.1}$\,eV after which
it decreases until approximately $10^{18.8}$\,eV. Beyond this energy,
\varlnA fluctuates around $0.5$ without exhibiting a clear trend, and
it remains compatible with a pure or more mixed composition within the
uncertainties.

Finally, we have used the measured $\Xmax$ distributions to examine
the energy evolution of four primary mass groups, protons, helium,
nitrogen, and iron nuclei. We observe similar general trends with the
proton contribution peaking around $10^{18.1}$\,eV and nearly
disappearing above the ankle in the cosmic-ray energy spectrum. As the
proton fraction decreases, the relative contribution of
intermediate-mass nuclei increases, with helium dominating near the
ankle in \Sibylle and nitrogen in \EPOSR. The difference is due to the deeper \Xmax\ scale of \EPOSR that is also favored by our analyses combining \Xmax\ and the SD signal~\cite{PierreAuger:2016qzj,PierreAuger:2024neu,PierreAuger:2025eiw,PierreAuger:2025zrl}. At higher energies, nitrogen becomes dominant, with the iron fraction remaining negligible in \Sibylle but gradually increasing toward the highest energies in \EPOSR.

The events selected for this analysis represent an unprecedented data
set of direct observations of the energy and depth of maximum of
air-shower profiles. As demonstrated in the paper, our measurements
provide unique insights into the mass composition of UHECRs across two
decades in energy. Importantly, this data set serves as the reference
for calibration of indirect \Xmax inferences from the surface detector
data~\cite{PierreAuger:2017tlx,PierreAuger:2024nzw,PierreAuger:2024flk,Stadelmaier:2025vty}. The
list of \Xmax and energy of each event used in this analysis, as well
as the derived measurements of \Xmax and \lnA moments and the
posterior samples of mass fractions are available for download from
Ref.~\cite{supplementaryMaterial}. Together with the information provided
in \cref{app:param} on the acceptance and resolution, this unique data
set can be used for self-contained studies of mass composition and
hadronic interactions.

\subsection*{Acknowledgments}
\section*{Acknowledgments}

\begin{sloppypar}
The successful installation, commissioning, and operation of the Pierre
Auger Observatory would not have been possible without the strong
commitment and effort from the technical and administrative staff in
Malarg\"ue. We are very grateful to the following agencies and
organizations for financial support:
\end{sloppypar}

\begin{sloppypar}
Argentina -- Comisi\'on Nacional de Energ\'\i{}a At\'omica; Agencia Nacional de
Promoci\'on Cient\'\i{}fica y Tecnol\'ogica (ANPCyT); Consejo Nacional de
Investigaciones Cient\'\i{}ficas y T\'ecnicas (CONICET); Gobierno de la
Provincia de Mendoza; Municipalidad de Malarg\"ue; NDM Holdings and Valle
Las Le\~nas; in gratitude for their continuing cooperation over land
access; Australia -- the Australian Research Council; Belgium -- Fonds
de la Recherche Scientifique (FNRS); Research Foundation Flanders (FWO),
Marie Curie Action of the European Union Grant No.~101107047; Brazil --
Minist\'erio da Ci\^encia, Tecnologia e Inova\c{c}\~ao (MCTI); Czech Republic --
GACR 24-13049S, CAS LQ100102401, MEYS LM2023032,
CZ.02.1.01/\allowbreak0.0/\allowbreak0.0/\allowbreak16{\textunderscore}013/\allowbreak0001402, CZ.02.1.01/\allowbreak0.0/\allowbreak0.0/\allowbreak18{\textunderscore}046/\allowbreak0016010
and CZ.02.1.01/\allowbreak0.0/\allowbreak0.0/\allowbreak17{\textunderscore}049/\allowbreak0008422 and
CZ.02.01.01/\allowbreak00/\allowbreak22{\textunderscore}008/\allowbreak0004632; France -- Centre de Calcul IN2P3/CNRS;
Centre National de la Recherche Scientifique (CNRS); Institut National
de Physique Nucl\'eaire et de Physique des Particules (IN2P3/CNRS);
Germany -- Bundesministerium f\"ur Forschung, Technologie und Raumfahrt
(BMFTR); Deutsche Forschungsgemeinschaft (DFG); Ministerium f\"ur Finanzen
Baden-W\"urttemberg; Helmholtz Alliance for Astroparticle Physics (HAP);
Hermann von Helmholtz-Gemeinschaft Deutscher Forschungszentren e.V.;
Ministerium f\"ur Kultur und Wissenschaft des Landes Nordrhein-Westfalen;
Ministerium f\"ur Wissenschaft, Forschung und Kunst des Landes
Baden-W\"urttemberg; Italy -- Istituto Nazionale di Fisica Nucleare
(INFN); Istituto Nazionale di Astrofisica (INAF); Ministero
dell'Universit\`a e della Ricerca (MUR); CETEMPS Center of Excellence;
Ministero degli Affari Esteri (MAE), ICSC Centro Nazionale di Ricerca in
High Performance Computing, Big Data and Quantum Computing, funded by
European Union NextGenerationEU, reference code CN{\textunderscore}00000013; M\'exico --
Consejo Nacional de Ciencia y Tecnolog\'\i{}a (CONACYT-SECHTI)
No.~CB-A1-S-46703, Universidad Nacional Aut\'onoma de M\'exico (UNAM)
PAPIIT-IN114924; Benem\'erita Universidad Aut\'onoma de Puebla (BUAP), VIEP
and Laboratorio Nacional de Superc\'omputo del Sureste de M\'exico (LNS);
and Benem\'erita Universidad Aut\'onoma de Chiapas (UNACH); The Netherlands
-- Ministry of Education, Culture and Science; Netherlands Organisation
for Scientific Research (NWO); Dutch national e-infrastructure with the
support of SURF Cooperative; Poland -- Ministry of Science and Higher
Education, grant No.~2022/WK/12; National Science Centre, grants
No.~2020/39/B/ST9/01398, and 2022/45/B/ST9/02163; Portugal -- Portuguese
national funds and FEDER funds within Programa Operacional Factores de
Competitividade through Funda\c{c}\~ao para a Ci\^encia e a Tecnologia
(COMPETE); Romania -- Ministry of Education and Research, contract
no.~30N/2023 under Romanian National Core Program LAPLAS VII, and grant
no.~PN 23 21 01 02; Slovenia -- Slovenian Research and Innovation
Agency, grants P1-0031, I0-0033; Spain -- Ministerio de Ciencia,
Innovaci\'on y Universidades/Agencia Estatal de Investigaci\'on MICIU/AEI
/10.13039/501100011033 (PID2022-140510NB-I00, PCI2023-145952-2,
CNS2024-154676, and Mar\'\i{}a de Maeztu CEX2023-001318-M), Xunta de Galicia
(CIGUS Network of Research Centers, Consolidaci\'on ED431C-2025/11 and
ED431F-2022/15) and European Union ERDF; USA -- Department of Energy,
Contracts No.~DE-AC02-07CH11359, No.~DE-FR02-04ER41300,
No.~DE-FG02-99ER41107 and No.~DE-SC0011689; National Science Foundation,
Grant No.~0450696, and NSF-2013199; The Grainger Foundation;
Astrophysics Centre for Multi-messenger studies in Europe (ACME) EU
Grant No 101131928; and UNESCO.
\end{sloppypar}

\bibliography{biblio_protected}
\appendix
\crefalias{section}{appendix}
\section{\label{sec:extreme_shower} Extreme shower events}

\subsection{Deepest event in the data sample}

\begin{figure}[t] \centering
  \includegraphics[width=\linewidth]{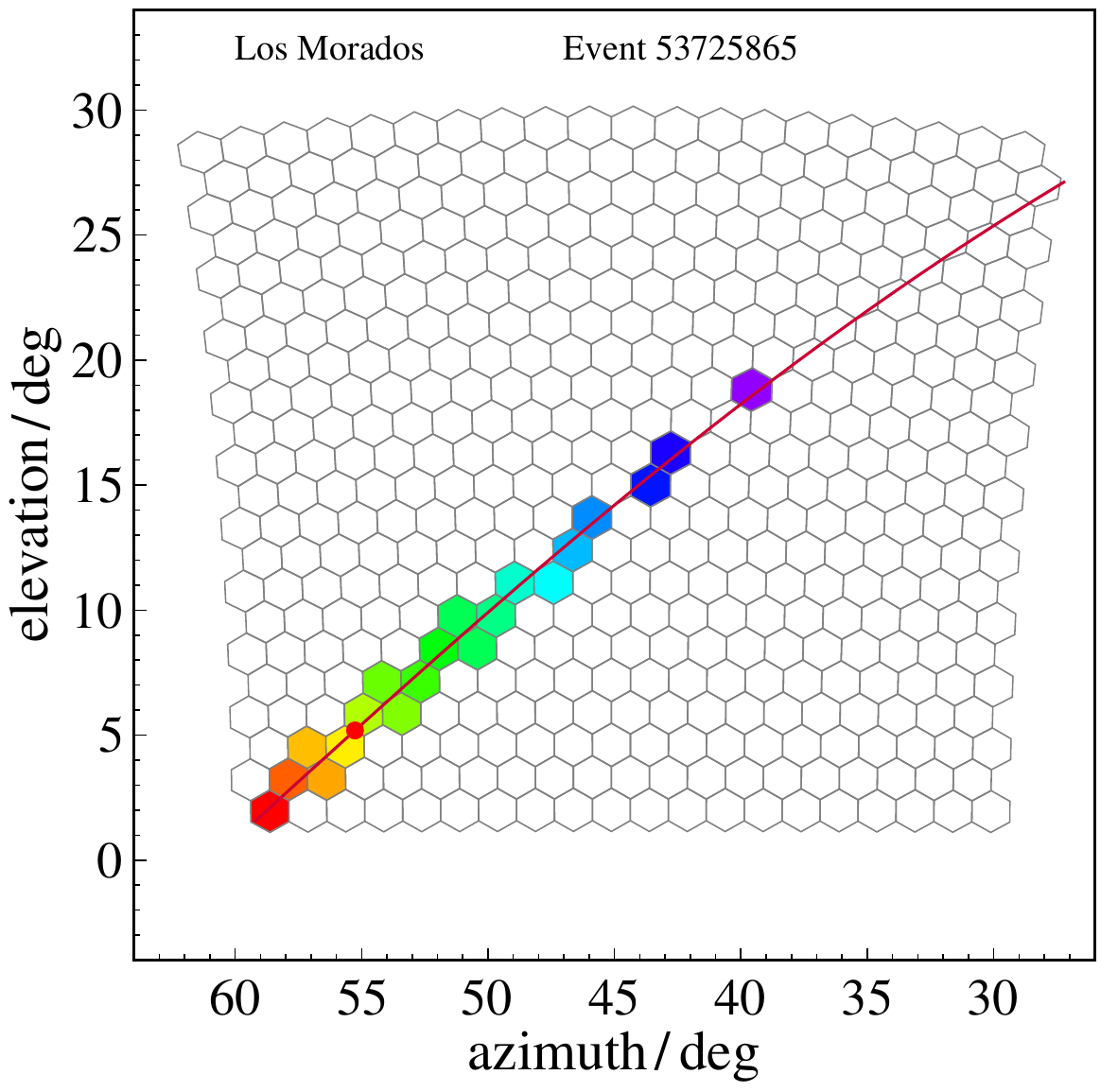}
   \includegraphics[width=\linewidth]{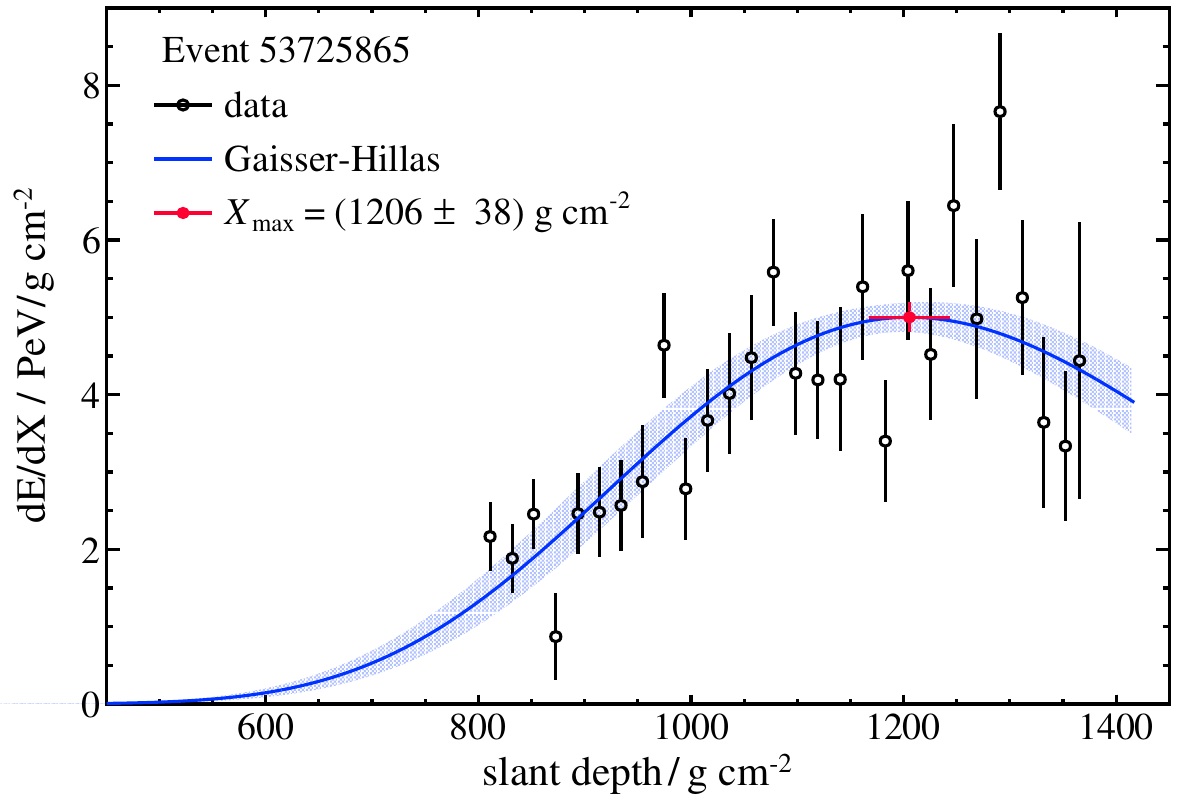}
    \caption{Reconstruction of the deep shower event
      \#53725865. Top panel: Camera view of the event. Colors
      denote the arrival time of photons in the hexagonal
      photomultipliers, with blue to red corresponding to early to
      late arrival times. The red line indicates the fitted
      shower--detector plane. Bottom panel: Longitudinal
      energy-deposit profile and fit with a Gaisser--Hillas function.}
    \label{fig:deep_shower}
\end{figure}

Event \#53725865 is the deepest event observed in this data set. Its
image in one of the cameras of the Los Morados site and the
corresponding longitudinal profile are shown in
\cref{fig:deep_shower}. The fit of the longitudinal profile yields
${\Xmax}_{,\,\rm LM} = 1206 \pm 38$\,\gcm and an energy of $E_{\rm
  LM}=4.2 \pm 0.6$~EeV. An independent reconstruction using data from
a second FD site (Loma Amarilla) yields compatible values of
${\Xmax}_{,\,\rm LA} = 1168 \pm 90$\,\gcm and $E_{\rm LA} = 3.9 \pm
0.5$~EeV, but this reconstruction does not pass the quality cuts.
In the following, we test whether the event, given its \Xmax, is
compatible with a deeply penetrating shower induced by a proton.

The corresponding probabilities, derived from air-shower simulations,
are listed in \cref{tab:deep_event_probabilities}. In all cases
discussed below, detector acceptance and \Xmax resolution are taken
into account, and the quoted uncertainties are statistical.

As a first step, we assume the event is a proton primary with energy
$E=E_{\rm LM}$ and compute the probability $p(E)$ that any single proton
shower is reconstructed with an \Xmax value at least as large as the one
observed. This probability is below $0.006$\,\%.

To account for the fact that multiple proton events are observed in the
same energy bin, we compute the penalized probability $\tilde p(E)$ by
multiplying $p(E)$ with the expected number of protons in that bin, using
the proton fraction from \cref{sec:fractions} for the relevant interaction
model. This yields a probability of order $1\%$.

Finally, we extend the calculation to the full dataset to obtain
$p_{\mathrm{full}}$. For this, we simulate the full energy range using the
measured composition from \cref{sec:fractions} and count how often a shower
is observed whose \Xmax exceeds the proton mean by at least as much as
event \#53725865 exceeds the proton mean at its energy. The resulting fully penalized probabilities range from about 0.2 to 0.6,
depending on the interaction model.

We therefore conclude that, while event \#53725865 is a remarkably deep
shower, it is entirely compatible with expectations, given the
fitted proton fractions from \cref{sec:fractions}.

\begin{table}[!htb]
    \centering
\caption{Probabilities to observe a proton shower at least as deep as
  event \#53725865. The first column gives the probability for
  protons at the energy of the observed event, $E$. The second column
  includes the trial factor, taking into account the number of proton
  events expected in the energy bin containing the
  event. The third column extends the trial factor to all protons in the full data set.}
\label{tab:deep_event_probabilities}
    \begin{tabular}{lccc}
      \toprule
        model &
        $p(E)$ / $10^{-5}$ &
        $\tilde p(E)$ &
        $p_\mathrm{full}$ \\
        \midrule
        \EPOSR      & $5.6 \pm 0.8$ & $0.025 \pm 0.003$ & $0.58 \pm 0.01$ \\
        \Sibylle    & $2.0 \pm 0.3$ & $0.007 \pm 0.001$ & $0.24 \pm 0.01$ \\
        \bottomrule
    \end{tabular}
\end{table}

\subsection{``Shallow'' showers}
Next, we characterize the number of shallow showers in the data set using the
condition
\begin{equation}
  X_{\max} < X_0 +  D_{10}\, \lg(E/\mathrm{EeV}),
  \label{eq:shallow}
\end{equation}
where $X_0 = 550~\gcm$ corresponds to a value
$100~\gcm$ below the expected $\Xmax$ for iron primaries at
1\,EeV, and $D_{10} = 60$~\gcm is the elongation rate
expected for a constant mass composition. With this definition, we find
seven shallow showers in the data set. For simulated events generated using the
\Sibylle\ hadronic interaction model, assuming an energy spectrum and mass
composition consistent with the data, we expect $5 \pm 2$ shallow events.
We therefore conclude that the data set does not contain events with an
unexpectedly shallow depth of shower maximum.

\section{\label{app:param} Parametrizations}

\subsection{\label{app:acceptance_E} Acceptance}

The energy dependence of the acceptance parameters \text{$\mathbf{p} = (X_1,\, X_2,\, \lambda_1,\, \lambda_2)$}, introduced in \cref{eq:accept}, can be described by second-order polynomials in $x_{18} = \lg(E / 10^{18}\,\mathrm{eV})$
\begin{equation}
  p_i = \sum_{j=0}^2 a_{ij} \, x_{18}^j.
\end{equation}
The fit parameters
are summarized in \cref{tab:acceptance_E} together with their
statistical uncertainties. Systematic uncertainties on $p_i$ are
estimated by assuming that the dominant uncertainty in the detector
simulations is related to modeling the production and transmission of
fluorescence light and the detector's optical response. This
uncertainty is equal to the uncertainty of the Auger energy scale of
14\%. A table of these parameters, including the fully correlated
systematic uncertainties, can be found in
Ref.~\cite{supplementaryMaterial}.

\begin{table}[!h]
  \centering
  \caption{\label{tab:acceptance_E} Parameters for $X_1$, $X_2$, $\lambda_1$ and $\lambda_2$.}
    \begin{tabular}{lcccc}
     \toprule
        & $a_0$ & $a_1$ & $a_2$ \\
        \midrule
        $X_1$ & $599.0 \pm 1.0$ & $\phantom{1}46.1 \pm 2.6$ & $-19.0 \pm 1.7$  \\
        $X_2$ & $882.5 \pm 0.9$ & $\phantom{1}49.9 \pm 1.9$ & $-21.8 \pm 1.0$  \\
        $\lambda_1$ & $122.8 \pm 0.9$ & $195.0 \pm 3.4$ & $\phantom{-}35.5 \pm 3.4$  \\
        $\lambda_2$ & $108.9 \pm 0.7$ & $\phantom{1}74.0 \pm 1.6$ & $\phantom{-1}9.1 \pm 1.0$  \\
        \bottomrule
    \end{tabular}
\end{table}

\subsection{Resolution}
\label{app:resotable}
\cref{tab:resolution} gives the parameters of \Xmax resolution, \cref{eq:Xmaxreso}, including all
effects discussed in \cref{sec:resoFromSim,sec:moreReso}.

\begin{table}[!h]
   \centering
  \caption[result]{Parameters of the \Xmax resolution, \cref{eq:Xmaxreso}.
    $\sigma_{1}$ and $\sigma_2$ are in \gcm. The uncertainties are systematic
   and fully correlated between  $\sigma_1$ and $\sigma_2$.}
  \label{tab:resolution}
  \begin{tabular}{cccc}
    \toprule
   $\lgE $ & $\sigma_1$ & $\sigma_2$ & $f$\\ \midrule
    17.7-17.8&        18.9 $\pm$         0.9&        32.5 $\pm$         1.6&        0.71\\
17.8-17.9&        17.7 $\pm$         0.9&        31.2 $\pm$         1.5&        0.72\\
17.9-18.0&        16.7 $\pm$         0.8&        30.0 $\pm$         1.5&        0.73\\
18.0-18.1&        15.8 $\pm$         0.8&        29.0 $\pm$         1.5&        0.74\\
18.1-18.2&        15.0 $\pm$         0.8&        28.0 $\pm$         1.5&        0.74\\
18.2-18.3&        14.3 $\pm$         0.8&        27.2 $\pm$         1.6&        0.75\\
18.3-18.4&        13.8 $\pm$         0.9&        26.4 $\pm$         1.6&        0.76\\
18.4-18.5&        13.3 $\pm$         0.9&        25.8 $\pm$         1.7&        0.77\\
18.5-18.6&        13.0 $\pm$         0.9&        25.2 $\pm$         1.8&        0.77\\
18.6-18.7&        12.6 $\pm$         0.9&        24.7 $\pm$         1.8&        0.78\\
18.7-18.8&        12.4 $\pm$         1.0&        24.2 $\pm$         1.9&        0.78\\
18.8-18.9&        12.2 $\pm$         1.0&        23.9 $\pm$         2.0&        0.79\\
18.9-19.0&        12.0 $\pm$         1.0&        23.5 $\pm$         2.0&        0.79\\
19.0-19.1&        11.8 $\pm$         1.1&        23.3 $\pm$         2.1&        0.79\\
19.1-19.2&        11.7 $\pm$         1.1&        23.0 $\pm$         2.1&        0.79\\
19.2-19.3&        11.6 $\pm$         1.1&        22.8 $\pm$         2.2&        0.79\\
19.3-19.4&        11.5 $\pm$         1.1&        22.6 $\pm$         2.2&        0.79\\
19.4-19.5&        11.4 $\pm$         1.1&        22.5 $\pm$         2.2&        0.79\\
19.5-19.6&        11.4 $\pm$         1.2&        22.3 $\pm$         2.3&        0.79\\
19.6-$\;\,\infty\;\,$&        11.3 $\pm$         1.2&        22.1 $\pm$         2.3&        0.79\\

  \end{tabular}
\end{table}

\section{Comparison of \texorpdfstring{\boldXmax}{Xmax} moments}
\begin{figure}[h!] \centering
    \includegraphics[width=\linewidth]{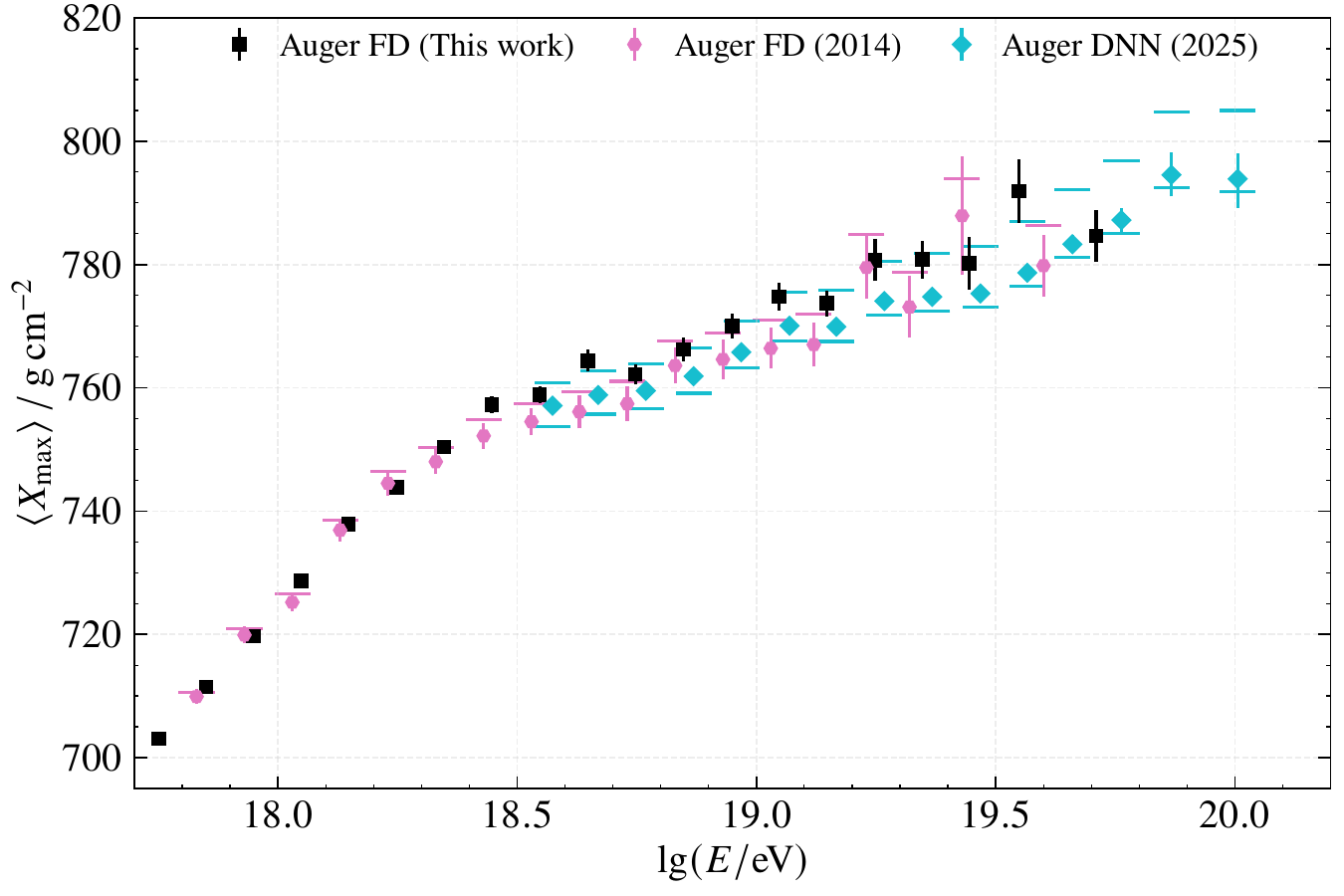}\\
    \includegraphics[width=\linewidth]{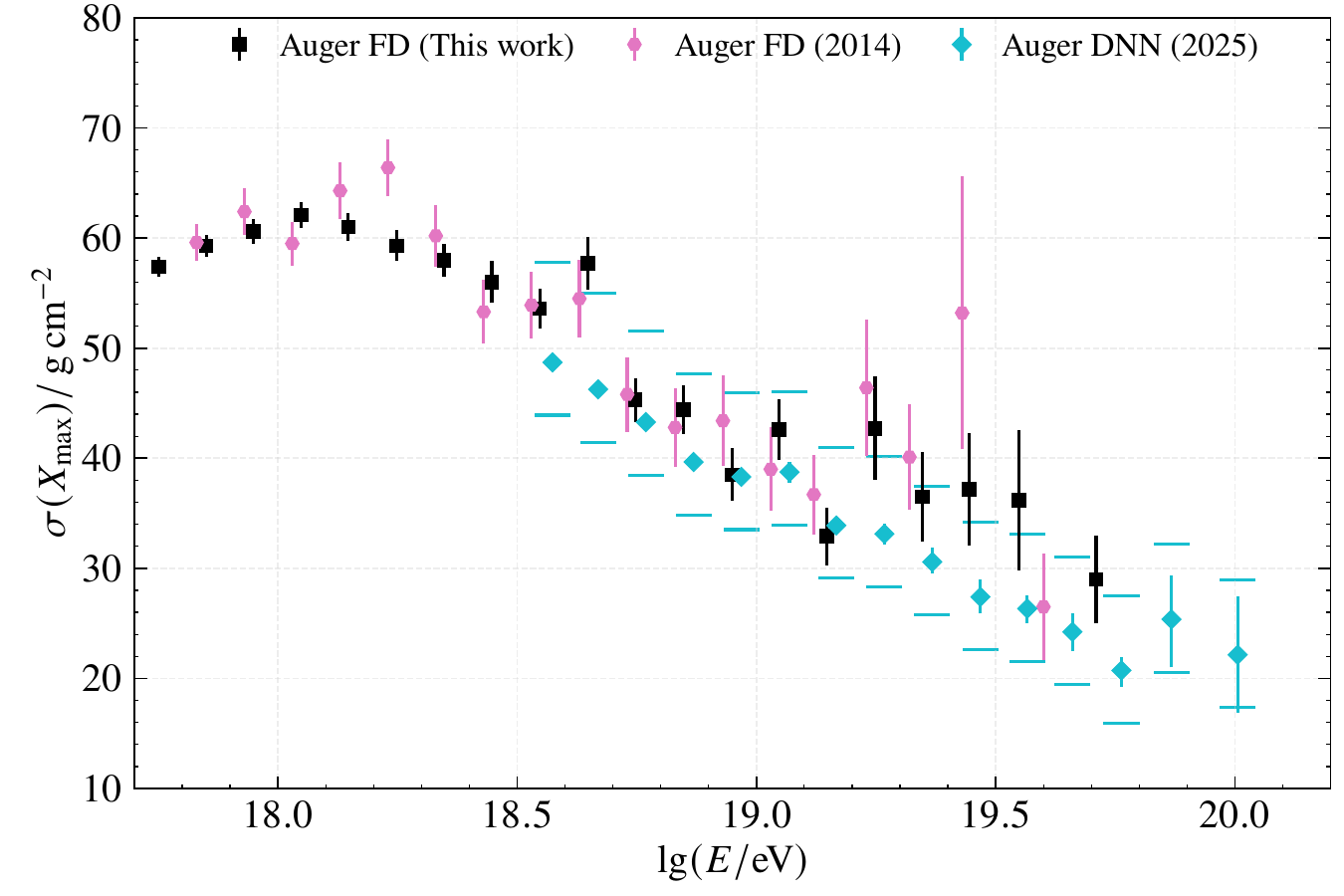}
    \caption{$\meanXmax$ and $\sigmaXmax$ versus energy for different results produced by the Pierre Auger Collaboration (PRD14 \cite{PierreAuger:2014sui}, DNN \cite{PierreAuger:2024nzw}). Error bars denote the statistical uncertainties, while the horizontal caps indicate the uncorrelated systematic uncertainties.}
    \label{fig:merged_moments}
\end{figure}

In \cref{fig:merged_moments}, we present the $\meanXmax$ and
$\sigmaXmax$ moments obtained in this work and compare them to
previous results from the Pierre Auger Collaboration: the earlier FD
$\Xmax$ analysis (``FD\,(2014)''~\cite{PierreAuger:2014sui}) and the deep neural
network analysis based on surface detector data
(``DNN\,(2025)''~\cite{PierreAuger:2024nzw}). An
energy-dependent $\meanXmax$-difference is observed relative to these
earlier results.  The shift is expected from the improved aerosol
treatment used in this analysis, in particular the correction for
residual aerosol contamination in the clear reference nights used for
data normalization~\cite{PierreAuger:2023nbk}. Previously, this effect
was accounted for as a systematic uncertainty and it is now corrected
explicitly, resulting in shifted central values and reduced systematic
uncertainties. Only the uncorrelated systematic uncertainties relevant
for the relative comparison are shown. For FD\,(2014), this is the
uncertainty associated with the VAOD treatment.  For DNN\,(2025), the
shown uncertainty is the quadratic sum of the same VAOD-related
contribution, inherited through the calibration on FD data
reconstructed prior to the aerosol-analysis update, and the
uncorrelated uncertainty associated with the inference of \Xmax from
the surface-detector data (see Fig.~11 of
\cite{PierreAuger:2024nzw}).

Good agreement with FD\,(2014) is found for $\sigmaXmax$. The fluctuations
derived in the DNN\,(2025) analysis are systematically smaller above
$10^{19.2}$~eV than those found in this work, but are consistent within the
systematic uncertainties of the DNN analysis, which amount to about
$^{+7}_{-5}$~\gcm.

The UHE results reported by the Telescope Array Collaboration cannot
be directly compared to ours, as their published $\Xmax$ moments are
not corrected for detector effects~\cite{TelescopeArray:2018xyi}.
Instead, we refer to the results of the Auger-TA composition working group, which
compares the data at detector level and finds good agreement between
the two experiments~\cite{PierreAuger:2023yym}. A comparison to TA
results below
$10^{18.5}$~eV~\cite{TelescopeArray:2020bfv,TelescopeArray:2026rdu}
remains to be carried out in a dedicated joint study.

\clearpage
\onecolumngrid

\section{\label{app:xmaxdistrfits}Fits of composition fractions to \texorpdfstring{$\boldXmax$}{Xmax} distributions}
\begin{figure*}[h!]
    \centering
    \includegraphics[width=\linewidth]{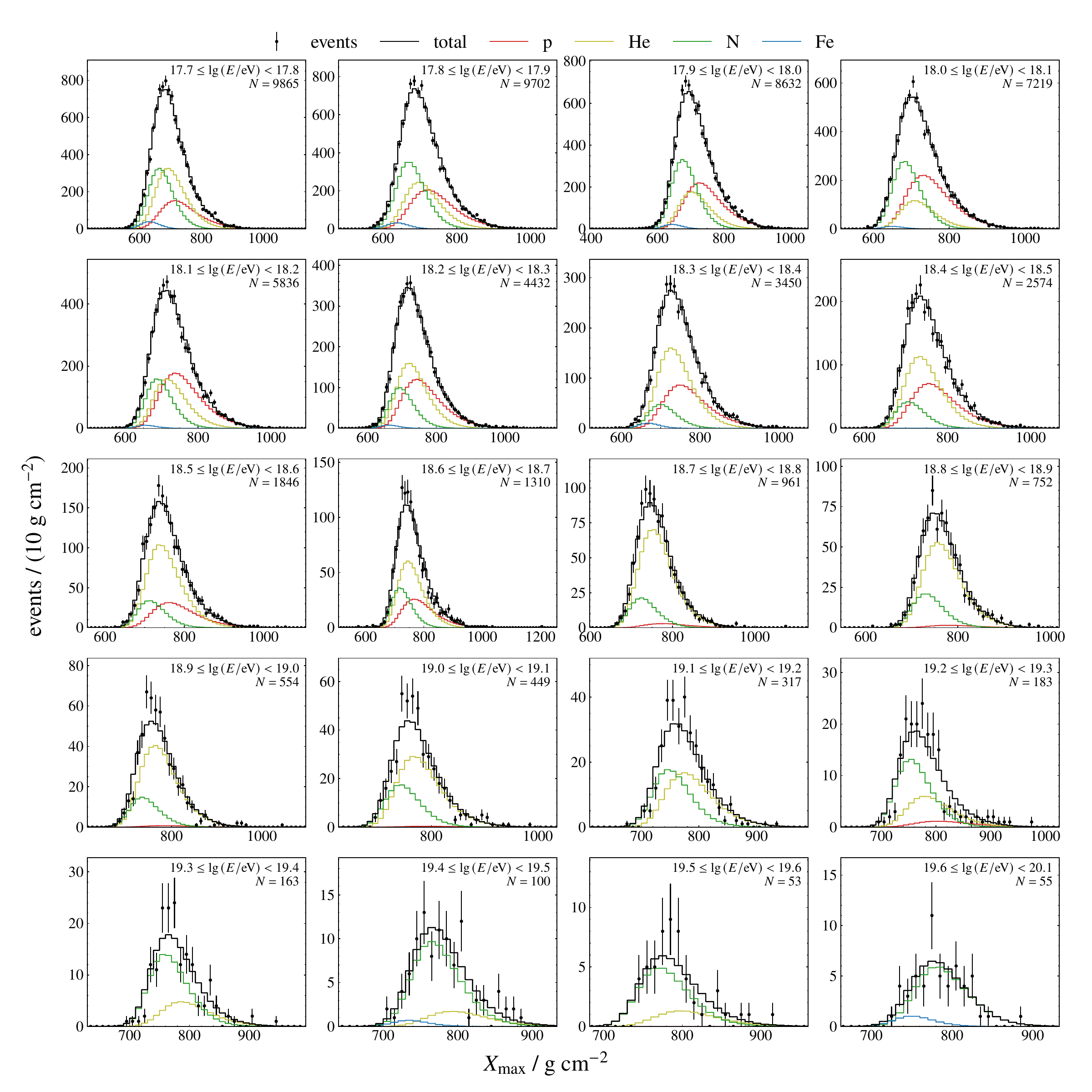}
    \caption{Fits of composition fractions of the mass
      groups (H, He, N, Fe) with \Sibylle in all energy bins: \Xmax distribution of the data (points with error bars) and composition templates (histograms).}
        \label{fig:XmaxDistrFitsAll_Sib}
\end{figure*}

\begin{figure*}[h!]
    \centering
     \includegraphics[width=\linewidth]{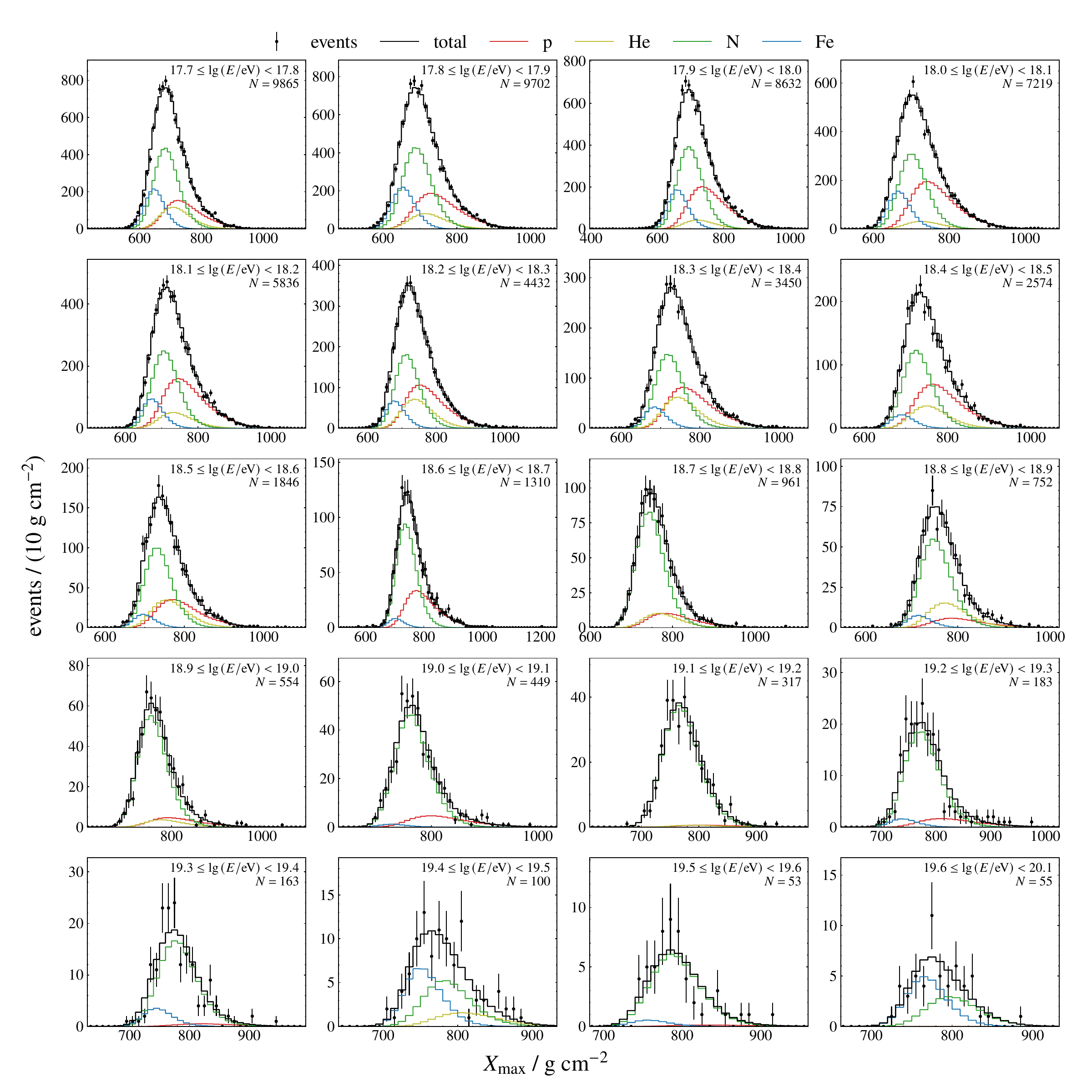}
    \caption{Same as \cref{fig:XmaxDistrFitsAll_Sib}, but for \EPOSR.}
    \label{fig:XmaxDistrFitsAll_EPOS}
\end{figure*}

The fitted $\Xmax$ distributions in each energy bin are shown in \cref{fig:XmaxDistrFitsAll_Sib,fig:XmaxDistrFitsAll_EPOS} for the \Sibylle and \EPOSR hadronic interaction models.

\clearpage

\section{\label{app:xmaxmoments}Moments of the \texorpdfstring{$\boldXmax$}{Xmax} distribution}
The bias- and resolution-corrected $\Xmax$ moments derived in this
analysis are listed in \cref{tab:Xmax_moments} together with their
statistical and systematic uncertainties.
\begin{table}[!th]
    \centering
    \caption{$\meanXmax$ and $\sigmaXmax$ with their statistical and systematic uncertainties. The third column shows the number of events $N$ in each energy bin.}
    \label{tab:Xmax_moments}
    \renewcommand{\arraystretch}{1.3}
    \setlength{\tabcolsep}{10pt}
 \begin{tabular}{c c r D{.}{.}{2.2}D{.}{.}{3.2}}
      \toprule
        $\lgE$ & $\left< \lgE\right>$ & $N$ & \multicolumn{1}{r}{$\meanXmax/(\gcm)$} & \multicolumn{1}{r}{$\sigmaXmax/(\gcm)$} \\
        \midrule
        $[17.7,17.8)$& 17.75&  9865&  703.1 \pm     0.7  ^{     +8.1}_{      -10.2} &   57.4 \pm     0.9  ^{     +1.7}_{      -1.6}\\
$[17.8,17.9)$& 17.85&  9702&  711.5 \pm     0.8  ^{     +7.9}_{      -10.2} &   59.3 \pm     1.0  ^{     +1.9}_{      -1.7}\\
$[17.9,18.0)$& 17.95&  8632&  719.7 \pm     0.8  ^{     +7.8}_{      -10.2} &   60.6 \pm     1.1  ^{     +2.0}_{      -1.8}\\
$[18.0,18.1)$& 18.05&  7219&  728.7 \pm     1.0  ^{     +7.6}_{      -10.2} &   62.1 \pm     1.2  ^{     +2.2}_{      -1.9}\\
$[18.1,18.2)$& 18.15&  5836&  737.9 \pm     1.0  ^{     +7.5}_{      -10.1} &   61.0 \pm     1.3  ^{     +2.4}_{      -2.1}\\
$[18.2,18.3)$& 18.25&  4432&  743.9 \pm     1.1  ^{     +7.4}_{      -9.9} &   59.3 \pm     1.4  ^{     +2.6}_{      -2.2}\\
$[18.3,18.4)$& 18.35&  3450&  750.4 \pm     1.1  ^{     +7.2}_{      -9.7} &   58.0 \pm     1.5  ^{     +2.3}_{      -2.0}\\
$[18.4,18.5)$& 18.45&  2574&  757.3 \pm     1.4  ^{     +7.1}_{      -9.4} &   56.0 \pm     1.9  ^{     +2.1}_{      -1.8}\\
$[18.5,18.6)$& 18.55&  1846&  758.9 \pm     1.4  ^{     +7.0}_{      -9.1} &   53.6 \pm     1.8  ^{     +1.9}_{      -1.7}\\
$[18.6,18.7)$& 18.65&  1310&  764.4 \pm     1.8  ^{     +7.0}_{      -8.8} &   57.7 \pm     2.4  ^{     +1.7}_{      -1.5}\\
$[18.7,18.8)$& 18.75&   961&  762.2 \pm     1.6  ^{     +6.9}_{      -8.4} &   45.3 \pm     2.0  ^{     +1.5}_{      -1.5}\\
$[18.8,18.9)$& 18.85&   752&  766.2 \pm     1.9  ^{     +6.9}_{      -8.1} &   44.4 \pm     2.2  ^{     +1.4}_{      -1.4}\\
$[18.9,19.0)$& 18.95&   554&  770.0 \pm     2.0  ^{     +6.9}_{      -7.8} &   38.5 \pm     2.4  ^{     +1.4}_{      -1.4}\\
$[19.0,19.1)$& 19.05&   449&  774.8 \pm     2.2  ^{     +6.9}_{      -7.5} &   42.6 \pm     2.8  ^{     +1.3}_{      -1.3}\\
$[19.1,19.2)$& 19.15&   317&  773.7 \pm     2.1  ^{     +6.9}_{      -7.3} &   32.9 \pm     2.6  ^{     +1.3}_{      -1.4}\\
$[19.2,19.3)$& 19.25&   183&  780.7 \pm     3.4  ^{     +7.0}_{      -7.2} &   42.7 \pm     4.7  ^{     +1.3}_{      -1.3}\\
$[19.3,19.4)$& 19.35&   163&  780.8 \pm     3.1  ^{     +7.0}_{      -7.0} &   36.5 \pm     4.1  ^{     +1.3}_{      -1.4}\\
$[19.4,19.5)$& 19.44&   100&  780.2 \pm     4.3  ^{     +7.1}_{      -7.0} &   37.2 \pm     5.1  ^{     +1.3}_{      -1.5}\\
$[19.5,19.6)$& 19.55&    53&  791.9 \pm     5.1  ^{     +7.1}_{      -6.9} &   36.2 \pm     6.4  ^{     +1.4}_{      -1.5}\\
$[19.6,\;\,\infty\;\,)$& 19.71&    55&  784.6 \pm     4.2  ^{     +7.2}_{      -6.9} &   29.0 \pm     4.0  ^{     +1.6}_{      -1.7}\\

    \end{tabular}
\end{table}

\newpage
\onecolumngrid
\section*{The Pierre Auger Collaboration}
\label{sec:PAcollaboration}
\noindent
\small\selectfont\setlength{\baselineskip}{0.9\baselineskip}\justifying
\begin{wrapfigure}[10]{l}{0.12\linewidth}
\vspace{-8ex}
\includegraphics[width=0.98\linewidth]{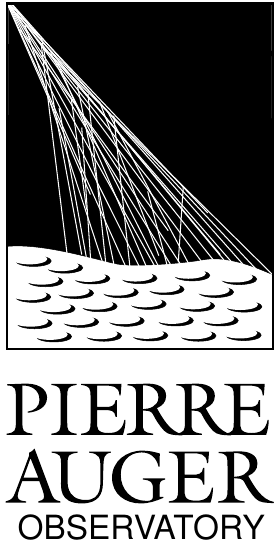}
\end{wrapfigure}
\begin{sloppypar}\noindent
A.~Abdul Halim$^{13}$,
P.~Abreu$^{67}$,
M.~Aglietta$^{50,49}$,
M.~Ahmed$^{31}$,
I.~Allekotte$^{1}$,
K.~Almeida Cheminant$^{75,74}$,
R.~Aloisio$^{42,43}$,
J.~Alvarez-Mu\~niz$^{73}$,
A.~Ambrosone$^{42,43}$,
J.~Ammerman Yebra$^{73}$
% ...

\begin{description}[labelsep=0.2em,align=right,labelwidth=0.7em,labelindent=0em,leftmargin=2em,noitemsep,before={\renewcommand\makelabel[1]{##1 }}]
\item[$^{1}$] Centro At\'omico Bariloche and Instituto Balseiro (CNEA-UNCuyo-CONICET), San Carlos de Bariloche, Argentina
\item[$^{2}$] Departamento de F\'\i{}sica and Departamento de Ciencias de la Atm\'osfera y los Oc\'eanos, FCEyN, Universidad de Buenos Aires and CONICET, Buenos Aires, Argentina
\item[$^{3}$] IFLP, Universidad Nacional de La Plata and CONICET, La Plata, Argentina
\item[$^{4}$] Instituto de Astronom\'\i{}a y F\'\i{}sica del Espacio (IAFE, CONICET-UBA), Buenos Aires, Argentina
\item[$^{5}$] Instituto de F\'\i{}sica de Rosario (IFIR) -- CONICET/U.N.R.\ and Facultad de Ciencias Bioqu\'\i{}micas y Farmac\'euticas U.N.R., Rosario, Argentina
\item[$^{6}$] Instituto de Tecnolog\'\i{}as en Detecci\'on y Astropart\'\i{}culas (CNEA, CONICET, UNSAM), and Universidad Tecnol\'ogica Nacional -- Facultad Regional Mendoza (CONICET/CNEA), Mendoza, Argentina
\item[$^{7}$] Instituto de Tecnolog\'\i{}as en Detecci\'on y Astropart\'\i{}culas (CNEA, CONICET, UNSAM), Buenos Aires, Argentina
\item[$^{8}$] International Center of Advanced Studies and Instituto de Ciencias F\'\i{}sicas, ECyT-UNSAM and CONICET, Campus Miguelete -- San Mart\'\i{}n, Buenos Aires, Argentina
\item[$^{9}$] Laboratorio Atm\'osfera -- Departamento de Investigaciones en L\'aseres y sus Aplicaciones -- UNIDEF (CITEDEF-CONICET), Argentina
\item[$^{10}$] Observatorio Pierre Auger, Malarg\"ue, Argentina
\item[$^{11}$] Observatorio Pierre Auger and Comisi\'on Nacional de Energ\'\i{}a At\'omica, Malarg\"ue, Argentina
\item[$^{12}$] Universidad Tecnol\'ogica Nacional -- Facultad Regional Buenos Aires, Buenos Aires, Argentina
\item[$^{13}$] Adelaide University, Adelaide, S.A., Australia
\item[$^{14}$] Universit\'e Libre de Bruxelles (ULB), Brussels, Belgium
\item[$^{15}$] Vrije Universiteit Brussels, Brussels, Belgium
\item[$^{16}$] Centro Brasileiro de Pesquisas Fisicas, Rio de Janeiro, RJ, Brazil
\item[$^{17}$] Centro Federal de Educa\c{c}\~ao Tecnol\'ogica Celso Suckow da Fonseca, Petropolis, Brazil
\item[$^{18}$] Universidade de S\~ao Paulo, Escola de Engenharia de Lorena, Lorena, SP, Brazil
\item[$^{19}$] Universidade de S\~ao Paulo, Instituto de F\'\i{}sica de S\~ao Carlos, S\~ao Carlos, SP, Brazil
\item[$^{20}$] Universidade de S\~ao Paulo, Instituto de F\'\i{}sica, S\~ao Paulo, SP, Brazil
\item[$^{21}$] Universidade Estadual de Campinas (UNICAMP), IFGW, Campinas, SP, Brazil
\item[$^{22}$] Universidade Estadual de Feira de Santana, Feira de Santana, Brazil
\item[$^{23}$] Universidade Federal do ABC, Santo Andr\'e, SP, Brazil
\item[$^{24}$] Universidade Federal do Paran\'a, Setor Palotina, Palotina, Brazil
\item[$^{25}$] Universidade Federal do Rio de Janeiro, Instituto de F\'\i{}sica, Rio de Janeiro, RJ, Brazil
\item[$^{26}$] Universidad de Medell\'\i{}n, Medell\'\i{}n, Colombia
\item[$^{27}$] Universidad Industrial de Santander, Bucaramanga, Colombia
\item[$^{28}$] Charles University, Faculty of Mathematics and Physics, Institute of Particle and Nuclear Physics, Prague, Czech Republic
\item[$^{29}$] Institute of Physics of the Czech Academy of Sciences, Prague, Czech Republic
\item[$^{30}$] Palacky University, Olomouc, Czech Republic
\item[$^{31}$] CNRS/IN2P3, IJCLab, Universit\'e Paris-Saclay, Orsay, France
\item[$^{32}$] Laboratoire de Physique Nucl\'eaire et de Hautes Energies (LPNHE), Sorbonne Universit\'e, Universit\'e de Paris, CNRS-IN2P3, Paris, France
\item[$^{33}$] Universit\'e Paris-Saclay, CNRS/IN2P3, IJCLab, Orsay, France
\item[$^{34}$] Bergische Universit\"at Wuppertal, Department of Physics, Wuppertal, Germany
\item[$^{35}$] Karlsruhe Institute of Technology (KIT), Institute for Experimental Particle Physics, Karlsruhe, Germany
\item[$^{36}$] Karlsruhe Institute of Technology (KIT), Institut f\"ur Prozessdatenverarbeitung und Elektronik, Karlsruhe, Germany
\item[$^{37}$] Karlsruhe Institute of Technology (KIT), Institute for Astroparticle Physics, Karlsruhe, Germany
\item[$^{38}$] RWTH Aachen University, III.\ Physikalisches Institut A, Aachen, Germany
\item[$^{39}$] TU Dortmund University, Department of Physics, Dortmund, Germany
\item[$^{40}$] Universit\"at Hamburg, II.\ Institut f\"ur Theoretische Physik, Hamburg, Germany
\item[$^{41}$] Universit\"at Siegen, Department Physik -- Experimentelle Teilchenphysik, Siegen, Germany
\item[$^{42}$] Gran Sasso Science Institute, L'Aquila, Italy
\item[$^{43}$] INFN Laboratori Nazionali del Gran Sasso, Assergi (L'Aquila), Italy
\item[$^{44}$] INFN, Sezione di Catania, Catania, Italy
\item[$^{45}$] INFN, Sezione di Lecce, Lecce, Italy
\item[$^{46}$] INFN, Sezione di Milano, Milano, Italy
\item[$^{47}$] INFN, Sezione di Napoli, Napoli, Italy
\item[$^{48}$] INFN, Sezione di Roma ``Tor Vergata'', Roma, Italy
\item[$^{49}$] INFN, Sezione di Torino, Torino, Italy
\item[$^{50}$] Osservatorio Astrofisico di Torino (INAF), Torino, Italy
\item[$^{51}$] Politecnico di Milano, Dipartimento di Scienze e Tecnologie Aerospaziali , Milano, Italy
\item[$^{52}$] Universit\`a del Salento, Dipartimento di Matematica e Fisica ``E.\ De Giorgi'', Lecce, Italy
\item[$^{53}$] Universit\`a dell'Aquila, Dipartimento di Scienze Fisiche e Chimiche, L'Aquila, Italy
\item[$^{54}$] Universit\`a di Catania, Dipartimento di Fisica e Astronomia ``Ettore Majorana``, Catania, Italy
\item[$^{55}$] Universit\`a di Milano, Dipartimento di Fisica, Milano, Italy
\item[$^{56}$] Universit\`a di Napoli ``Federico II'', Dipartimento di Fisica ``Ettore Pancini'', Napoli, Italy
\item[$^{57}$] Universit\`a di Palermo, Dipartimento di Fisica e Chimica ''E.\ Segr\`e'', Palermo, Italy
\item[$^{58}$] Universit\`a di Roma ``Tor Vergata'', Dipartimento di Fisica, Roma, Italy
\item[$^{59}$] Universit\`a Torino, Dipartimento di Fisica, Torino, Italy
\item[$^{60}$] Benem\'erita Universidad Aut\'onoma de Puebla, Puebla, M\'exico
\item[$^{61}$] Unidad Profesional Interdisciplinaria en Ingenier\'\i{}a y Tecnolog\'\i{}as Avanzadas del Instituto Polit\'ecnico Nacional (UPIITA-IPN), M\'exico, D.F., M\'exico
\item[$^{62}$] Universidad Aut\'onoma de Chiapas, Tuxtla Guti\'errez, Chiapas, M\'exico
\item[$^{63}$] Universidad Michoacana de San Nicol\'as de Hidalgo, Morelia, Michoac\'an, M\'exico
\item[$^{64}$] Universidad Nacional Aut\'onoma de M\'exico, M\'exico, D.F., M\'exico
\item[$^{65}$] Institute of Nuclear Physics PAN, Krakow, Poland
\item[$^{66}$] University of \L{}\'od\'z, Faculty of High-Energy Astrophysics,\L{}\'od\'z, Poland
\item[$^{67}$] Laborat\'orio de Instrumenta\c{c}\~ao e F\'\i{}sica Experimental de Part\'\i{}culas -- LIP and Instituto Superior T\'ecnico -- IST, Universidade de Lisboa -- UL, Lisboa, Portugal
\item[$^{68}$] ``Horia Hulubei'' National Institute for Physics and Nuclear Engineering, Bucharest-Magurele, Romania
\item[$^{69}$] Institute of Space Science, Bucharest-Magurele, Romania
\item[$^{70}$] Center for Astrophysics and Cosmology (CAC), University of Nova Gorica, Nova Gorica, Slovenia
\item[$^{71}$] Experimental Particle Physics Department, J.\ Stefan Institute, Ljubljana, Slovenia
\item[$^{72}$] Universidad de Granada and C.A.F.P.E., Granada, Spain
\item[$^{73}$] Instituto Galego de F\'\i{}sica de Altas Enerx\'\i{}as (IGFAE), Universidade de Santiago de Compostela, Santiago de Compostela, Spain
\item[$^{74}$] IMAPP, Radboud University Nijmegen, Nijmegen, The Netherlands
\item[$^{75}$] Nationaal Instituut voor Kernfysica en Hoge Energie Fysica (NIKHEF), Science Park, Amsterdam, The Netherlands
\item[$^{76}$] Stichting Astronomisch Onderzoek in Nederland (ASTRON), Dwingeloo, The Netherlands
\item[$^{77}$] Case Western Reserve University, Cleveland, OH, USA
\item[$^{78}$] Colorado School of Mines, Golden, CO, USA
\item[$^{79}$] Department of Physics and Astronomy, Lehman College, City University of New York, Bronx, NY, USA
\item[$^{80}$] Michigan Technological University, Houghton, MI, USA
\item[$^{81}$] New York University, New York, NY, USA
\item[$^{82}$] University of Chicago, Enrico Fermi Institute, Chicago, IL, USA
\item[$^{83}$] University of Delaware, Department of Physics and Astronomy, Bartol Research Institute, Newark, DE, USA
\item[] -----
\item[$^{a}$] Max-Planck-Institut f\"ur Radioastronomie, Bonn, Germany
\item[$^{b}$] also at Kapteyn Institute, University of Groningen, Groningen, The Netherlands
\item[$^{c}$] School of Physics and Astronomy, University of Leeds, Leeds, United Kingdom
\item[$^{d}$] Fermi National Accelerator Laboratory, Fermilab, Batavia, IL, USA (Affiliation for identification purposes only)
\item[$^{e}$] Pennsylvania State University, University Park, PA, USA
\item[$^{f}$] Colorado State University, Fort Collins, CO, USA
\item[$^{g}$] Louisiana State University, Baton Rouge, LA, USA
\item[$^{h}$] Universit\'e Grenoble Alpes, CNRS, Grenoble Institute of Engineering, LPSC-IN2P3, Grenoble, France
\item[$^{i}$] now at Graduate School of Science, Osaka Metropolitan University, Osaka, Japan
\item[$^{j}$] Institut universitaire de France (IUF), France
\end{description}

\end{sloppypar}

\end{document}